\documentclass[twocolumn,tighten]{aastex6} 
\usepackage{rotating}

\def\m31{{M\,31}}

\def\chandra{{\it Chandra~}}
\def\chandrak{{\it Chandra}}
\def\swift{{\it Swift~}}
\def\swiftk{{\it Swift}}
\def\xmm{{\it XMM-Newton~}}

\def\nova{{M31N\,2008-12a~}}
\def\novak{{M31N\,2008-12a}}

\newcommand{\nh}{\hbox{$N_{\rm H}$}~}

\newcommand{\hcm}[1]{$\times 10^{#1}$\,cm$^{-2}$}
\newcommand{\ohcm}[1]{$10^{#1}$\,cm$^{-2}$}

\newcommand{\cts}[1]{$\times 10^{#1}$ ct s$^{-1}$}

\newcommand{\tpower}[1]{$\times 10^{#1}$}
\newcommand{\power}[1]{$10^{#1}$}

\newcommand{\eton}{t_{\mbox{\small{on}}}}
\newcommand{\etoff}{t_{\mbox{\small{off}}}}

\defcitealias{2014A&A...563L...9D}{DWB14}
\defcitealias{2014A&A...563L...8H}{HND14}
\defcitealias{2014ApJ...786...61T}{TBW14}
\defcitealias{2015A&A...580A..45D}{DHS15}
\defcitealias{2015A&A...580A..46H}{HND15}
\defcitealias{2015A&A...582L...8H}{HDK15}

\newcommand{\oonek}{\citetalias{2014A&A...563L...9D}}
\newcommand{\xonek}{\citetalias{2014A&A...563L...8H}}
\newcommand{\ponek}{\citetalias{2014ApJ...786...61T}}
\newcommand{\otwok}{\citetalias{2015A&A...580A..45D}}
\newcommand{\xtwok}{\citetalias{2015A&A...580A..46H}}
\newcommand{\halfk}{\citetalias{2015A&A...582L...8H}}

\begin{document} 

\received{2016 July 27}
\revised{2016 August 26}
\accepted{2016 August 29}
\slugcomment{The Astrophysical Journal, \today, Accepted version}

\title{\novak\ --- the remarkable recurrent nova in \m31:\\ Pan-Chromatic observations of the 2015 eruption.}
\shorttitle{\novak: The 2015 eruption}
\shortauthors{Darnley \& Henze et al.\ 2016}

\author{	
M. J. Darnley,\altaffilmark{1}
M. Henze,\altaffilmark{2}
M. F. Bode,\altaffilmark{1}
I. Hachisu,\altaffilmark{3}
M. Hernanz,\altaffilmark{2}
K. Hornoch,\altaffilmark{4}
R. Hounsell,\altaffilmark{5}
M. Kato,\altaffilmark{6}\\
J.-U. Ness,\altaffilmark{7}
J. P. Osborne,\altaffilmark{8}
K. L. Page,\altaffilmark{8}
V. A. R. M. Ribeiro,\altaffilmark{9}
P. Rodr\'\i guez-Gil,\altaffilmark{10,11}
A. W. Shafter,\altaffilmark{12}\\
M. M. Shara,\altaffilmark{13}
I. A. Steele,\altaffilmark{1}
S. C. Williams,\altaffilmark{14,1}
A. Arai,\altaffilmark{15}
I. Arcavi,\altaffilmark{16,17}
E. A. Barsukova,\altaffilmark{18}
P. Boumis,\altaffilmark{19}\\
T. Chen,\altaffilmark{20}
S. Fabrika,\altaffilmark{18,21}
J. Figueira,\altaffilmark{22,23}
X. Gao,\altaffilmark{24}
N. Gehrels,\altaffilmark{25}
P. Godon,\altaffilmark{26}
V. P. Goranskij,\altaffilmark{27}
D. J. Harman,\altaffilmark{1}\\
D. H. Hartmann,\altaffilmark{28}
G. Hosseinzadeh,\altaffilmark{16,29}
J. Chuck Horst,\altaffilmark{12}
K. Itagaki,\altaffilmark{30}
J. Jos\'{e},\altaffilmark{22,23}
F. Kabashima,\altaffilmark{31}
A. Kaur,\altaffilmark{28}\\
N. Kawai,\altaffilmark{32}
J. A. Kennea,\altaffilmark{33}
S. Kiyota,\altaffilmark{34}
H. Ku\v{c}\'akov\'a,\altaffilmark{35}
K. M. Lau,\altaffilmark{20}
H. Maehara,\altaffilmark{36}
H. Naito,\altaffilmark{37}
K. Nakajima,\altaffilmark{37,38}\\
K. Nishiyama,\altaffilmark{31}
T. J. O'Brien,\altaffilmark{39}
R. Quimby,\altaffilmark{12}
G. Sala,\altaffilmark{22,23}
Y. Sano,\altaffilmark{37,40}
E. M. Sion,\altaffilmark{26}
A. F. Valeev,\altaffilmark{18,21}\\
F. Watanabe,\altaffilmark{37}
M. Watanabe,\altaffilmark{41,42}
B. F. Williams,\altaffilmark{43}
Z. Xu\altaffilmark{44}
}
\email{M.J.Darnley@ljmu.ac.uk}
\email{henze@ice.cat}

\altaffiltext{1}{Astrophysics Research Institute, Liverpool John Moores University, IC2 Liverpool Science Park, Liverpool, L3 5RF, UK} 
\altaffiltext{2}{Institut de Ci\`encies de l'Espai (CSIC-IEEC), Campus UAB, C/Can Magrans s/n, 08193 Cerdanyola del Valles, Spain} 
\altaffiltext{3}{Department of Earth Science and Astronomy, College of Arts and Sciences, University of Tokyo, 3-8-1 Komaba, Meguro-ku, Tokyo 153-8902, Japan}  
\altaffiltext{4}{Astronomical Institute, Academy of Sciences, CZ-251 65 Ond\v{r}ejov, Czech Republic}  
\altaffiltext{5}{Astronomy Department, University of Illinois at Urbana-Champaign, 1002 W.\ Green Street, Urbana, IL 61801, USA}  
\altaffiltext{6}{Department of Astronomy, Keio University, Hiyoshi, Yokohama 223-8521, Japan}  
\altaffiltext{7}{European Space Astronomy Centre, Camino Bajo del Castillo s/n, Urb.\ Villafranca del Castillo, 28692 Villanueva de la Ca\~{n}ada, Madrid, Spain}  
\altaffiltext{8}{X-Ray and Observational Astronomy Group, Department of Physics \& Astronomy, University of Leicester, LE1 7RH, UK}  
\altaffiltext{9}{Department of Astrophysics/IMAPP, Radboud University, PO Box 9010, 6500 GL Nijmegen, The Netherlands}  
\altaffiltext{10}{Instituto de Astrof\'\i sica de Canarias, V\'\i a L\'actea, s/n, La Laguna, E-38205, Santa Cruz de Tenerife, Spain}  
\altaffiltext{11}{Departamento de Astrof\'\i sica, Universidad de La Laguna, La Laguna, E-38206, Santa Cruz de Tenerife, Spain}  
\altaffiltext{12}{Department of Astronomy, San Diego State University, San Diego, CA 92182, USA}  
\altaffiltext{13}{American Museum of Natural History, 79th Street and Central Park West, New York, NY 10024, USA}  
\altaffiltext{14}{Physics Department, Lancaster University, Lancaster, LA1 4YB, UK}  
\altaffiltext{15}{Koyama Astronomical Observatory, Kyoto Sangyo University, Motoyama, Kamigamo, Kita-ku, Kyoto, Kyoto 603-8555, Japan}  
\altaffiltext{16}{Las Cumbres Observatory Global Telescope Network, 6740 Cortona Dr., Suite 102, Goleta, CA 93117, USA}  
\altaffiltext{17}{Kavli Institute for Theoretical Physics, University of California, Santa Barbara, CA 93106-4030, USA}  
\altaffiltext{18}{Special Astrophysical Observatory of Russian Academy of Sciences, Nizhnij Arkhyz, Karachai-Cherkessian Republic 369167, Russia}  
\altaffiltext{19}{Institute for Astronomy, Astrophysics, Space Applications and Remote Sensing, National Observatory of Athens, 15236 Penteli, Greece}  
\altaffiltext{20}{Corona Borealis Observatory, Kunsha Town, Ngari, Tibet, PR China}  
\altaffiltext{21}{Kazan Federal University, Kazan 420008, Russia}  
\altaffiltext{22}{Departament de F\'\i sica, EUETIB, Universitat Polit\`{e}cnica de Catalunya, c/ Compte d'Urgell 187, 08036 Barcelona, Spain}  
\altaffiltext{23}{Institut d'Estudis Espacials de Catalunya, c/ Gran Capit\`{a} 2-4, Ed. Nexus-201, 08034, Barcelona, Spain}  
\altaffiltext{24}{Xingming Observatory, Mt. Nanshan, Urumqi, Xinjiang, PR China}  
\altaffiltext{25}{Astrophysics Science Division, NASA Goddard Space Flight Center, Greenbelt, MD 20771, USA}  
\altaffiltext{26}{Department of Astronomy and Astrophysics, Villanova University, 800 Lancaster Avenue, Villanova, PA 19085, USA}  
\altaffiltext{27}{Sternberg Astronomical Institute, Moscow University, Universitetsky
Prospect, 13, Moscow 119899, Russia}  
\altaffiltext{28}{Department of Physics and Astronomy, Clemson University, Clemson, SC 29634, USA}  
\altaffiltext{29}{Department of Physics, University of California, Santa Barbara, CA 93106-9530, USA}  
\altaffiltext{30}{Itagaki Astronomical Observatory, Teppo, Yamagata 990-2492, Japan}  
\altaffiltext{31}{Miyaki-Argenteus Observatory, Miyaki, Saga 840-1102, Japan}  
\altaffiltext{32}{Department of Physics, Tokyo Institute of Technology, 2-12-1 Ookayama,
Meguro-ku, Tokyo 152-8551, Japan}  
\altaffiltext{33}{Department of Astronomy and Astrophysics, 525 Davey Lab, Pennsylvania State University, University Park, PA 16802, USA}  
\altaffiltext{34}{Variable Stars Observers League in Japan (VSOLJ), 7-1 Kitahatsutomi, Kamagaya 273-0126, Japan}  
\altaffiltext{35}{Astronomical Institute of the Charles University, Faculty of Mathemathics and Physics, V Hole\v{s}ovi\v{c}k\'ach 2, 180 00 Praha 8, Czech Republic}  
\altaffiltext{36}{Okayama Astrophysical Observatory, NAOJ, NINS, 3037-5 Honjo, Kamogata, Asakuchi, Okayama 719-0232, Japan}  
\altaffiltext{37}{Nayoro Observatory, 157-1 Nisshin, Nayoro, Hokkaido 096-0066, Japan}  
\altaffiltext{38}{Rikubetsu Space and Earth Science Museum, Uenbetsu, Rikubetsu-cho, Ashoro, Hokkaido 089-4301, Japan}  
\altaffiltext{39}{Jodrell Bank Centre for Astrophysics, Alan Turing Building, University of Manchester, Manchester, M13 9PL, UK}  
\altaffiltext{40}{Observation and Data Center for Cosmosciences, Faculty of Science,
Hokkaido University, Kita-ku, Sapporo, Hokkaido 060-0810, Japan}  
\altaffiltext{41}{Department of Cosmosciences, Hokkaido University, Kita 10, Nishi 8, Kita-ku, Sapporo, Hokkaido 060-0810, Japan}  
\altaffiltext{42}{Department of Applied Physics, Okayama University of Science, 1-1 Ridai-cho, Kita-ku, Okayama, Okayama 700-0005, Japan}   
\altaffiltext{43}{Department of Astronomy, Box 351580, University of Washington, Seattle, WA 98195, USA}  
\altaffiltext{44}{Nanjing Putian Telecommunications Co., Ltd., 1 Putian Rd, Qinhuai, Nanjing, 210012 Jiangsu, PR China}  

\begin{abstract}
The Andromeda Galaxy recurrent nova \novak\ had been observed in eruption ten times, including yearly eruptions from 2008--2014.  With a measured recurrence period of $P_{\mathrm{rec}}=351\pm13$\,days (we believe the true value to be half of this) and a white dwarf very close to the Chandrasekhar limit, \novak\ has become the leading pre-explosion supernova type Ia progenitor candidate.  Following multi-wavelength follow-up observations of the 2013 and 2014 eruptions, we initiated a campaign to ensure early detection of the predicted 2015 eruption, which triggered ambitious ground and space-based follow-up programs.  In this paper we present the 2015 detection; visible to near-infrared photometry and visible spectroscopy; and ultraviolet and X-ray observations from the \swift observatory.  The LCOGT 2\,m (Hawaii) discovered the 2015 eruption, estimated to have commenced at Aug.\ $28.28\pm0.12$ UT.  The 2013--2015 eruptions are remarkably similar at all wavelengths.  New early spectroscopic observations reveal short-lived emission from material with velocities $\sim13000$\,km\,s$^{-1}$, possibly collimated outflows.  Photometric and spectroscopic observations of the eruption provide strong evidence supporting a red giant donor.  An apparently stochastic variability during the early super-soft X-ray phase was comparable in amplitude and duration to past eruptions, but the 2013 and 2015 eruptions show evidence of a brief flux dip during this phase.  The multi-eruption \swiftk/XRT spectra show tentative evidence of high-ionization emission lines above a high-temperature continuum.  Following \citet{2015A&A...582L...8H}, the updated recurrence period based on all known eruptions is $P_\mathrm{rec}=174\pm10$\,d, and we expect the next eruption of \nova to occur around mid-Sep.\ 2016.
\end{abstract}


\section{Introduction}

Novae are the powerful eruptions resulting from a brief thermonuclear runaway (TNR) occurring at the base of the surface layer of an accreting white dwarf (WD; see \citealp{1949AnAp...12..281S,1951AnAp...14..294S,1959ApJ...130..916C,1957IAUS....3...77G,1972ApJ...176..169S}, and \citealp{Sta08,2016PASP..128e1001S,JS08,Jos16}, for recent reviews).  Belonging to the group of cataclysmic variables \citep{1949ApJ...109...81S,1954ApJ...120..377J,1964ApJ...139..457K}, the companion star in these interacting close-binary systems transfers hydrogen-rich material to the WD usually via an accretion disk around the WD.  The TNR powers an explosive ejection of the accreted material, with a rapidly expanding pseudo-photosphere initially increasing the visible luminosity of the system by up to eight orders of magnitude \citep[see][for recent reviews]{2008clno.book.....B,2010AN....331..160B,2014ASPC..490.....W}.  Following the TNR the nuclear fusion enters a period of short-lived, approximately steady-state, burning until the accreted fuel is exhausted, partly because it has been ejected and partly as that remaining has been burned to helium \citep{1978A&A....62..339P}.  As the optical depth of the expanding ejecta becomes progressively smaller, the pseudo-photosphere begins to recede back toward the WD surface, subsequently shifting the peak of the emission back to higher energies until ultimately a supersoft X-ray source (SSS) may emerge \citep[see, for example,][]{2006ApJS..167...59H,2008ASPC..401..139K,2015JHEAp...7..117O}.  The `turn-off' of the SSS indicates the end of the nuclear burning, after which the system eventually returns to its quiescent state.

All nova eruptions are inherently recurrent, with the WD and companion surviving each eruption, and accretion reestablishing or continuing shortly afterwards.  By definition, the Classical Novae (CNe) have had a single {\it observed} eruption, whereas Recurrent Novae (RNe) have been detected in eruption at least twice.  Observed intervals between eruptions range from $\sim1$\,yr \citep[for \novak]{2014A&A...563L...9D} up to 98\,yrs \citep[for V2487~Ophiuchi]{2010ApJS..187..275S}, with the shortest predicted recurrence period -- albeit derived from incomplete observational data -- being just six months \citep{2015A&A...582L...8H}.  The theoretical limits on the recurrence period of all novae may be as short as 50\,days \citep{2015MNRAS.446.1924H} or even 25\,days \citep{2016ApJ...824...22H}\footnote{In both the \citet{2015MNRAS.446.1924H} and \citet{2016ApJ...824...22H} studies, accretion is assumed to completely stop during the eruption period.}, and as high as mega-years \citep[see, for example,][]{1985ApJ...291..136S,1994ApJ...424..319K,2005ApJ...623..398Y}.  The shorter recurrence periods are driven by a combination of a high mass WD and a high mass accretion rate.  Such high accretion rates are typically driven by an evolved companion star, such as a Roche lobe overflowing sub-giant star (SG-novae; also the U~Scorpii type of RNe), or the stellar wind from a red giant companion \citep[RG-novae; symbiotic novae; or the RS~Ophiuchi type RNe; see][for recent reviews]{2012ApJ...746...61D,2014ASPC..490...49D}.

With the most luminous novae reaching peak visible magnitudes $M_{V}<-10$ \citep{2009ApJ...690.1148S,2016WillLN}, novae are readily observable out to the distance of the Virgo Cluster and beyond \citep[see, for example,][]{2015ApJ...811...34C,2016arXiv160200758S}.  But it is the nearby Andromeda Galaxy (\m31), with an annual nova rate of $65^{+16}_{-15}$\,yr$^{-1}$ \citep{2006MNRAS.369..257D}, that provides the leading laboratory for the study of galaxy-wide nova populations \citep[see, for example,][]{1987ApJ...318..520C,1990ApJ...356..472C,2001ApJ...563..749S,2004MNRAS.353..571D,2006MNRAS.369..257D,2008A&A...477...67H,2010A&A...523A..89H,2011A&A...533A..52H,2014A&A...563A...2H,2011ApJ...727...50S,2011ApJ...734...12S,2015ApJS..216...34S,2014ApJS..213...10W,2016ApJ...817..143W}.  Since the discovery of the first \m31 nova by \citet[also spectroscopically confirmed]{1917PASP...29..210R} and the pioneering work of \citet{1929ApJ....69..103H} more than 1000 nova candidates have been discovered \citep[see][and their on-line database\footnote{\url{http://www.mpe.mpg.de/~m31novae/opt/m31/index.php}}]{2007A&A...465..375P,2010AN....331..187P} with over 100 now spectroscopically confirmed \citep[see, for example,][]{2011ApJ...734...12S}.

Recently, pioneering X-ray surveys with \xmm and \chandra have revealed that novae are the major class of SSSs in \m31 \citep[][]{2005A&A...442..879P,2007A&A...465..375P}. A dedicated multi-year follow-up program with the same telescopes studied the multi-wavelength population properties of \m31 novae in detail \citep[][]{2010A&A...523A..89H,2011A&A...533A..52H,2014A&A...563A...2H}. A major result of this work was the discovery of strong correlations between various observable parameters: indicating that novae with a faster visible decline tend to show a shorter SSS phase with a higher temperature \citep{2014A&A...563A...2H}. This is consistent with the trends seen in Galactic novae \citep[see][]{2011ApJS..197...31S}. Theoretical models indicate that a shorter SSS phase corresponds to a higher mass WD \citep[e.g.][]{2006ApJS..167...59H,2010ApJ...709..680H,2013ApJ...777..136W}. Thus, the \m31 nova population provides a unique framework within which to understand the properties of individual novae and their ultimate fate.

Supernovae Type Ia (SNe\,Ia) are the outcome of a thermonuclear explosion of a carbon--oxygen (CO) WD as it reaches and surpasses the \citet{1931ApJ....74...81C} mass limit \citep[see, for example,][]{1973ApJ...186.1007W,1999ApJ...522..487H,1999ApJ...519..314H,2000ARA&A..38..191H}.  An accreting oxygen--neon (ONe) WD however, is predicted to undergo electron capture and subsequent neutron star formation \citep[see, for example,][]{1996ApJ...459..701G}.  It seems increasingly likely that there is not a single progenitor pathway producing all observed SNe\,Ia; but a combination of different double-degenerate (WD--WD) and single-degenerate (SD; WD--donor) binary systems, with the metallicity and age of the parent stellar population possibly determining the weighting of those pathways \cite[see, for example,][]{2014ARA&A..52..107M}.  Novae, in particular the RNe with their already high mass WDs, are potentially a leading SD pathway.  Recent studies have indicated that the mass of the WD can indeed grow over time in RN systems \citep[see, for example,][]{2008NewAR..52..386H,2012BASI...40..419S,2016ApJ...819..168H}.  A number of questions remain over the size of their contribution to the SN\,Ia rate; including the composition of the WD in the RN systems, the feasibility of growing a CO WD from their formation mass to the Chandrasekhar limit, and the size of the population of high mass WD novae.  Of course, the lack of observational signatures of hydrogen following {\it most} SN\,Ia explosions still provides a significant hurdle for the SD scenario \citep[see, for example,][]{2012NewAR..56..122W,2014ARA&A..52..107M}.  But the unmistakable presence of hydrogen in PTF\,11kx \citep{2012Sci...337..942D} and the possible presence of hydrogen in SN\,2013ct \citep{2016MNRAS.457.3254M} supports the view that at least some SN\,Ia arise in SD systems.

At the time of writing, there have been around 450 detected eruptions of nova candidates in the Milky Way \citep{2014ASPC..490...49D} from which just ten confirmed RN systems are known \citep{2010ApJS..187..275S} accounting for $\sim3\%$ of known Galactic nova systems or $\sim9\%$  of detected Galactic eruptions.  A number of recent detailed studies of archival observations have uncovered new results relating to the RN populations of both the Milky Way and \m31, these are summarized below:

\citet{2014ApJ...788..164P} used a combination of three different methods to estimate that the RN nova population (essentially $10\le P_{\mathrm{rec}}\le100$\,yrs; A.~Pagnotta, priv.\ comm.) of the Milky Way is $25\pm10\%$ of the Galactic nova population.  However, the range of methodologies employed predicted a wide range of contributions, from $9 - 38\%$, with the authors themselves indicating that the statistical errors were ``likely being much too small'' \citep{2014ApJ...788..164P}.

\citet{2015ApJS..216...34S} uncovered multiple eruptions of 16 RN systems in \m31, the subsequent analysis predicted an historic \m31 RN discovery efficiency of just 10\% and that as many as 33\% of \m31 nova eruptions may arise from RN systems ($P_{\mathrm{rec}}\le100$\,yrs).  

\citet{2014ApJS..213...10W,2016ApJ...817..143W} employed a different approach, by recovering the progenitor systems of 11 \m31 RG-novae, they determined that $30^{+13}_{-10}\%$ of all \m31 nova eruptions occur in RG-nova systems, a sub-population that also appears strongly associated with the \m31 disk.  

Additionally, other recent results for the Milky Way \citep{2016arXiv160602358S}, Magellanic Clouds \citep{2016ApJS..222....9M}, \m31 \citep{2016MNRAS.458.2916C,2016MNRAS.455..668S}, and M\,87 \citep{2016arXiv160200758S} all indicate that the luminosity specific nova rate \citep[see, for example,][]{1990AJ.....99.1079C} may be much higher than previously thought.   Together, all these results  boost the size of the available `pool' of novae that may contribute to the SN\,Ia population by a factor of $>5$.  

\section{A remarkable recurrent nova}\label{sec:thenova}

\novak\ was originally discovered far out in the disk of \m31 in visible observations while undergoing an eruption in 2008 \citep{2008Nis}.  Subsequent eruptions were discovered in each of the next six years; 2009 \citep[first reported in 2013]{2014ApJ...786...61T}, 2010 \citep[only recovered in 2015]{2015A&A...582L...8H}, 2011 \citep{2011Kor}, 2012 \citep{2012Nis}, 2013 \citep{2013ATel.5607....1T}, and 2014 \citep{2014ATel.6527....1D}, see Table~\ref{eruption_history} for a summary of all detected eruptions.  \citet[hereafter HDK15]{2015A&A...582L...8H} calculated that the mean recurrence period, based only on these seven consecutive eruptions, is $P_{\mathrm{rec}}=351\pm13$\,days.

\begin{table*}
\caption{List of all observed eruptions of \novak.\label{eruption_history}}
\begin{center}
\begin{tabular}{lllll}
\hline\hline
Eruption date\tablenotemark{a} & SSS-on date\tablenotemark{b} & Days since & Detection wavelength & References\\
(UT) & (UT) & last eruption\tablenotemark{c} & (observatory) & \\
\hline
(1992 Jan.\ 28) & 1992 Feb.\ 03 & \nodata & X-ray ({\it ROSAT}) & 1, 2 \\
(1993 Jan.\ 03) & 1993 Jan.\ 09 & 341 & X-ray ({\it ROSAT}) & 1, 2 \\
(2001 Aug.\ 27) & 2001 Sep.\ 02 & \nodata & X-ray ({\it Chandra}) & 2, 3 \\
2008 Dec.\ 25 & \nodata & \nodata & Visible (Miyaki-Argenteus) & 4 \\
2009 Dec.\ 02 & \nodata & 342 & Visible (PTF) & 5 \\
2010 Nov.\ 19 & \nodata & 352 & Visible (Miyaki-Argenteus) & 2 \\
2011 Oct.\ 22.5 & \nodata & 337.5 & Visible (ISON-NM) & 5, 6--8 \\
2012 Oct.\ 18.7 & $<2012$ Nov.\ 06.45 & 362.2 & Visible (Miyaki-Argenteus) & 8--11 \\
2013 Nov.\ $26.95\pm0.25$ & $\le2013$ Dec.\ 03.03 & 403.5 & Visible (iPTF); UV/X-ray (\swiftk) & 5, 8, 11--14 \\
2014 Oct.\ $02.69\pm0.21$ & 2014 Oct.\ $08.6\pm0.5$ & $309.8\pm0.7$ & Visible (LT); UV/X-ray (\swiftk) & 8, 15 \\
2015 Aug.\ $28.28\pm0.12$ & 2015 Sep.\ $02.9\pm0.7$ & $329.6\pm0.3$ & Visible (LCOGT); UV/X-ray (\swiftk) & 14, 16--18\\
\hline
\end{tabular}
\end{center}
\catcode`\&=12
\tablenotetext{a}{Estimated times of the visible eruption, those in parentheses have been extrapolated from the X-ray data \protect \citep[see][]{2015A&A...582L...8H}. The rapid evolution of the eruption (see Figure \ref{optical_lc}) limits any  associated uncertainties to just a few days.}
\tablenotetext{b}{Turn-on time of the SSS emission. The ROSAT detections from 1992 and 1993 permit accurate estimates of SSS-on.  There was only a single \chandra data point obtained on 2001 Sep.\ 08, sometime during the 12\,d SSS phase (cf.\ Figure~\ref{fig:xray_lc}), as such we take Sep.\ 08 as the mid point of the SSS phase (with an uncertainty of $\pm6$\,d) to extrapolate the eruption date and SSS-on.}
\tablenotetext{c}{Time since last eruption only quoted when consecutive detections occurred in consecutive years, under the assumption of $P_\mathrm{rec}\simeq1$\,year.  Time is taken as the period between estimated eruption dates.}
\tablecomments{Modified and updated version of Table~1 from \protect \citet{2014ApJ...786...61T}, \protect \citet{2015A&A...580A..45D}, and \protect \citet{2015A&A...582L...8H}.}
\tablerefs{(1)~\citet{1995ApJ...445L.125W}, (2)~\citet{2015A&A...582L...8H}, (3)~\citet{2004ApJ...609..735W}, (4)~\citet{2008Nis}, (5)~\citet{2014ApJ...786...61T}, (6)~\citet{2011Kor}, (7)~\citet{2011ATel.3725....1B}, (8)~\citet{2015A&A...580A..45D}, (9)~\citet{2012Nis}, (10)~\citet{2012ATel.4503....1S}, (11)~\citet{2014A&A...563L...8H}, (12)~\citet{2013ATel.5607....1T}, (13)~\citet{2014A&A...563L...9D}, (14)~this paper, (15)~\citet{2015A&A...580A..46H}, (16)~\citet{2015ATel.7964....1D}, (17)~\citet{2015ATel.7965....1D}, (18)~\citet{2015ATel.7984....1H}.}
\end{table*}

\citet[hereafter \xonek]{2014A&A...563L...8H} and \citet[hereafter \ponek]{2014ApJ...786...61T} independently uncovered earlier eruptions from 1992, 1993, and 2001.  These were based on archival X-ray data from ROSAT and \textit{Chandra} first reported by  \citet{1995ApJ...445L.125W} and \citet{2004ApJ...609..735W}, respectively.  Using these additional eruptions, \halfk\ predicted that the actual mean recurrence period of \novak\ is only $P_{\mathrm{rec}}=175\pm11$\,days, and subsequently predicted that the next {\it observable} eruption would occur between early September and mid October 2015.

The shortest observed inter-eruption period seen in the Galactic nova population is eight years between the 1979 and 1987 eruptions of U~Scorpii \citep[respectively]{1979IAUC.3341....1B,1987IAUC.4395....1O}.  The Large Magellanic Cloud recurrent nova LMCN~1968-12a \citep{1991ApJ...370..193S} may have an eruption-cycle of only six years \citep{2016ATel.8587....1D}. Furthermore, a five-year cycle has been observed for the \m31 nova M31N\,1963-09c \citep{2015ApJS..216...34S,2015ATel.8234....1W}, and a four-year cycle for M31N\,1997-11k \citep{2015ApJS..216...34S}.  Nevertheless, the discovery of a nova with a recurrence period as short as one year, or even six months, presents an unprecedented and significant advance over any of these objects.  The implications of such a short recurrence period suggest the presence of a WD with a mass very close to the Chandrasekhar mass \citep[see, for example,][]{1995ApJ...445..789P,2005ApJ...623..398Y,2013ApJ...777..136W,2014ApJ...793..136K}. Based on population synthesis models, \citet{2016MNRAS.458.2916C} predicted that the nova rate for systems with $P_{\mathrm{rec}}<1$\,yr in `\m31-like' galaxies should be $\sim4$\,yr$^{-1}$, of which \novak\ {\it could} account for  2\,yr$^{-1}$ in \m31.  But the question of the true population size of such ultra-short cycle RNe remains an open one. 

The 2012 eruption of \novak\ was chronologically the third to be discovered but the first to be spectroscopically confirmed \citep{2012ATel.4503....1S}, and provided the first hint of the true nature and short recurrence period of this system.  Subsequently, the 2013 eruption was expected and results of visible, UV, and X-ray observations were published by \citet[][hereafter \oonek]{2014A&A...563L...9D}, \xonek, and independently by \ponek.  By employing the technique developed by \citet{2009ApJ...705.1056B}, \citet{2013ATel.5611....1W} recovered the progenitor system from archival {\it Hubble Space Telescope (HST)} data.  These {\it HST} visible and near-UV (NUV) photometric data indicated the presence of a bright accretion disk, similar in luminosity to that seen around RS~Oph (\oonek, \ponek; also see \citealt{2008ASPC..401.....E}, for detailed reviews of the RS~Oph system).  \swift X-ray observations began six days after the 2013 discovery and immediately revealed the presence of SSS emission (\xonek).  Black body fits to the X-ray spectra indicated a particularly hot source ($\sim100$\,eV) compared to the \m31 nova population \citep[see][]{2014A&A...563A...2H}.  The SSS-phase lasted for only twelve days; at the time \novak\ had the fastest SSS turn-on and turn-off ever observed \citep[these were both surpassed by the 2014 eruption of the Galactic RN V745~Scorpii, a RG-nova, see][and Section~\ref{sec:v745}]{2015MNRAS.454.3108P}.  The X-ray properties pointed to a combination of a high mass WD and low ejected mass, with the {\it HST} data indicating a high mass accretion rate.  Modeling of the system reported by \ponek\ pointed toward $M_{\mathrm{WD}}>1.3\,M_{\odot}$ and $\dot{M}>1.7\times10^{-7}\,M_{\odot}$\,yr$^{-1}$.

A successful campaign to discover the predicted 2014 eruption was reported by \citet[][hereafter \otwok]{2015A&A...580A..45D}.  The discovery triggered a swathe of pre-planned high-cadence visible, UV, and X-ray observations, led by the fully-robotic 2m Liverpool Telescope \citep[LT;][]{2004SPIE.5489..679S} from the ground, and \swift from low-Earth orbit.  \otwok\ reported a visible light curve that evolved faster than all known Galactic RNe ($t_{3}\left(V\right)=3.84\pm0.24$\,days; also see Section~\ref{sec:initial-decline}), before entering a short-lived `plateau' phase; as seen in other RNe \citep[see, for example,][]{2014ApJ...788..164P}.  The plateau coincided approximately with the start of the SSS phase (see \xtwok).  A series of visible spectra were collected, the first just 1.27\,days after the eruption, these showed modest expansion velocities ($\bar{v_{\mathrm{ej}}}=2600\pm100$\,km\,s$^{-1}$) for such a fast nova, which significantly decreased over the course of just a few days.  Such an inferred deceleration is reminiscent of the interaction of the ejecta with pre-existing circumbinary material \citep[such as the red giant wind in the case of RS~Oph;][]{1985MNRAS.217..205B,2006ApJ...652..629B}.  

Independently of any eruptions from the system, \otwok\ also reported that deep H$\alpha$ imaging of \novak\ at quiescence uncovered a vastly extended elliptical shell centered on the system; the structure is larger than most Galactic supernova remnants.  Serendipitous spectra of the shell obtained during the 2014 eruption revealed strong H$\alpha$, [N\,{\sc ii}] (6584\,\AA), and [S\,{\sc ii}] (6716/6731\,\AA) emission (\otwok).  The measured [S\,{\sc ii}]/H$\alpha$ ratio, and the lack of any [O\,{\sc iii}] emission suggest a non-SN origin, and hence a possible association with \novak.

\citet[][hereafter \xtwok]{2015A&A...580A..46H} reported the fruits of an intensive X-ray follow-up campaign of the 2014 eruption using \swiftk. Their main results included a precise measurement of the SSS turn-on time ($5.9\pm0.5$~d), a fast effective temperature evolution during the SSS phase, and a strong aperiodic X-ray variability that decreased significantly around day 14 after eruption. \xtwok\ found the 2014 SSS properties to be remarkably similar to those of the 2013 eruption. 

Theoretical studies of hypothetical systems similar to \novak\ (before such a short recurrence period system was discovered) consistently show that a combination of a high mass WD and high mass accretion rate are required to achieve a short recurrence period and drive the rapid turn-on of a short-lived SSS phase \citep[see, for example][]{2005ApJ...623..398Y}.  Based on a recurrence period of 1\,year, \citet{2015ApJ...808...52K} determined that the \novak\ eruptions are consistent with a WD mass of $1.38\,\mathrm{M}_\odot$, an accretion rate $\dot{M}=1.6\times10^{-7}\,\mathrm{M}_\odot\,\mathrm{yr}^{-1}$, and an ejected mass of $\sim0.6\times10^{-7}\,\mathrm{M}_\odot$ leading to a mass accumulation efficiency of the WD of $\eta\simeq0.63$ -- i.e.\ the WD retains 63\% of the accreted material and therefore is expected to be increasing in mass.

Overall, the striking similarities between the past eruptions facilitated the development of a detailed observing strategy for the detection and follow-up of the expected 2015 eruption.

\section{Quiescent Monitoring and Detection of the 2015 Eruption}\label{sec:monitor_and_detect}

Following the 2014 eruption of \novak, a dedicated quiescent monitoring campaign was again put in place to detect the next eruption, as had been employed to discover the 2014 eruption (\otwok).  For the 2015 eruption detection campaign a large array of telescopes were employed.  These included the Kiso Schmidt Telescope, the Okayama Telescope, the Miyaki-Argenteus Observatory, all in Japan; the Xingming Observatory, China; the Ond\v{r}ejov Observatory, Czech Republic; Montsec Observatory, Spain; and the Kitt Peak Observatory, USA.  The majority of the quiescent monitoring was performed by three facilities, the sister telescopes the LT and Las Cumbres Observatory
Global Telescope Network (LCOGT) 2-meter telescope on Haleakala, Hawaii (formally the Faulkes Telescope North), and the \swift observatory.

The LT began monitoring the system immediately after the cessation of the 2014 eruption, although these observations were tempered by the diminishing visibility of \m31.  From 2015 May 27 onward, the LT obtained nightly (weather permitting) observations at the position of \novak\ using the IO:O visible CCD camera\footnote{\url{http://telescope.livjm.ac.uk/TelInst/Inst/IOO}} (a $4096\times4112$ pixel e2v detector which provided a $10^{\prime}\times10^{\prime}$ field of view).  From 2015 Jun.\ 10 onward, the LT data were supplemented by observations from LCOGT \citep[2\,m, Hawaii;][]{2013PASP..125.1031B}, which employed the Spectral visible CCD camera\footnote{\url{http://lcogt.net/observatory/instruments/spectral}} (a $4\mathrm{k}\times4\mathrm{k}$ pixel detector providing a $10^{\prime}\!.5\times10^{\prime}\!.5$ field of view).

Each LT and LCOGT observation consisted of a single 60\,s exposure taken through a Sloan-like $r'$-band filter, with a target cadence of 24\,hours; although this was decreased to 2\,hours within the $\sim1\sigma$ eruption prediction window (from the night beginning 2015 Jul.\ 30 onward; \halfk).  The LT and LCOGT data were automatically pre-processed by a pipeline running at the LT and LCOGT, respectively, and were automatically retrieved, typically within minutes of the observation.  An automatic data analysis pipeline \citep[based on a real-time \m31 difference image analysis pipeline, see][]{2007ApJ...661L..45D,2010MNRAS.409..247K} then further processed the data and searched for transient objects in real-time.  Any object detected with significance $\ge5\sigma$ above the local background, within one seeing-disk of the position of \novak, would generate an automatic alert.

An ambitious \swift program to monitor the quiescent system with the aim of detecting the predicted initial X-ray flash of the eruption \citep{2015ApJ...808...52K} was also active.  Full details of this campaign are to be reported in a companion paper \citep{XrayFlash}.  While focusing on X-ray emission, the \swift UV/optical telescope \citep[UVOT;][]{2005SSRv..120...95R} was also employed to monitor the system.  To complement the UVOT observations the LT monitoring program included additional Sloan $u'$-band observations from 2015 August 16.121 UT onward.

A transient was detected with high significance in LCOGT $r'$-band data taken on 2015 August $28.425\pm0.001$ UT by the automated pipeline at a position of $\alpha=0^{\mathrm{h}}45^{\mathrm{m}}28\fs82$, $\delta=41\degr54\arcmin10\farcs0$ (J2000), with separations of $0\farcs09\pm0\farcs07$ and $0\farcs16\pm0\farcs07$ from the position of the 2013 (\oonek) and 2014 (\otwok) eruptions, respectively.  Preliminary photometry at the time (see Section~\ref{lt_photometry} for detailed photometric analysis) indicated that this object had a magnitude of $r'=19.09\pm0.04$, around one magnitude below  the peak brightness of previous eruptions of \novak\ (\oonek, \otwok, \ponek).  Our pre-planned follow-up observations were immediately triggered, and a request for further observations was released \citep{2015ATel.7964....1D}.

A transient was also detected in the \swift UVOT uvw1 data with $m_{\mathrm{w1}}=17.7\pm0.1$ at the position of \novak\ taken on 2015 August 28.41 UT -- marginally before the LCOGT detection.  However, the longer data retrieval time for \swift meant these data were received and processed after the LCOGT data.

No object was detected at the position of \novak\ in an LT IO:O observation $0.265\pm0.001$\,days earlier down to a $3\sigma$ limiting magnitude of $r'>21.8$.  Additional LT IO:O observations $0.353\pm0.001$ and $0.444\pm0.001$\,days before detection also detected no sources down to $r'>21.8$.  LT Sloan $u'$-band observations taken $0.452\pm0.001$, $0.442\pm0.001$, and $0.264\pm0.001$\, days before the LCOGT detection found no source at the position of \novak\ down to limits of $u'>19.8$, $>21.6$, and $>21.5$, respectively.  Similarly, no object was detected in the \swift UVOT uvw1 data on 2015 August 28.01 down to a $3\sigma$ limit of $m_{\mathrm{w1}}>20.3$.

A full analysis of all the inter-eruption -- quiescent -- data will be published in a later paper.

\section{Observations of the 2015 Eruption}\label{sec:observations}

In this section we will describe the strategy and various data analysis techniques employed for the near infrared (NIR), visible, UV, and X-ray follow-up observations of the 2015 eruption of \novak. 

\subsection{Visible and Near Infrared Photometry}\label{sec:optical_photometry}

\begin{figure*}
\begin{center}
\includegraphics[width=0.95\textwidth]{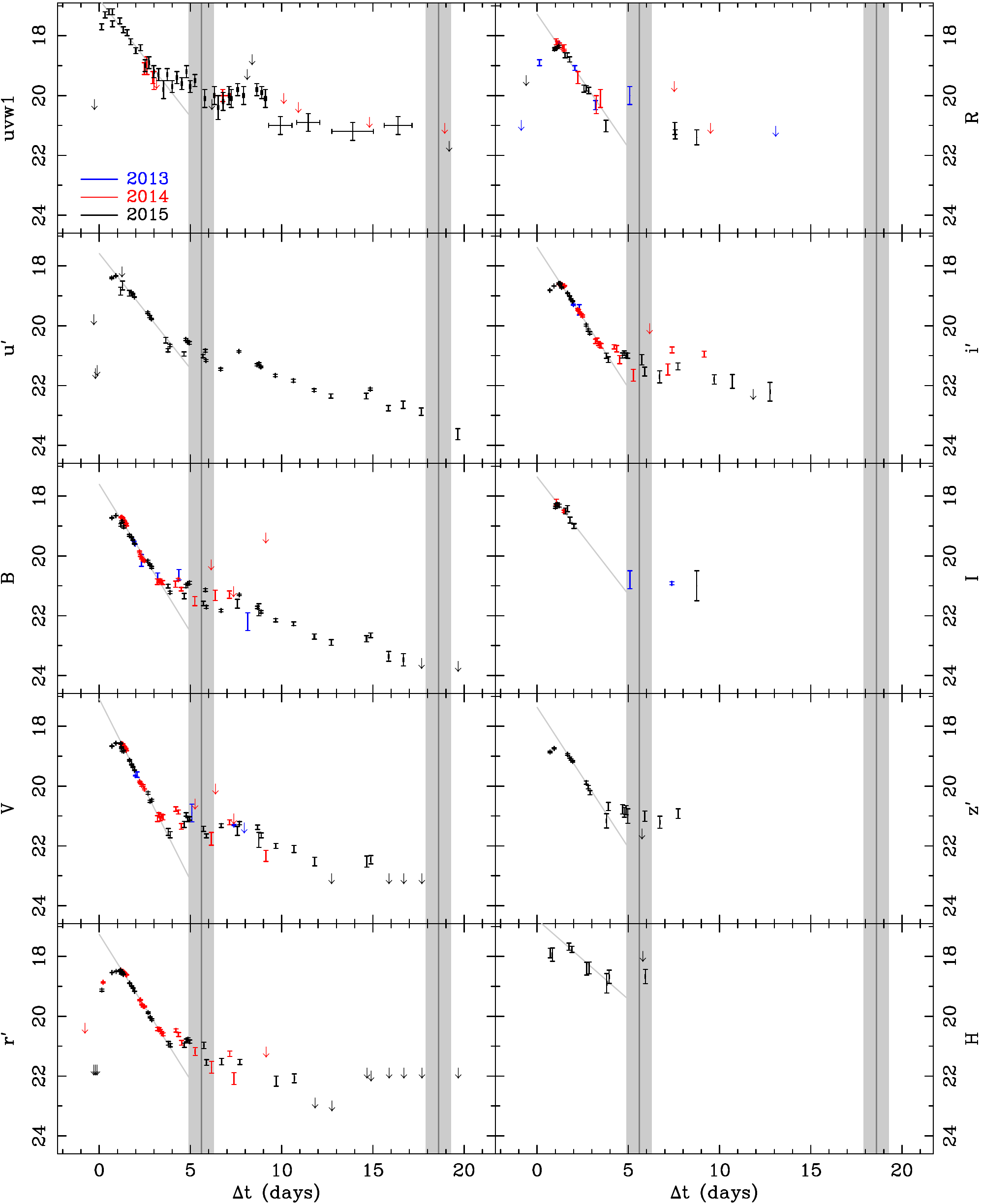}
\end{center}
\caption{Near-ultraviolet through near-infrared photometry of the 2013--2015 the eruptions of \novak.  Black points indicate the 2015 eruption, all data taken from Tables~\ref{tab:obs_swift} and \ref{optical_photometry}, red points indicate data from the 2014 eruption (\otwok; \xtwok), and blue points the 2013 eruption (\oonek; \ponek).  The vertical gray lines indicate the turn-on and turn-off times of the SSS from the 2015 eruption (the shaded areas their associated uncertainties).  The gray lines show a linear fit (exponential decay in luminosity) to each light curve between $1\le\Delta t\le4$\,days (see Table \ref{t2_table} for the decline times and other characteristics). \label{optical_lc}}
\end{figure*}

\begin{figure*}
\includegraphics[width=\textwidth]{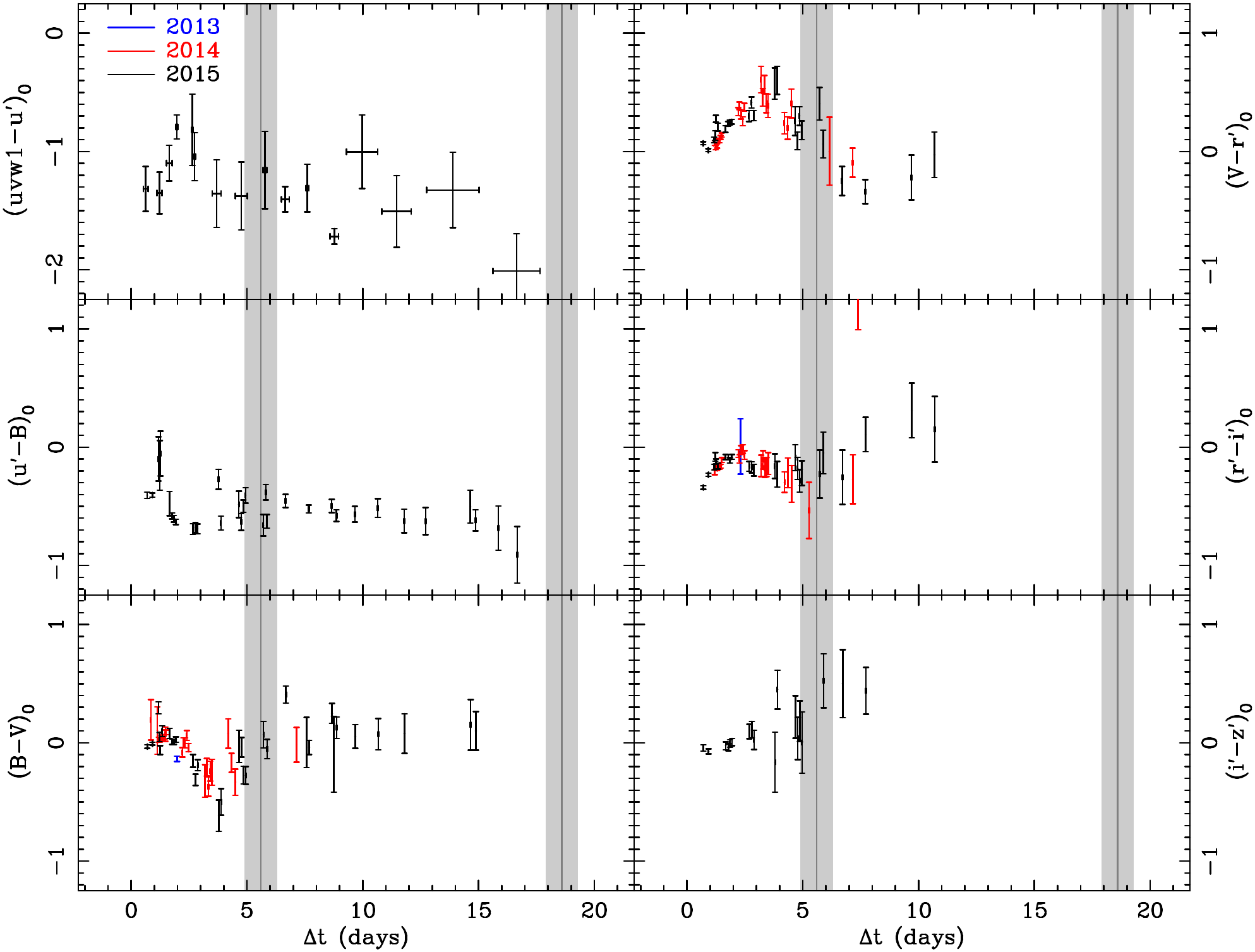}
\caption{As Figure~\ref{optical_lc} but showing the dereddened color evolution of the eruptions of \novak, assuming $E_{B-V}=0.096\pm0.026$ \citep{HST2016}.  The (uvw1--$u'$)$_0$ plot uses a different y-axis and mixes 2014 and 2015 uvw1 data.\label{color_lc}}
\end{figure*}

The 2015 eruption of \novak\ was followed photometrically by a large number of ground-based visible/NIR facilities.  These include the aforementioned LT and LCOGT, the Mount Laguna Observatory (MLO) 1.0\,m, the Ond\v{r}ejov Observatory 0.65\,m, the Bolshoi Teleskop Alt-azimutalnyi (BTA) 6.0\,m, the Corona Borealis Observatory (CBO) 0.3\,m, the Nayoro Observatory of Hokkaido University 1.6\,m Pirka telescope, the Okayama Astrophysical Observatory (OAO) 0.5\,m MITSuME telescope, and the iTelescope.net T24. The data acquisition and analysis for each of these facilities is described in detail in Appendix~\ref{app:optical_photometry}.  The resulting photometric data are presented in Table~\ref{optical_photometry}, and the subsequent light curves are shown in Figure~\ref{optical_lc}.  Where near-simultaneous multi-color observations are available {\it from the same facility}, the color data are presented in Table~\ref{colour_table}, and the color evolution plots are shown in Figure~\ref{color_lc}.

\subsection{Visible Spectroscopy}\label{optical_spectroscopy}

The primary aim of spectroscopy of the 2015 eruption was to obtain the earliest spectra post-eruption and to confirm the nature of the apparent ejecta deceleration reported by \otwok.  Spectroscopy was obtained by the LT, LCOGT, and Kitt Peak National Observatory 4\,m telescope.  The text in Appendix~\ref{app:optical_spectroscopy} describes the resulting data acquisition and processing, a log of the spectroscopic observations is provided in Table~\ref{spec_log}.

\begin{table}
\caption{Log of spectroscopic observations of the 2015 eruption of \novak.\label{spec_log}}
\begin{center}
\begin{tabular}{llll}
\tableline\tableline
Date & $\Delta t^{\dag}$ & Telescope & Exp.\ time\\
(2015 UT) & (days)&  & (s)\\
\tableline
Aug.\,28.95 & $0.67\pm0.02$ & LT  & $3\times900$\\
Aug.\,29.24 & $0.96\pm0.02$ & LT  & $3\times900$\\
Aug.\,29.38 & $1.10\pm0.01$ & KPNO 4\,m & 1200 \\
Aug.\,29.42 & $1.14\pm0.02$ & LCOGT 2\,m & 3600\\
Aug.\,30.07 & $1.79\pm0.11$ & LT  & $6\times900^{\ddag}$\\
Aug.\,30.41 & $2.03\pm0.02$ & LCOGT 2\,m & 3600\\
Aug.\,31.12 & $2.84\pm0.11$ & LT  & $6\times900^{\ddag}$\\ 
Sep.\,01.12 & $3.84\pm0.02$ & LT  & $3\times1,200$\\
Sep.\,02.19 & $4.91\pm0.02$ & LT  & $3\times1,200$\\
\tableline
\end{tabular}
\end{center}
\tablenotetext{\dag}{The quoted uncertainty on $\Delta t$ relates to the total elapsed time during each observation.}
\tablenotetext{\ddag}{Two epochs of spectroscopy (both with $3\times900$\,s exposure time) were collected by the LT on each of the nights of 2015 Aug 30 and 31, these were combined into single `nightly' spectra to improve overall signal-to-noise ratio.}
\end{table}

\subsection{\swift X-ray and UV observations}\label{swift_data}

The high-cadence \swift observations employed for the initial X-ray flash monitoring of \novak\ \citep[see][]{XrayFlash} were continued for a further 20 days following the eruption to study the UV and X-ray light curves of the eruption.  The observations are summarized in Table~\ref{tab:obs_swift}.

\begin{table*}
\caption{Stacked \swift UVOT images and magnitudes. \label{tab:uvot_merge} }
\begin{center}
\begin{tabular}{lllllll}
\tableline\tableline
ObsIDs$^a$ & Exp$^b$ & Date$^c$ & MJD$^c$ & $\Delta t^c$ & Duration$^d$ & \texttt{uvw1}\\
 & (ks) & (UT) & (d) & (d) & (d) & (mag)\\
\tableline
00032613115/121 & 6.3 & 2015 Sep.\ 01.76 & 57266.76 & 4.48 & 1.52 & $19.5\pm0.1$ \\
00032613123/127 & 4.4 & 2015 Sep.\ 03.56 & 57268.56 & 6.28 & 1.00 & $20.3\pm0.2$ \\
00032613128/135 & 6.1 & 2015 Sep.\ 05.29 & 57270.29 & 8.01 & 1.79 & $19.9\pm0.1$ \\
00032613137/142 & 6.2 & 2015 Sep.\ 07.21 & 57272.21 & 9.93 & 1.28 & $21.0\pm0.3$ \\
00032613142/148 & 6.4 & 2015 Sep.\ 08.61 & 57273.61 & 11.33 & 1.52 & $20.6\pm0.3$ \\
00032613151/160 & 4.4 & 2015 Sep.\ 11.17 & 57276.17 & 13.89 & 2.27 & $21.2\pm0.3$ \\
00032613162/168 & 5.5 & 2015 Sep.\ 13.66 & 57278.66 & 16.38 & 1.52 & $21.0\pm0.3$ \\
00032613171/182 & 9.5 & 2015 Sep.\ 16.46 & 57281.46 & 19.18 & 2.73 & $<21.6$\\
\tableline
\end{tabular}
\end{center}
\tablenotetext{a}{First and last observation of the stack (cf.\ Table~\ref{tab:obs_swift}).}
\tablenotetext{b}{Combined exposure time.}
\tablenotetext{c}{Midpoint of the stack with $\Delta t$ referring to the eruption date on 2015 Aug.\ 28.28 UT (MJD 57262.28; see Section~\ref{sec:time}).}
\tablenotetext{d}{Time between the first and last observations of the stack.}
\end{table*}

The decline of the UV light curve and the early SSS phase received a high-cadence coverage with
on average a single 1\,ks pointing obtained every six hours (see Table~\ref{tab:obs_swift}).  However, the coverage of the later SSS light curve was occasionally interrupted by higher-priority observations such as $\gamma$-ray bursts. This resulted in the omission of certain ObsIDs in the otherwise consecutive list in Table~\ref{tab:obs_swift}. Some other ObsIDs were not included because they collected less than 20\,s of exposure. In the text of this paper, individual \swift observations are referred to by their segment ID (i.e.\ `ObsID 123' is shorthand for ObsID 00032613123).

All our \swift data analysis is based on the cleaned level 2 files locally reprocessed at the \swift UK Data Centre\footnote{\url{http://www.swift.ac.uk}} with HEASOFT (v6.15.1). For our higher level analysis we used the \swift software packages included in HEASOFT (v6.16) together with \texttt{XIMAGE} (v4.5.1), \texttt{XSPEC} \citep[v12.8.2;][]{1996ASPC..101...17A}, and \texttt{XSELECT} (v2.4c).

Before extracting light curves and spectra we carefully inspected the level 2 event files for the \swift X-ray telescope \citep[XRT;][]{2005SSRv..120..165B} and UVOT. We found that five observations were affected by the star trackers not being continuously `locked on' during some observations. Of those, ObsIDs 109, 122, and 178 corresponded to non-detections and the intermittent tracking did not affect the derived upper limits. During ObsIDs 149 and 161 the source was detected and the loss of tracking might have resulted in somewhat larger count rate uncertainties than those given in Table~\ref{tab:obs_swift}. Both ObsIDs were excluded from the X-ray variability analysis in Section \ref{sec:lc_xvar}. In case of the UVOT, the ObsIDs 122, 149, and 178 showed strong indications of unstable pointing and were excluded from the UVOT analysis and light curve. In other UVOT images the point spread functions (PSFs) were slightly elongated, but still acceptable for photometry.

Furthermore, we inspected the XRT exposure maps for bad columns and bad pixels. As a result, we excluded a small number of ObsIDs from the X-ray variability analysis because those observations had bad detector columns going through the source counts extraction region. The excluded ObsIDs were: 128, 137, 140, 151, and 160.  All of the excluded observations except 151, which has the most severe bad column issue, are included in the overall X-ray light curve described in Section \ref{sec:vis_lc}.

All XRT data were obtained in photon counting (PC) mode. We applied the standard charge distribution grade selection (0--12) for XRT/PC data. The XRT count rates and upper limits presented here were extracted using the \texttt{ximage} \texttt{sosta} tool, which applies corrections for vignetting, dead-time, and PSF losses. The PSF model used is the same as for the 2014 observations (see \xtwok) and was based on all merged XRT detections of the 2014 eruption. We visually inspected all XRT images and confirmed that the detections were realistic.

The X-ray spectra were extracted with the \texttt{XSELECT} software (v2.4c) and fitted for energies above 0.3\,keV using \texttt{XSPEC} \citep[v12.8.2;][]{1996ASPC..101...17A}. Our \texttt{XSPEC} models assumed the ISM abundances from \citet{2000ApJ...542..914W}, the T\"ubingen-Boulder ISM absorption model (\texttt{TBabs} in \texttt{XSPEC}), and the photoelectric absorption cross-sections from \citet{1992ApJ...400..699B}. The spectra were binned to include at least one count per bin and fitted in \texttt{XSPEC} assuming Poisson statistics according to \citet{1979ApJ...228..939C}. We describe the fitting of black body models, some of which include additional emission or absorption features, in Section~\ref{sec:xspec}.

For the UVOT data, we examined all the individual sky images by eye. We found that ObsIDs 126 and 180 had no aspect correction and we manually adjusted the source and background regions for consistent UVOT photometry.

We optimized the \texttt{uvotsource} source and background extraction regions, with respect to the 2013/14 analysis, based on a stacked image of all 2015 observations. The new source region has a $3\farcs6$, radius and \texttt{uvotsource} was operated with a curve-of-growth aperture correction. The background is derived from a number of smaller regions in the vicinity of the source that show a similar background luminosity as the source region in the deep image. All magnitudes assume the UVOT photometric system \citep{2008MNRAS.383..627P} and have not been corrected for extinction.

The statistical analysis was performed using the \texttt{R} software environment \citep{R_manual}. All uncertainties correspond to 1$\sigma$ confidence and all upper limits to 3$\sigma$ confidence unless otherwise noted.

\subsection{Time of eruption}\label{sec:time}

For all observations of the 2015 eruption of \novak\ we use the reference date ($\Delta t=0$) defined as 2015 Aug 28.28 UT ($\mathrm{MJD}=57262.28$) as the epoch of the eruption.  This date is defined as the midpoint between the last non-detection by the LT visible monitoring (2015 Aug.\ 28.16 UT) and the first detection of the eruption by \swift UVOT (Aug.\ 28.41), with an uncertainty of 0.12\,d.  We draw direct comparison to data from the 2014 and 2013 eruptions by assuming reference dates of 2014 Oct 2.69 UT ($\mathrm{MJD}=56932.69$) and 2013 Nov 26.95 UT ($\mathrm{MJD}=56622.95$)\footnote{The epoch of the 2013 eruption has been updated from Nov 26.60 UT in \xonek\ by fitting of the linear early-decline (see Section~\ref{sec:vis_lc}) of the light curve to the 2014 and 2015 data}, respectively.  

\section{Panchromatic eruption light curve (soft X-ray to near-infrared)}\label{sec:vis_lc}

The NIR/visible light curve of the 2015 eruption, obtained via an array of ground-based telescopes, matched the high-cadence achieved in 2014.  However, the 2015 data surpass those from previous eruptions by virtue of their broader wavelength coverage ($H$--$u'$-band) and depth -- extending the light curve from $\sim9$~days (2014) to just under 20~days, and following the decline through almost 6 magnitudes ($u'$-band).  The 2015 light curve data alone are the most extensive visible data compiled for a nova beyond the Milky Way and Magellanic Clouds.  When combined with data from past eruptions, the light curve data are now comparable in detail to many Galactic novae.

The multi-color, high-cadence, light curves of the eruption of \novak\ are presented in Figure~\ref{optical_lc}.  Here the black data points are the new 2015 data, red show 2014 data, and blue 2013, all plotted relative to their respective eruption times (see Table~\ref{eruption_history}).  It is clear from inspecting these plots that the agreement between the light curves of the last three eruptions is indeed remarkable.  

\begin{figure}
\includegraphics[width=\columnwidth]{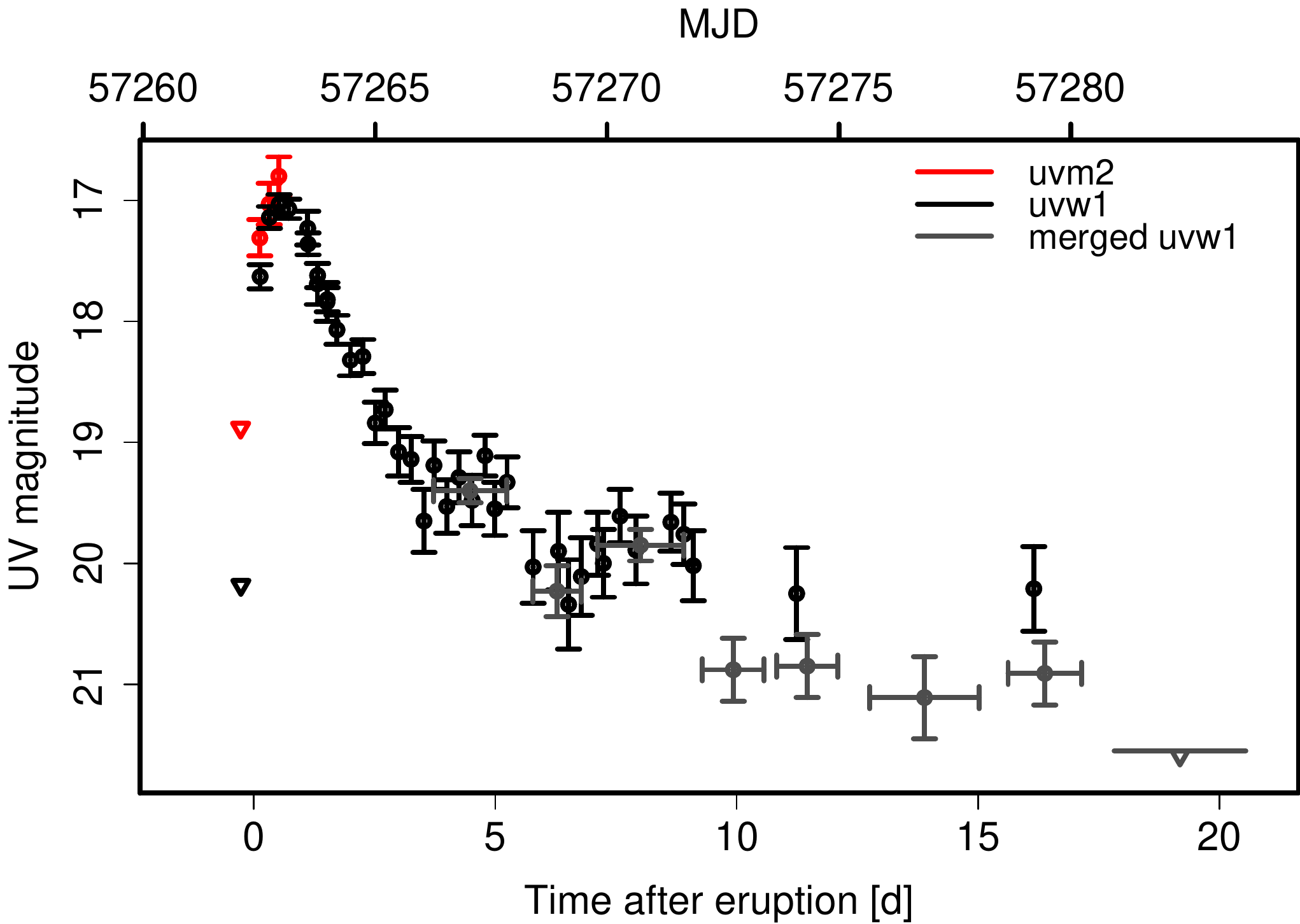}
  \caption{Unprecedentedly detailed \swift UVOT light curve for the 2015 eruption of \novak, showing for the first time the rising phase, the rapid smooth decline, and various plateaus. Black data points are individual \texttt{uvw1} snapshots. Gray points are based on stacked images (see Table~\ref{tab:uvot_merge}). For clarity, no individual \texttt{uvw1} upper limits after day zero are shown (see Table~\ref{tab:obs_swift} for those). Red data points show a few initial \texttt{uvm2} snapshots. Open triangles indicate $3\sigma$ upper limits. Uncertainties are combined $1\sigma$ statistical and systematic. Day zero is defined as MJD = $57262.28\pm0.12$ (see Section~\ref{sec:time}).\label{fig:uv}}
\end{figure}

The unprecedentedly detailed and complete UV light curve of the 2015 eruption of \nova is the focus of Figure~\ref{fig:uv} (the combined 2014/2015 UV light curve is shown in Figure~\ref{optical_lc}). The corresponding magnitudes are given in Table~\ref{tab:obs_swift}. For the first time, we observed the rise of the UV flux to the maximum and can put very tight constraints on the time of the UV peak. We followed the UV light curve for almost 20 days with a high cadence, until the UV flux finally dropped below our sensitivity limit. The result is by far the best UV light curve yet recorded for \nova or indeed for any \m31 nova.

Our observations used the UVOT \texttt{uvw1} filter throughout, which has a central wavelength of 2600\,\AA\ (FWHM 693\,\AA) and the highest throughput of the three UV filters \citep{2008MNRAS.383..627P}. On the rise to maximum, these were accompanied by occasional \texttt{uvm2} filter measurements. Those magnitudes, for a shorter central wavelength of 2250\,\AA\ (FWHM 657\,\AA), appear slightly brighter than the quasi-simultaneous \texttt{uvw1} values.

\begin{figure}
\includegraphics[width=\columnwidth]{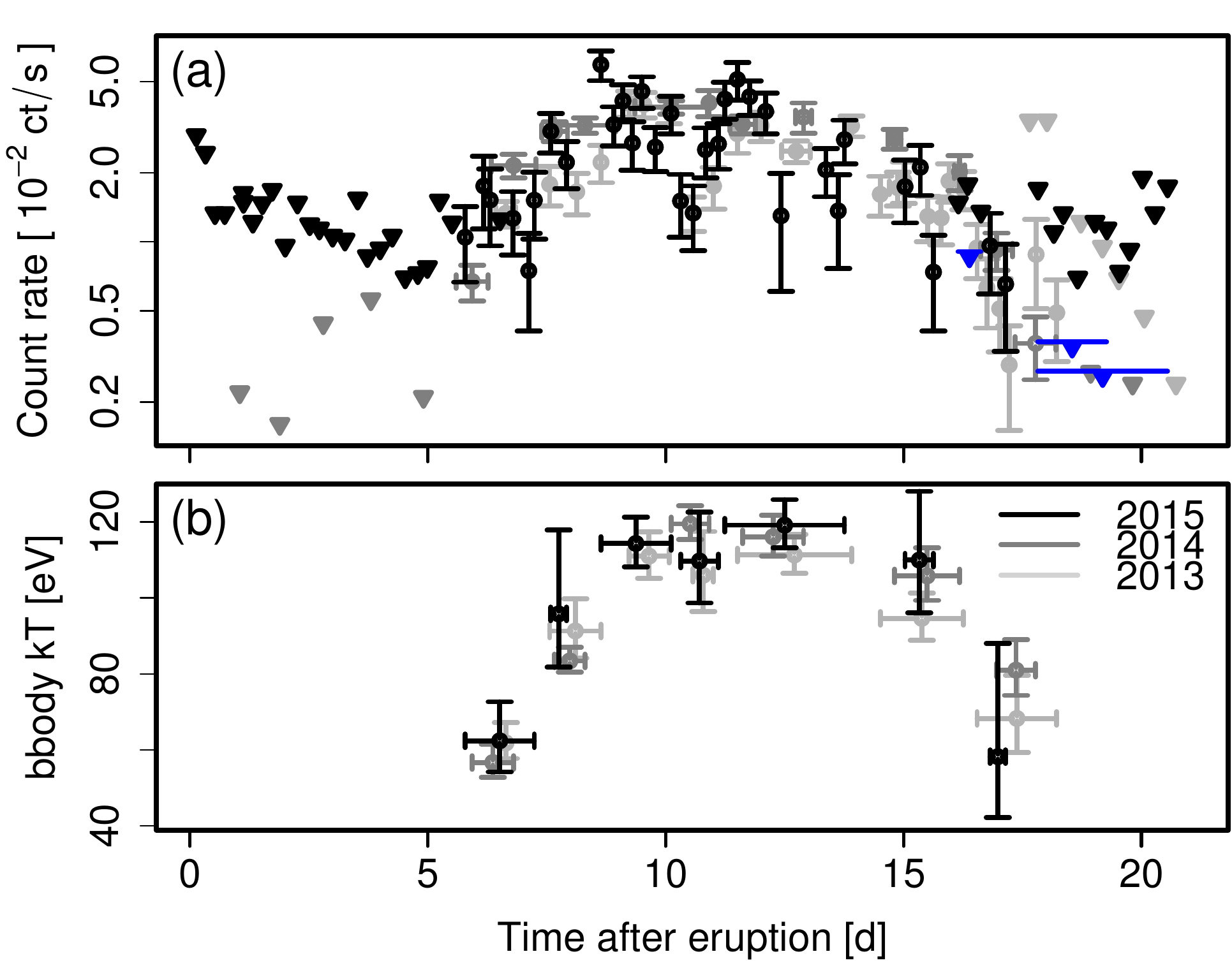}
  \caption{\swift XRT (a) count rate light curve (0.2--10\,keV) and (b) effective black body temperature evolution of \nova during the 2015 eruption (black). In light/dark gray we show the corresponding data of the 2013/14 eruption. Time is in days after 2015 Aug.\ 28.28 UT for the 2015 data. {\it Panel a:} Triangles indicate upper limits. The blue points are merged detections and upper limits. {\it Panel b:} Sets of observations with similar spectra have been fitted simultaneously assuming a fixed \nh = $1.4$ \hcm{21}. The error bars in time represent either (a) the duration of the observation or (b) the time covering the sets of observations. The three eruptions show very similar time scales and luminosity/temperature evolution. \label{fig:xray_lc}}\end{figure}

In Figure~\ref{fig:xray_lc} we show the 0.2--10.0\,keV X-ray light curve of the 2015 eruption compared to the 2014/13 results. Also shown is the evolution of the effective temperature based on a simple black body parametrization with a constant \nh $= 1.4$ \hcm{21}. A more detailed spectral analysis is the subject of Section \ref{sec:xspec}. Overall, the count rate and temperature evolution are very similar between the three eruptions. The X-ray count rate was initially very variable as the effective temperature rose to maximum. After around day 13, we observed a decrease in the variability amplitude (see discussion below) although our observations became more sparse in the second part of the SSS phase.

For the further X-ray variability and spectral analysis we assume that the last three eruptions evolved sufficiently similarly to warrant a combined treatment. Figure~\ref{fig:xray_lc} indicates that this assumption is justified. The combined data provide improved statistics and signal-to-noise ratio to further explore the initial result presented by \xtwok\ and investigate features such as the `dip' in the X-ray light curve around day eleven.

The far blue ($u'$-band) to NIR light curves, and the early UV evolution, can be separated into four distinct phases on the basis of their rate of change of flux.  The final rise, from $t=0$ to $t\simeq1$\,d; the initial decline, $1\la t\la4$\,d; the `plateau' and SSS onset, from $4\la t\la8$\,d; and the SSS-peak and decline, $8\la t\la19$\,d (when the SSS is still detected).  Here we define and discuss each of these four phases in turn.

\subsection{The final rise (Day 0--1)}

As in 2014, the 2015 eruption was discovered before the peak in the visible light curves and, for the first time, detailed pre-visible peak data have been compiled, particularly in the $r'$ and $i'$ bands.  However, other than single-filter initial detections, there are still limited data before the final magnitude of the rise to peak -- a regime which must be a target for future eruptions.  These eruptions appear characterized by a relatively slow rise to maximum light, compared to CNe of similar speed class \citep[see][and below]{2010ApJ...724..480H,2016ApJ...820..104H}, with the final magnitude of the rise taking around 1\,day.  

\begin{figure}
\includegraphics[width=\columnwidth]{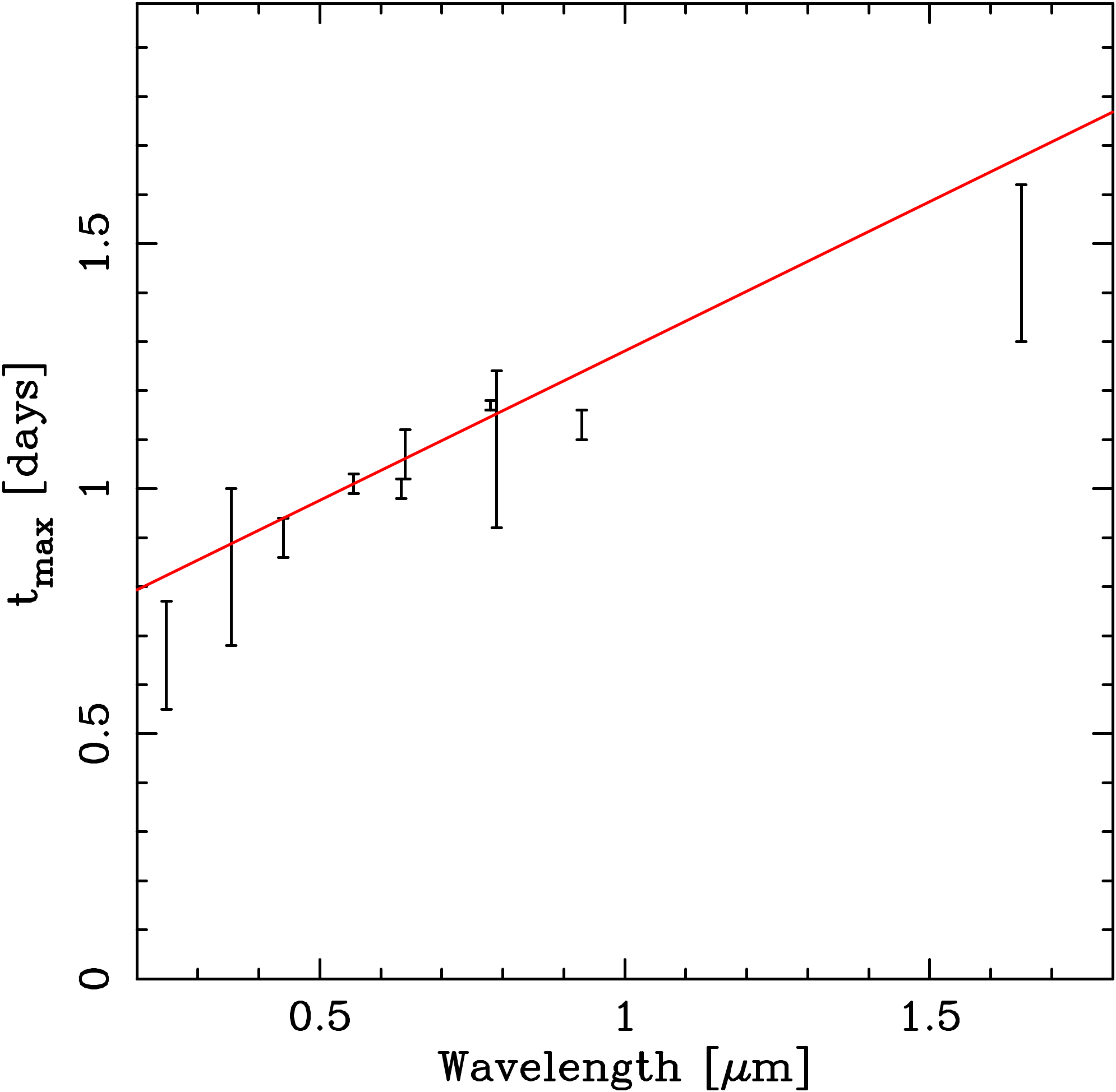}
\caption{Evolution of the time of maximum light with wavelength.  These data are consistent with the time of maximum increasing linearly with wavelength with a gradient of $0.61\pm0.11\,\mathrm{days}\,\mu\mathrm{m}^{-1}$ (indicated by the red line).\label{fig:max}}
\end{figure}

Using data from the 2013--2015 eruptions, the time of maximum in each filter ($t_\mathrm{max}$) was estimated by fitting a quadratic function to the data around peak ($0\le t\le 2$\,d), in all cases the peak data were well fit by this simple model.  The resulting $t_\mathrm{max}$ estimates are reported in Table~\ref{t2_table} and are shown in Figure~\ref{fig:max}, here any systematic uncertainties arising from the estimation of the eruption time for the separate years are ignored.  As expected, the uncertainties on the values of $t_\mathrm{max}$ are dominated by the sampling around the peak of the light curves.  The time of maximum shows a strong trend of increasing with wavelength, and is consistent with increasing linearly with wavelength with a gradient of $0.61\pm0.11\,\mathrm{days}\,\mu\mathrm{m}^{-1}$ (within the range $0.25\la\lambda\la1.6\,\mu\mathrm{m}$; $\chi^2_{/dof}=3.4$).

\begin{table*}
\caption{Light curve parameters of the eruption of \novak, based on combined data from the 2013, 2014, and 2015 eruptions.\label{t2_table}}
\begin{center}
\begin{tabular}{llllllll}
\tableline\tableline
Filter & $t_\mathrm{max}$ & $m_\mathrm{max}$ & $t_{2}$ & $t_{3}$ & \multicolumn{3}{c}{Decline rates (mag\,day$^{-1}$)}\\
& (days) & (mags) & (days) & (days) & Early decline & `Plateau' & Final decline \\
& & & & & $t_\mathrm{max}\le t\le 4$\,d & $4\le t \le 8$\,d & $t>8$\,d \\
\tableline
$uvw1$ & $0.66\pm0.11$ & $17.34\pm0.08$ & $2.55\pm0.16$ & $5.65^{+0.22}_{-0.40}$ & $0.78\pm0.05$ & $0.19\pm0.05$ & $0.08\pm0.03$ \\ 
$u'$ & $0.84\pm0.16$ & $18.35\pm0.03$ & $2.60\pm0.08$ & $5.74^{+1.02}_{-0.84}$ & $0.77\pm0.03$ & $0.11\pm0.08$ & $0.16\pm0.02$ \\ 
$B$  & $0.90\pm0.04$ & $18.67\pm0.02$ & $2.02\pm0.07$ & $3.03\pm0.10^{\ddag}$ & $0.99\pm0.03$ & $0.14\pm0.06$ & $0.18\pm0.02$ \\ 
$V$ & $1.01\pm0.02$ & $18.55\pm0.01$ & $1.65\pm0.04$ & $2.47\pm0.06^{\ddag}$ & $1.21\pm0.03$ & $0.09\pm0.03$ & $0.17\pm0.05$ \\ 
$R$ & $1.07\pm0.05$ & $18.38\pm0.02$ & $2.24\pm0.13$ & \nodata & $0.89\pm0.05$ & \nodata & \nodata \\ 
$r'$ & $1.00\pm0.02$ & $18.45\pm0.01$ & $2.05\pm0.04$ & $4.72^{+0.26}_{-0.15}$ & $0.97\pm0.02$ & $0.30\pm0.05$ & \nodata\\ 
$i'$ & $1.17\pm0.01$ & $18.60\pm0.01$ & $2.13\pm0.05$ & $3.40^{+0.53}_{-0.31}$ & $0.94\pm0.02$ & $0.11\pm0.06$ & \nodata \\ 
$I$ & $1.08\pm0.16$ & $18.31\pm0.03$ & $2.54\pm0.28$ & \nodata & $0.79\pm0.09$ & \nodata & \nodata \\ 
$z'$ & $1.13\pm0.03$ & $18.73\pm0.02$ & $2.13\pm0.09$ & \nodata & $0.94\pm0.04$ & $0.06\pm0.04$ & \nodata \\ 
$H$ & $1.46\pm0.16$ & $17.66\pm0.13$ & $3.75\pm0.45^{\dag}$ & \nodata & $0.53\pm0.06$ & \nodata & \nodata \\ \tableline
\end{tabular}
\end{center}
\tablenotetext{\dag}{$H$-band $t_{2}$ is determined by extrapolation of a linear fit to the data as less than two magnitudes of decline were recorded.} 
\tablenotetext{\ddag}{A linear fit to the data between $t_\mathrm{max}\le\Delta t\le4$\,days includes a decline of $\ge3$\,magnitudes.}
\end{table*}

The UV flux rose quickly from a $>20.5$\,mag upper limit (\texttt{uvw1}) on day $-0.27$ to a $17.7\pm0.1$\,mag detection on day 0.13 (see Table\,\ref{tab:obs_swift}). The maximum of $17.13\pm0.08$\,mag and $17.17\pm0.09$\,mag was reached on days 0.52 and 0.73, respectively.  The preceding observation on day 0.32 had shown $17.3\pm0.01$\,mag. In the next observation, on day 1.12, the nova had declined to $17.3\pm0.1$\,mag. 

\subsection{Initial decline (Day 1--4)}\label{sec:initial-decline}

In all filters (\texttt{uvw1}--$H$), the combined three-eruption light curves between $t_\mathrm{max}$ and 4 days post-eruption are well fit by a linear decline (exponential decline in luminosity; see the diagonal gray lines in Figure~\ref{optical_lc}; as also noted by \ponek\ and \otwok).  We use this simple model to determine the $t_{2}$ decline times for each filter (see Table~\ref{t2_table}).  The derived $t_{2}$ values are consistent with those determined based on the 2014 data alone (\otwok), but better constrained.  With the light curve decaying fastest in the $B$ and $V$-bands ($t_2\la2$\,d) this method also allowed the determination of $t_3$ (in both cases $t_3\la3$\,d).  For the remaining filters, $t_3$ was determined by linear interpolation between the light curve points bracketing $\Delta m=3$\,mags from peak, where the data were available.  The redder filters suffer more severely from crowding of nearby bright sources in \m31 (typically red giants; see \oonek), as such it was not possible to follow the $z'$-band light curve down to $t_3$, and the $H$-band light curve could only be followed for around 1\,mag from peak (here $t_2$ is extrapolated, but poorly constrained, under the assumption that the linear behavior seen in other bands would be replicated).  

To date, at least 14 Galactic novae (including three confirmed and one suspected RN) have been observed with a decline time $t_2\la4$\,d \citep[see][]{2010AJ....140...34S,2010ApJ...724..480H,2011MNRAS.410..525M,2015MNRAS.448L..35O}, and these are summarized in Table~\ref{nova_speeds}.  \novak\ resides at the faster end of this rapidly declining sample, exhibiting very similar decline times to the RN U~Sco (marginally slower to $t_2$, but faster to $t_3$; assuming $V$-band luminosities), and exhibits the only known decline with $t_3<3$\,d.  The decline time of a nova is fundamentally linked to the WD mass and the accretion rate, with the shortest decline times corresponding to the combination of a high mass WD and a high accretion rates \citep[see, for example][their Figure~2d]{2005ApJ...623..398Y}.  From the extremely short $t_3$ of \novak, we can infer that the WD in this system must be among the most massive yet observed.

\begin{table}
\caption{Galactic novae with $V$-band decline times $t_2\la4$\,d.\label{nova_speeds}}
\begin{center}
\begin{tabular}{llll}
\tableline\tableline
Nova & $t_2$ (days) & $t_3$ (days) & $P_\mathrm{rec}$ (years)$^\dag$ \\
\tableline
T CrB & 4 & 6 & 80$^\mathrm{a}$ \\
V1500 Cyg & 2 & 4 & \nodata \\
V2275 Cyg & 3 & 8 & \nodata \\
V2491 Cyg$^\ddag$ & 4 & 16 &  ($\ga100^\mathrm{b}$)\\
V838 Her & 1 & 4 & \nodata \\
LZ Mus & 4 & 12 & \nodata \\
V2672 Oph$^\mathrm{c}$ & 2.3 & 4.3 & \nodata \\
CP Pup & 4 & 8 & \nodata \\
V598 Pup$^\mathrm{d}$ & 4 & \nodata & \nodata \\
V4160 Sgr & 2 & 3 & \nodata \\
V4643 Sgr & 3 & 6 & \nodata \\
V4739 Sgr & 2 & 3 & \nodata \\
U Sco & 1 & 3 & 10.3$^\mathrm{a}$ \\
V745 Sco$^\mathrm{e}$ & 2 & 4 & 25$^\mathrm{a}$ \\
\tableline
\end{tabular}
\end{center}
\tablecomments{Unless otherwise indicated, all decline data are from \citet{2010AJ....140...34S}.}
\tablenotetext{\dag}{A recurrence period is quoted only if the system is known, or suspected (in parentheses), to be a RN.}
\tablenotetext{\ddag}{V2491~Cygni is a suspected RN \citep{2010MNRAS.401..121P,2011A&A...530A..70D}, but only a single eruption has been observed from this system.}
\tablerefs{(a)~\citet{2010ApJS..187..275S}, (b)~\citet{2010MNRAS.401..121P}, (c)~\citet{2011MNRAS.410..525M}, (d)~\citet{2010ApJ...724..480H}, (e)~\citet{2015MNRAS.448L..35O}.}
\end{table}

\subsection{The `plateau' and SSS onset (Day 4--8)}

\citet{2010AJ....140...34S} define a nova light curve plateau as an approximately flat interval occurring within an otherwise smooth decline.  Those authors also point out that observed plateaus often include some scatter and that the light curve may still decline slightly during such times.  

Following the linear decline from peak to $t\simeq4$\,days, the visible light curves appear to enter such a plateau phase lasting until at least day 8.  This plateau phase is observed in the $u'$-band through $i'$-band, but the nova was already too faint to be detected above the crowded un-resolved stellar background of \m31\ in the $z'$ and $H'$-band observations (see above).  The plateau phase in the combined light curves shows a small decrease in brightness over this period.  The plateau occurs around $2.5-3$\,mags below peak and, in the combined light curves, displays apparent variability with an amplitude of up to 1\,mag.  

Following the 2014 eruption, \otwok\ also noted the plateau phase, but the more limited data led them to conclude that the light curve was essentially flat during this stage.  Hence the `up turn' in brightness at the end of the plateau noted by \otwok\ is likely to be related to the variability of this stage seen in the combined data.  The onset of the quasi-plateau occurs around 1--2\,days before the SSS is unveiled and may be related \citep[see, for example,][]{2008ASPC..401..206H}.  

The UV light curve also shows similar behavior around this time, albeit these data have larger associated uncertainties (see Figure~\ref{optical_lc}).  However, an alternative interpretation could include a series of two shorter-lived UV plateaus (see, particularly the gray, combined, points in, Figure~\ref{fig:uv}):  A first plateau at $19.5\pm0.1$\,mag, centered around day 4.5, lasting about 1.5 days. The end of this plateau, around day 5.5, roughly coincided with the appearance of the SSS in X-rays (cf.\ Figure~\ref{fig:xray_lc}). The UV magnitude then dropped to a second plateau around $20.4\pm0.3$\,mag for about 1.5\,d around day 6.5, but soon showed indications for another slight rebrightening to $19.9\pm0.1$\,mag. This phase lasting for another 1.3\,d until around day 9.0. 

\begin{figure}
\includegraphics[width=\columnwidth]{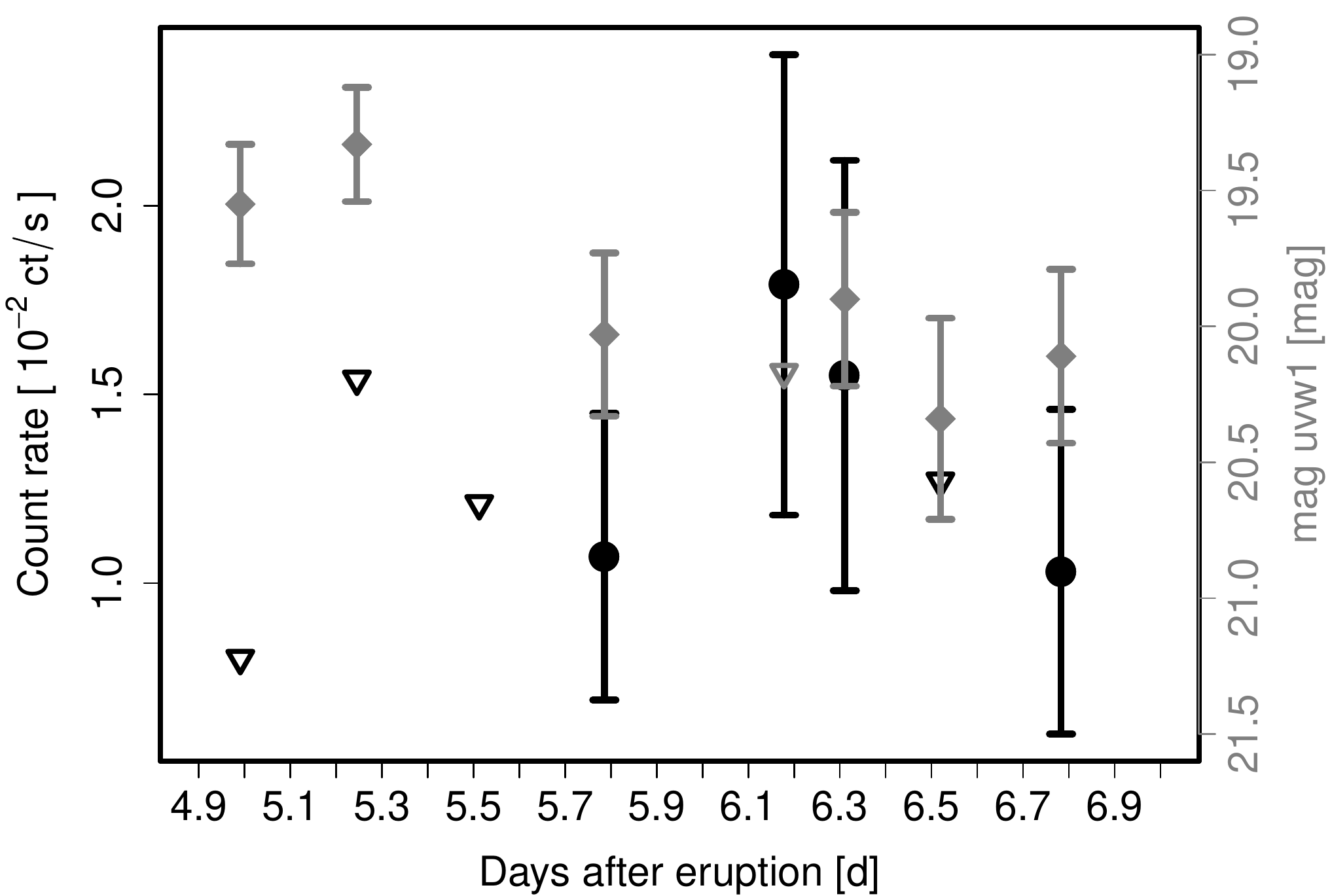}
  \caption{\swift XRT count rates (black circles) and UVOT \texttt{uvw1} magnitudes (gray diamonds) used for estimating the SSS turn-on time as day $5.6\pm0.7$ after eruption (cf. Figures \ref{fig:uv} and \ref{fig:xray_lc}. XRT/UVOT upper limits are shown as open triangles in black/gray. \label{fig:ton})}
\end{figure}

The X-ray and UV light curves around the time of the SSS turn-on are shown in Figure~\ref{fig:ton}. As the SSS flux gradually emerged, the UV magnitude was seen to drop from the first plateau seen in Figure~\ref{fig:uv}. Based on the last deep XRT upper limit on day 5.0 (ObsID 120) and the second detection on day 6.2 (ObsID 124), with a count rate significantly above this upper limit, we estimate the SSS turn-on time as $\eton = 5.6\pm0.7$~d (cf.\ Table\,\ref{tab:obs_swift}). This supersedes the initial estimate by \citet{2015ATel.7984....1H} and includes the uncertainty of the eruption date.

This uncertainty range is a conservative estimate. By day 5.8 (ObsID 123), a clear concentration of source counts can already be seen at the nova position. Note that ObsID 122 (day 5.5) had severe star tracking issues that would have affected the XRT PSF (see Section \ref{swift_data}). However, the XRT event file showed no indication of an increase in counts within a generous radius of the source position. Therefore, it is unlikely that the SSS was already visible before day 5.8.

We emphasize that the early SSS flux of \nova is highly variable, which limits the precision of turn-on time estimates. Nevertheless, the observed turn-on time scale is consistent with the 2013 ($6\pm1$~d) and 2014 ($5.9 \pm 0.5$~d) eruptions (see \xonek, \xtwok).

\subsection{The `SSS peak and decline' ($>$8\,\lowercase{d})}

Once the plateau phase ended at day $\sim8$ the far blue--NIR light curve entered a second phase of apparent linear decline in magnitude (exponential decline in flux).  With the nova again fading rapidly, the system only remained visible through the $i'$, $r'$ and $V$-band filters for a few more days.  However, this near-linear decline was followed until day $\sim17$ in the $B$-band and day $\sim20$ in the $u'$-band (just beyond SSS turn-off).  The decay-rate of this decline phase of the light curve was steeper than that of the quasi-plateau phase, but much shallower than the early decline.  The mean decline rate in the $u'$, $B$, and $V$-bands during this phase was 0.2\,mag\,day$^{-1}$ (five times slower than the mean early-decline in the same bands).  

The end of the SSS phase was first announced by \citet{2015ATel.8062....1H}. Based on the overall X-ray light curve in Figure~\ref{fig:xray_lc} we estimated the SSS turn-off time as $\etoff = 18.6\pm0.7$~d after eruption. This is a conservative estimate that takes into account the last X-ray detection on day 17.1 (ObsID 168) and the midpoint of the merged deep upper limit on day 19.2. The uncertainty includes the eruption date range. Within the errors, the estimate is consistent with the 2013 ($19 \pm 1$) and 2014 ($18.4\pm0.5$) eruptions as well as the 2012 X-ray non-detection on day 20 (see \xonek).

However, in the UV the decline after day 9 reached $\sim21$\,mag where the source remained, with typical uncertainties of 0.3 mag, until about day 17--18. All average magnitudes are based on sets of stacked UVOT images which are summarized in Table~\ref{tab:uvot_merge} and shown in Figure~\ref{fig:uv}.

The extent of the UV light curve is consistent with the duration of the SSS phase. After day 10, there were only two detections in individual images: around the time of the possible X-ray dip around day 11 and during the SSS decline around day 16 (cf.\ Figure~\ref{fig:xray_lc} and Table~\ref{tab:obs_swift}). After the SSS turn-off the UV flux dropped sharply and nothing was detected in 10 merged observations (9.5\,ks covering 2.7\,d) around day 19.2 with a $3\sigma$ upper limit of 21.7\,mag.

We were able to follow the light curve in the $u'$-band until $\sim1$\,mag above quiescence \citep[as determined from {\it HST} photometry; \oonek; \ponek;][]{HST2016}.  Utilizing the pre-existing {\it HST} quiescent photometry of this system (see \oonek\ and \ponek),  we can estimate that -- assuming a continuation of this linear decline -- the time to return to quiescent luminosity would be only 25--30\,d post-eruption.  This, of course, assumes that there is no dip below quiescence as seen in, for example, RS~Oph as the accretion disk in that system reestablishes post-eruption \citep{2007MNRAS.379.1557W,2008ASPC..401..203D}.  

It is worth noting that this decline phase can also be fit with a power-law decline in flux (providing a marginally better fit than a linear decline).  The index of the best fit power law ($f\propto t^\alpha$) to the $u'$-band data is $\alpha=-1.8\pm0.2$.  This is inconsistent with the late-time decline predicted by the universal nova decline law of \citet[$\alpha=-3.5$]{2006ApJS..167...59H,2007ApJ...662..552H}, but is consistent with the `middle' part of their decline law ($\alpha=-1.75$).  Based on such a power law decline in this phase we would predict a timescale of 30--35\,days to return to the quiescent luminosity.

\subsection{Light curve color evolution}

In Figure~\ref{color_lc} we present the dereddened color evolution of the 2013--2015 eruptions of \novak\ (blue, red, and black points, respectively).  With the exception of the UV data, here, color data are only provided where there are near-simultaneous multi-color observations available from the same facility.  We also note that the $(u'-B)_0$ and $(V-r')_0$ plots contain a mix of photometric systems (Vega and AB); no attempt was made to correct between the photometric systems due to the non-black body nature of the \novak\ spectra.  In order to provide better temporal matches with ground-based 2015 data, the UV data from all eruptions are combined here.

Ground-based $u'$ and $z'$-band data were only collected during the 2015 eruption, as such the coverage in the $(u'-B)_0$ and $(i'-z')_0$ colors is less complete; the $(i'-z')_0$ plot is also compounded by crowding.  The color plots all cover the final rise of the eruption (from $t\simeq0$ until $t\simeq1$\,d).  The $(u'-B)_0$, $(B-V)_0$, $(V-r')_0$, and $(r'-i')_0$ plots all indicate that the emission from the system is becoming redder during this phase -- as might be expected if the pseudo-photosphere was still expanding at this stage (however, the SED snapshots during the final rise do not show evidence of a change in slope, see Section~\ref{sed_sec}).   However, as will be discussed in Sections~\ref{sec:vis_spec} and \ref{sed_sec}, even at these early times line emission in the visible spectra is already important and this may significantly affect the  color behavior.

From $t\simeq1$ to $t\simeq4$\,d -- during the linear early-decline phase, the $(B-V)_0$ and $(V-r')_0$ plots exhibit a linear  evolution in the color; although, interestingly, in $(B-V)_0$ the emission becomes significantly bluer, whereas the opposite is true for $(V-r')_0$.  The $(V-r')_0$ evolution is almost certainly affected by the change in the H$\alpha$ line profile and flux, see again Section~\ref{sec:vis_spec}.  The $(u'-B)_0$ data, albeit sparser, initially become bluer, but appears to stabilize around day~3.  Whereas $(r'-i')_0$ continues to redden until day~2 and then becomes systematically bluer.  In general, the very uniform pan-chromatic linear early-decline seen from the NIR to the NUV in this phase, is not replicated in the color data, probably due to the additional complications of line emission.

The color behavior during the plateau phase ($4\la t\la8$\,d) is again varied.  The $(u'-B)_0$ color remains approximately constant, although there is some variability.  Again $(B-V)_0$ and $(V-r')_0$ colors have opposing behavior, the former becoming redder, the latter bluer.  This behavior may again be related to line emission, but with no spectra beyond day 5 (see Section~\ref{sec:vis_spec}) we can only speculate; the trends seen in these colors may be due to diminishing Balmer emission with increased nebular line emission (e.g.\ [O\,{\sc iii}]\,4959/5007\AA).  If we compare to the behaviour observed from the 2006 eruption of RS~Oph, \citet{2009A&A...505..287I} reported that between days 50 and 71 a broad component of the [O\,{\sc iii}] lines began to grow, peaking in intensity around day 90, a similar analysis was reported by \citet{2009ARep...53..203T}.  These time-scales are roughly consistent with the SSS evolution as reported by \citet[also see references therein]{2011ApJ...727..124O}, with the SSS being roughly constant in luminosity between days 45 and 60.  We also note that somewhat of a plateau phase is observed between days 50--76 \citep[in $B$- and $V$-band data;][]{2010ApJS..187..275S}.  The effective consistency of these three time-scales in RS~Oph supports our prediction of nebular emission driving the color evolution during the plateau phase in \novak.

As the color plots enter the SSS-decline $t\ga8$\,d, where there are data, $(\mathrm{uvw1}-u')_0$, $(u'-B)_0$, and $(B-V)_0$, the color of the system remains approximately constant during the later part of the SSS phase.  However, the $(\mathrm{uvw1}-u')_0$ and $(u'-B)_0$ color plots show a marked shift to the blue as the SSS begins to turn-off.

\subsection{Color--magnitude evolution}\label{col:mag}

\begin{figure*}
\begin{center}
\includegraphics[width=\textwidth]{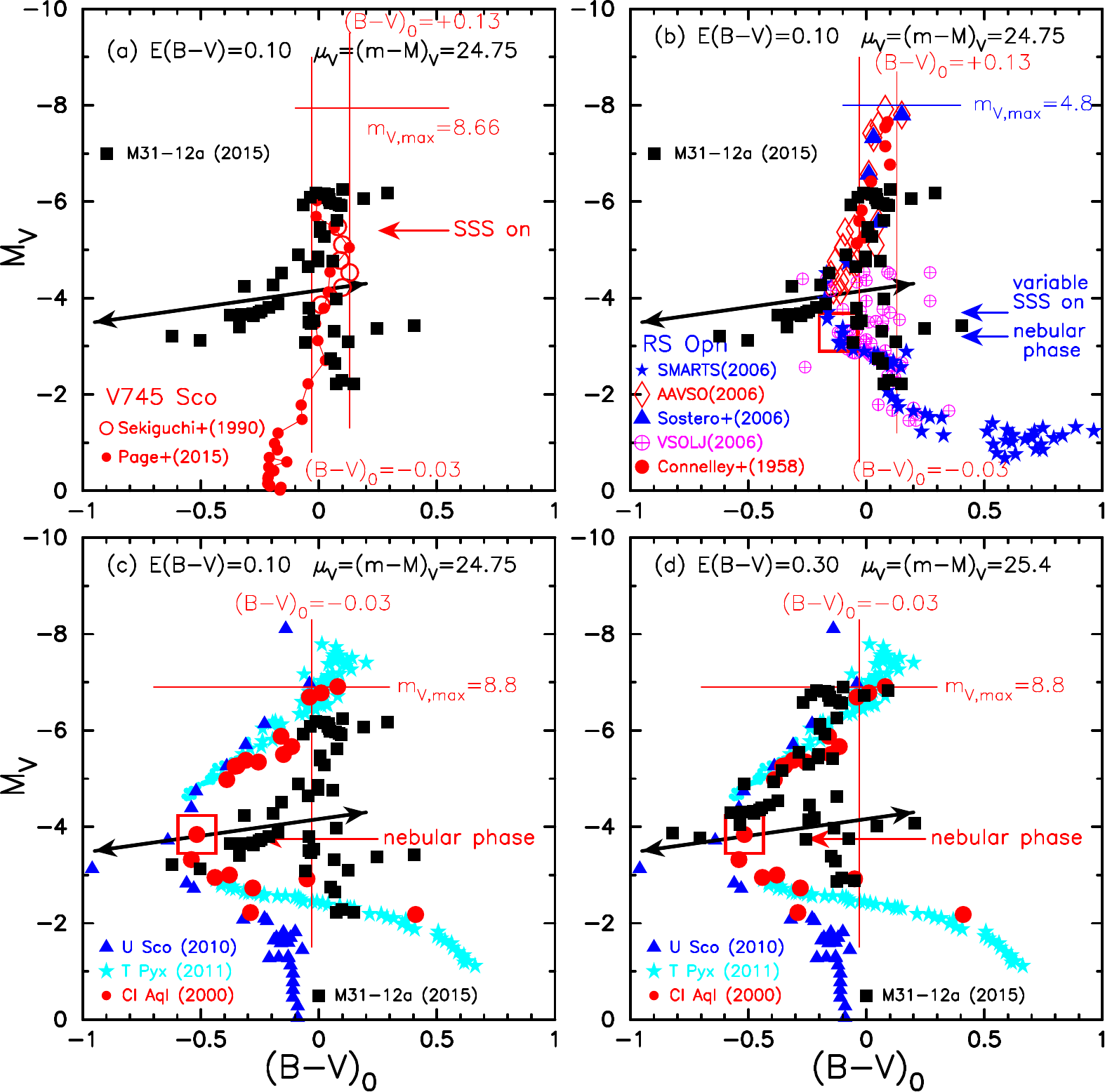}
\end{center}
\caption{Color-magnitude diagrams of \novak\ for various extinctions and {\it apparent} distance moduli (indicated at the top of each plot) are compared with those of Galactic RNe.  Throughout, the filled black squares denote the color-magnitude points of \novak\ from the 2013, 2014, and 2015 eruptions.  In each plot the evolution of \novak\ is compared directly to (a) V745~Sco (RG-nova), (b) RS~Oph (RG-nova), (c-d) U~Sco and CI~Aql (SG-novae), and T~Pyx (MS-nova); see the plot keys and text for further details.  The horizontal lines labeled ``$m_\mathrm{v,max}$'' show the maximum brightnesses of (a) V745 Sco, (b) RS Oph, and (c-d) CI Aql.  The vertical red lines show intrinsic colors of optically thick ($(B-V)_0=-0.03$) and optically thin ($(B-V)_0=+0.13$) winds \citep[see][for further detail]{2014ApJ...785...97H}.  In panel (a) the red arrow labeled ``SSS on'' indicates the optical luminosity at the SSS turn-on of V745~Sco. In panel (b) the blue arrows labeled ``variable SSS on'' and ``nebular phase'' indicate the onsets of the variable SSS phase and nebular phase of RS~Oph. In panel (c) and (d) the red arrows indicate where the nebular phase of CI~Aql started.  The two-headed black arrows indicate where the color-magnitude tracks of some novae show a turning from toward blue to toward red \citep[see][for more detail]{2016ApJS..223...21H}.  In plot (d) the extinction toward \novak\ was allowed to vary for illustrative purposes.  Given the known extinction, the track of \novak\ is closer to those of the RG-novae (a-b) than the other RNe (c-d), consistent with the interpretation from the eruption spectra that the companion is a red giant.\label{col-mag-evo}}
\end{figure*}

Color-magnitude diagrams of RNe are useful to distinguish the evolutionary stage of the companion star.  \citet{2016ApJS..223...21H} demonstrated a clear difference between the color-magnitude tracks of eruptions from systems hosting a red giant companion (RG-nova) and those having a sub-giant or main sequence companion (SG- or MS-nova).  The color-magnitude track evolves almost vertically along the line of $(B-V)_0=-0.03$ (the intrinsic color of optically thick free-free emission; as shown in Figure~\ref{col-mag-evo}, along with $(B-V)_0=0.13$ for optically thin free-free emission; also see the discussion in Section~\ref{sed_sec}) for RNe harboring a red giant companion, such as V745~Sco \citep[2014 eruption, data from][also see Section~\ref{sec:v745}]{2015MNRAS.454.3108P} and RS~Oph \citep[1958, 1985, and 2006 eruptions, data from][AAVSO\footnote{American Association of Variable Star Observers, \url{https://www.aavso.org}}, VSOLJ\footnote{Variable Star Observers League in Japan, \url{http://vsolj.cetus-net.org/}}, and SMARTS\footnote{The Stony Brook/SMARTS Spectral Atlas of Southern Novae, \url{http://www.astro.sunysb.edu/fwalter/SMARTS/NovaAtlas}, see \citet{2012PASP..124.1057W}}; see \citealt{2016ApJS..223...21H} for full details]{1958PASP...70..600C,2006IAUC.8673....3S,2006IAUC.8681....4S,2006CBET..502....2S,2008ASPC..401..206H}; see Figures~\ref{col-mag-evo}(a) and (b), respectively.  On the other hand, the track goes blueward and then turns back redward near the two-headed arrow, as shown in Figure~\ref{col-mag-evo}(c), for RNe with a sub-giant or main sequence companion, for example, U~Sco, CI~Aquilae, and T~Pyxidis \citep[data from][VSOLJ, and AAVSO/SMARTS, respectively]{2015ApJ...811...32P}.  The tracks of these three RNe are very similar to each other and clearly different from those for V745~Sco and RS~Oph. Color-magnitude diagrams are plotted for only five Galactic RNe due of the general lack of pan-chromatic (X-ray/UV/visible) eruption data for Galactic RNe \citep[see discussion in][]{2016ApJS..223...21H}.

If we adopt the newly constrained extinction of $E_{B-V}=0.10$ \citep{HST2016} and therefore the {\it apparent} distance modulus $\mu_V=24.75$ \citep{1990ApJ...365..186F}, the track of \novak\ appears closer to those of V745~Sco and RS~Oph rather than those of U~Sco, CI~Aql, and T~Pyx, as shown in Figure~\ref{col-mag-evo}(a-c).  This is consistent with the interpretation, drawn in this paper from the eruption spectroscopy, that the companion in \novak\ is a red giant.

However, it should be noted that the position of the color-magnitude track depends strongly on the assumed extinction (and distance).  If we increase the value of the extinction, for example, $E_{B-V}=0.30$ (as originally proposed in \otwok; see Figure~\ref{col-mag-evo}(d)), the track moves closer to those of U~Sco, CI~Aql, and T~Pyx.

The conclusion reached here differs from that reported in \citet{XrayFlash}, who favored \citep[based, partly on the \novak\ color--magnitude diagram published by][]{2016ApJS..223...21H} that the companion is a sub-giant.  This earlier analysis used the less detailed data available at the time, but also had no strong constraint on the extinction.  It should be noted that the color--magnitude analysis presented in this paper therefore supersedes that for \novak\ presented by \citet{2016ApJS..223...21H}.

\subsection{The X-ray variability}\label{sec:lc_xvar}

\begin{figure}
\includegraphics[width=\columnwidth]{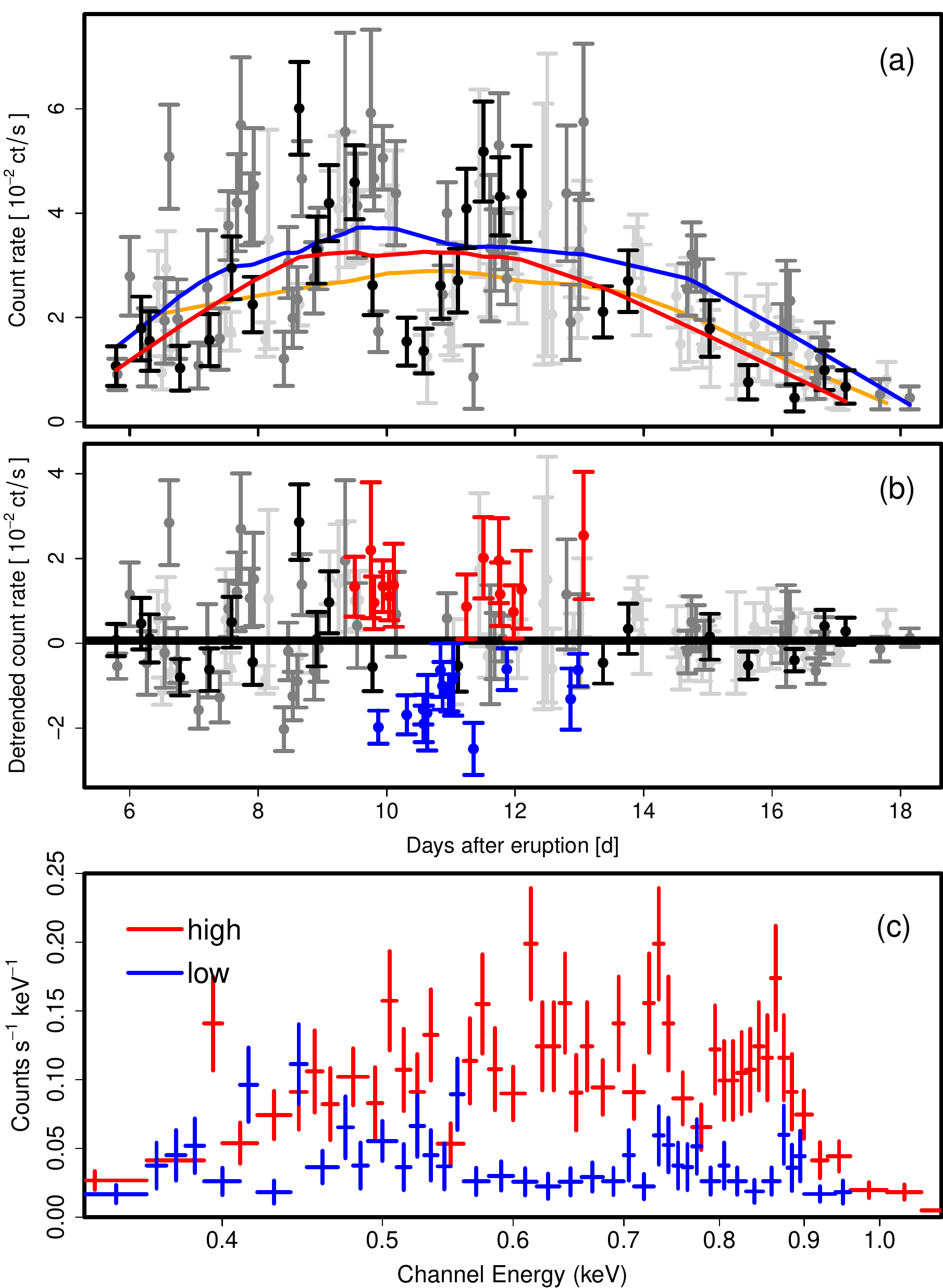}
  \caption{\textit{Panel a}: The short-term X-ray light curve of \nova based on the individual XRT snapshots. Data points with error bars show the XRT count rates and corresponding errors for the 2015 (black), 2014 (dark gray), and 2013 (light gray) eruptions. Note that the count rate axis uses a linear scale in contrast to the logarithmic scale in Figure~\ref{fig:xray_lc}. Solid lines represent smoothed fits, based on local regression, on the 2015 (red), 2014 (blue), and 2013 (orange) data. The three eruptions display a very similar behavior \textit{Panel b}: The light curves from panel a have been de-trended by subtracting the smoothed fits from the respective data. The red data points mark the count rates that are at least $1\sigma$ above the smoothed fit for the 2015 (2014/13) data during the temperature maximum. The blue data are at least $1\sigma$ below the average 2015 (2014/13) count rate for the same time range. The drop in variability amplitude around day 13 is clearly visible in all three eruptions. \textit{Panel c}: Binned XRT spectra for all the high- (red colors) and low-luminosity (blue colors) snapshots of the 2015/14/13 monitoring that are indicated in panel b with corresponding colors. There are indications that the spectra are different in more than overall luminosity (see Sect \ref{sec:disc_xvar} and Table \ref{tab:xspec}). \label{fig:xray_split}}
\end{figure}

The SSS phase variability is examined in detail in Figure~\ref{fig:xray_split}. There, we show the 2013--2015 XRT count rates based on the individual XRT snapshots. In the case of the 2015 data, there is no difference between the count rates binned by ObsID (see Figure~\ref{fig:xray_lc}) because all detections during the SSS phase only consisted of single snapshots.

The early high-amplitude variability is clearly visible in particular in Figure~\ref{fig:xray_split}b. During the 2015 campaign we collected only a few observations during the late SSS phase. However, the combined light curve of the last three eruptions suggests a relatively sudden drop in variability after day 13.

As in \xtwok, we identified snapshots with count rates significantly above or below the (smoothed) average for the time around the SSS maximum. Those measurements are marked in Figure~\ref{fig:xray_split}b in red (high rate) or blue (low rate). The combined XRT spectra of these data points for all three eruptions are shown in Figure~\ref{fig:xray_split}c using the same color scheme. Those spectra are discussed in the context of spectral variability in Section~\ref{sec:disc_xvar} below. All three eruptions show a consistent factor of 2.6 in difference between high- and low-count rate snapshots.

We note that in 2015 there appears to be less variability during the first two days of the SSS phase than in 2013 and 2014. This is reflected in a less significant statistical difference between the X-ray count rate before and after day 13. An F-test results in a p-value of $0.03$ which, while still significant at the 95\% confidence level, is considerably reduced with respect to the 2013 (2.1\tpower{-6}) and 2014 (1.8\tpower{-5}) results (see \xtwok).

\begin{figure}
\includegraphics[width=\columnwidth]{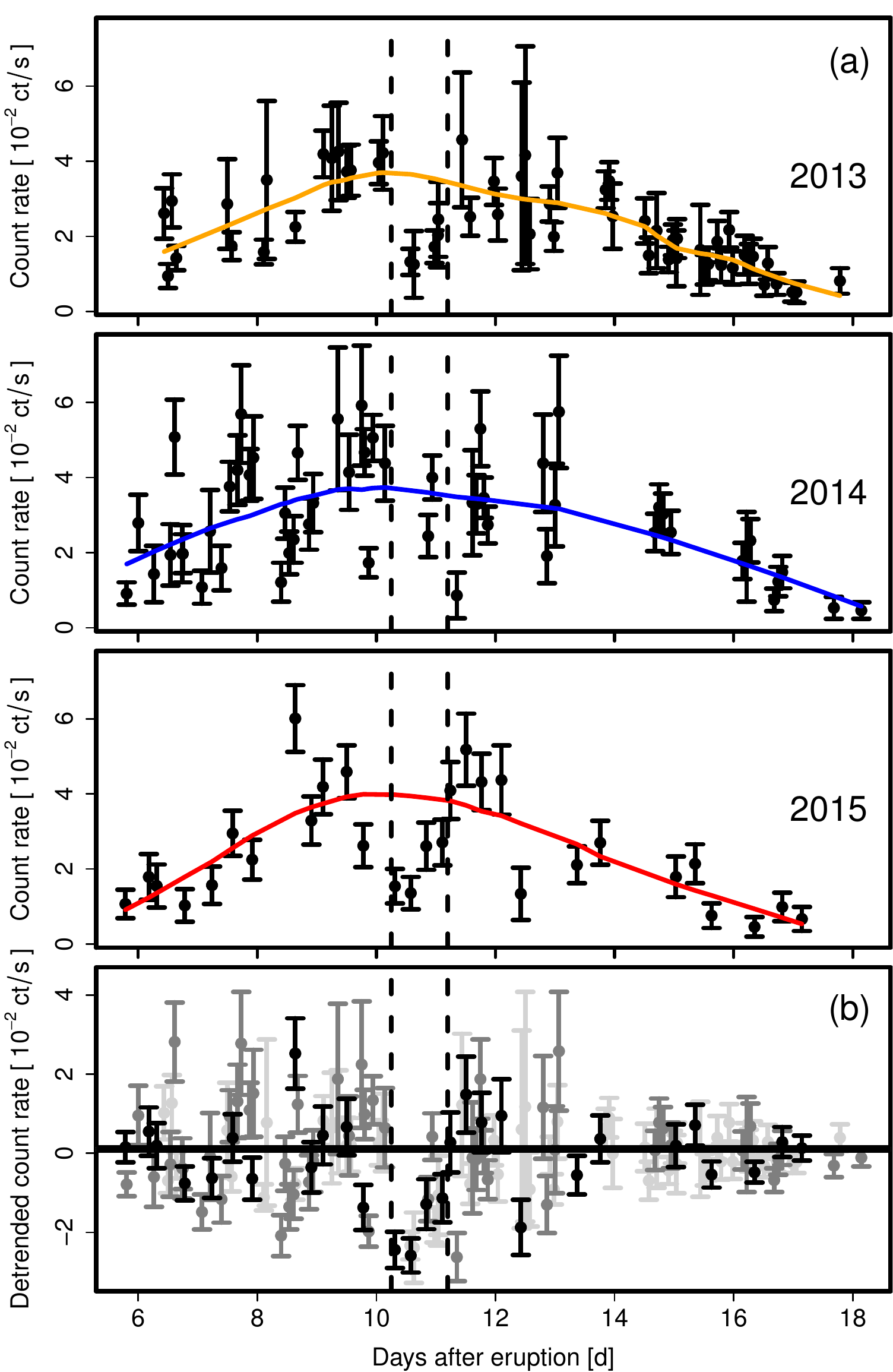}
  \caption{Same as Figure~\ref{fig:xray_split} for the individual XRT snapshot light curves (\textit{panel a}) and the combined de-trended light curves (\textit{panel b}). Here, the smoothed fits (solid lines with the same colors as in Figure~\ref{fig:xray_split}a) were determined excluding the data points within the time range indicated by the vertical dashed lines. Panel b shows the 2015 data in black overlaid on the 2013/2014 data in light/dark gray. A possible dip on days 10--11 is clearly visible in 2013 and 2015, with 2014 having insufficient data to reject such a feature. \label{fig:xray_dip}}
\end{figure}

In fact, the SSS variability in 2015 might be almost entirely explained by a dip in flux on day 10--11. To investigate this possibility we plot the three XRT snapshot light curves separately in Figure~\ref{fig:xray_dip}a. The smoothed fits now exclude a 1\,day window centered on day 10.75, during which the 2015 dip occurred. Interestingly, there seems to be a similar feature in the 2013 light curve. For both years the X-ray flux dropped by a factor of $\sim2$ during this window. In 2014 there is no clear dip during this time. However, there were only two snapshots within the 1\,d window.

The 2014 X-ray light curve might instead show a dip between days eight and nine, during which time there were fewer observations in 2013 and 2015 (see Figure~\ref{fig:xray_dip}a). In any case, all light curves display additional variability besides the potential dip features. This can be seen in Figure~\ref{fig:xray_dip}b where we subtracted the smoothed fits in Figure~\ref{fig:xray_dip}a to highlight the potential dip and residual variability.

If the potential dip as the main source of variability is removed then the light curves of the 2013 and, in particular, the 2015 SSS phase appear to show significantly less residual variability (see Figure~\ref{fig:xray_dip}). However, the 2014 light curve does not seem to show the same behavior. Clearly, high-cadence coverage of several future observations is needed for a proper statistical treatment of this peculiar variability.

Interestingly, the ROSAT light curve of the 1993 detection in \citet{1995ApJ...445L.125W} might also show a tentative, one-bin dip between day 9 and 10 (days 10 and 11 in the lower panel of their Figure 2). The ROSAT data of the preceding 1992 detection only extend to about day 8 after eruption, but shows significant variability over its coverage.

As in 2013 and 2014 there is no evidence for any periodicities during the SSS variability phase (Figure \ref{fig:xray_split}b), according to a Lomb--Scargle test \citep{1976Ap&SS..39..447L,1982ApJ...263..835S}. The apparently aperiodic variability is present on all accessible time scales (hours to several days) and the amplitude shows no significant relation with the frequency. We have been granted a 100\,ks \xmm (RGS) ToO observation to study the (spectral) variability of \nova with higher time resolution in a future eruption. However, because of the stringent (anti-)sun constraints of the \xmm observatory there are only two possible observing windows in Jan. - mid Feb. and Jul. - mid Aug., respectively. Given the remaining uncertainty in predicting future eruption dates (see Section \ref{sec:prec}) a successful \xmm trigger might take several years time.

Due to the very short duration of the SSS plateau phase and the XRT count rate of the source our analysis was only sensitive to periods of a few hours to a few days \citep[and sensitive to amplitudes larger than 1.5\cts{-2} on the 99\% confidence level; following][]{1982ApJ...263..835S}. This time range includes typical orbital periods of Roche lobe-overflow RNe, e.g.\ U~Sco with $\sim 1.2$\,d \citep{2012ApJ...745...43N} or Nova LMC~2009a also with $\sim 1.2$\,d \citep{2016ApJ...818..145B}. Spin periods of high-mass WDs in CVs without strong magnetic fields (i.e.\ not polars) are typically shorter \citep[several 100--1000\,s see, for example,][]{2004ApJ...614..349N}. For instance, a period of $\sim 1110$\,s was reported for the suspected intermediate polar \citep[and suggested RN; see][]{2009ApJ...705.1056B} M31N\,2007-12b \citep{2011A&A...531A..22P}. Polars, like the old nova V1500 Cyg \citep[see, for example,][]{2011Ap.....54...36L} have generally longer spin cycles of several hours due to a magnetic synchronization of the orbital and spin periods that slow down the WD rotation \citep[see, for example,][]{2004ApJ...614..349N}. Even shorter transient periods $<100$\,s have been found in the RNe RS~Oph (35\,s) and LMC~2009a (33\,s), as well as in a few other CNe and the canonical SSS Cal~83 by \citet{2015A&A...578A..39N}, who discuss pulsation mechanisms as the possible origin.

\section{Panchromatic eruption spectroscopy}\label{sec:vis_spec}

\begin{figure*}
\includegraphics[width=\textwidth]{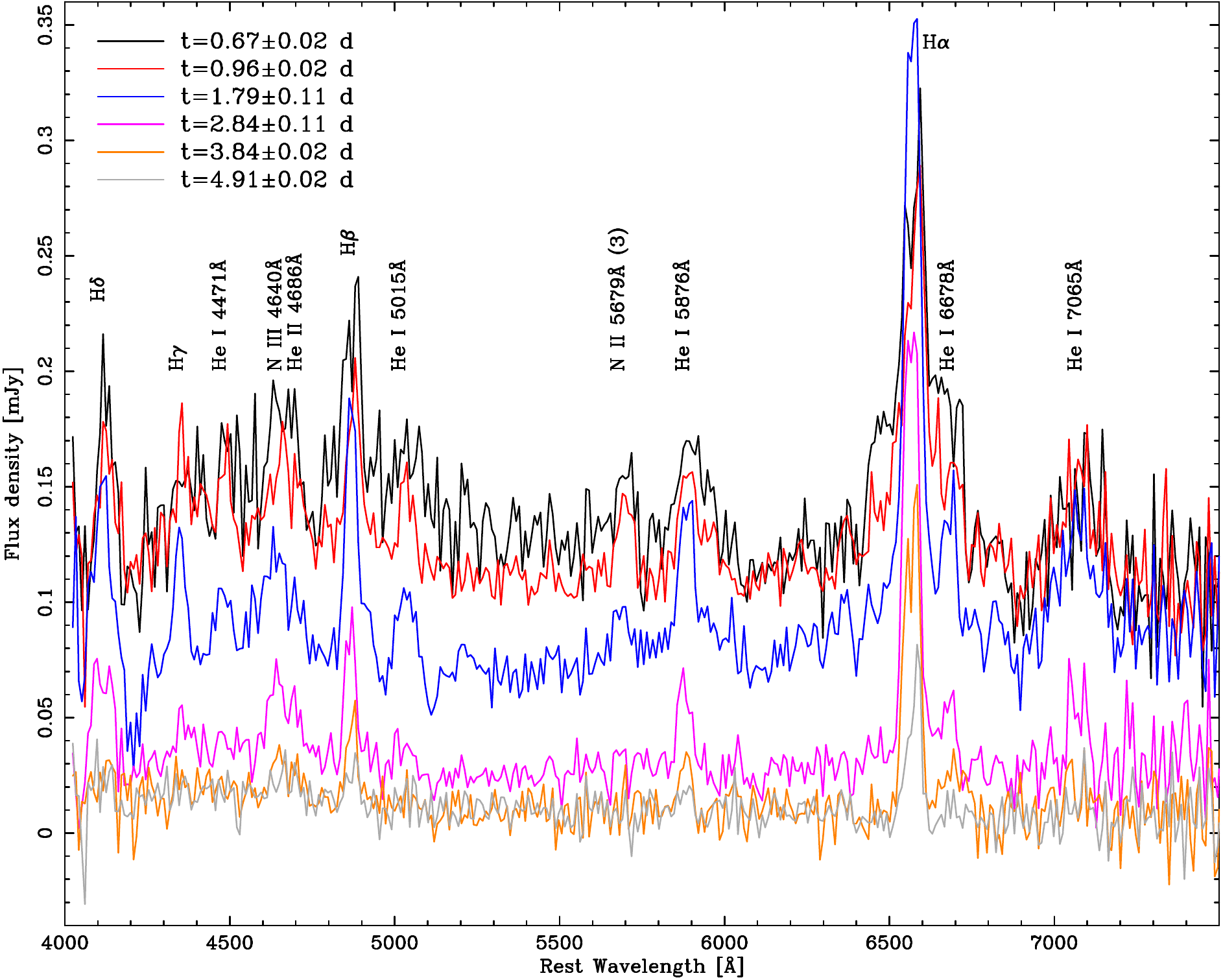}\\
\includegraphics[width=\textwidth]{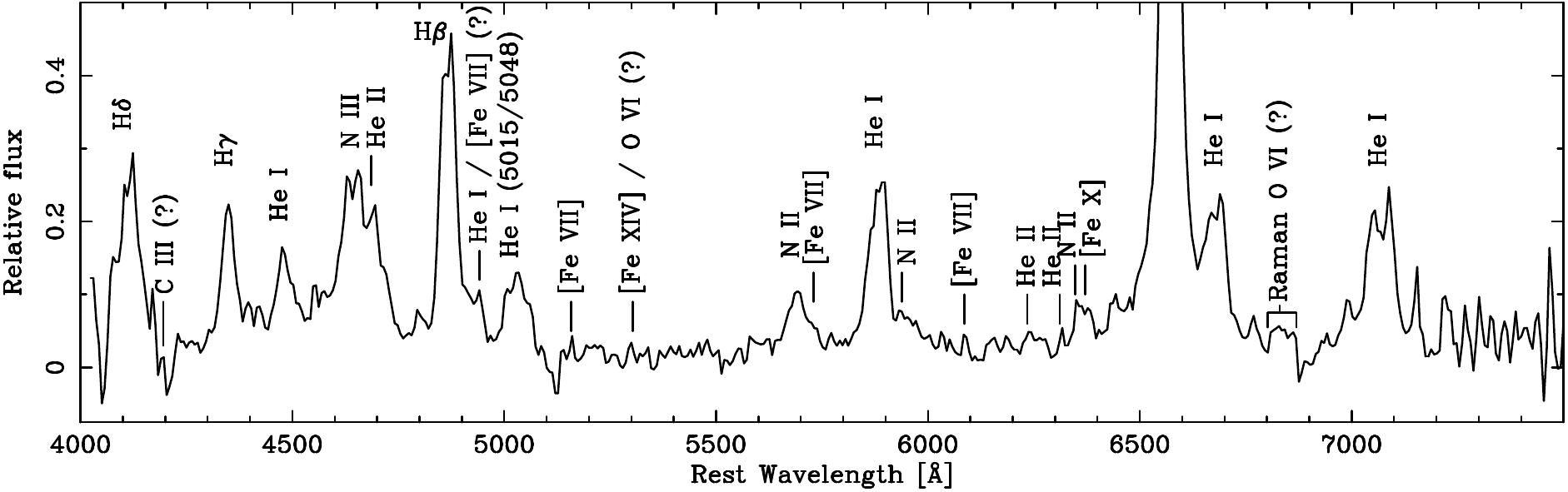}
\caption{{\bf Top:} Liverpool Telescope SPRAT flux calibrated spectra of the 2015 eruption of \novak, see the key for line identifiers; the continuum flux decreases for each successive spectrum.  Spectra are consistent with the He/N canonical class of nova. {\bf Bottom:} Combined spectra from the 2012, 2014, and 2015 eruptions; see text for details. Additional features, not identifiable in the individual spectra are indicated.  These include tentative detections of the coronal [Fe\,{\sc vii}], [Fe\,{\sc x}], and [Fe\,{\sc xiv}] emission lines, typically associated with shocks between the ejecta and surrounding material, and possibly the Raman scattered O\,{\sc vi} emission band, a signature of symbiotic stars.  The spectra shown in the top panel of this figure are available as the Data behind the Figure.\label{optical_spec}}
\end{figure*}

The earliest spectroscopic observations of \novak\ prior to the 2015 eruption were obtained by the William Herschel Telescope (WHT) 1.27\,days after the 2014 eruption (\otwok).  Following the 2015 eruption, the first three visible spectra were obtained at 0.67\,days, 0.96\,days, and 1.10\,days post-eruption.  For a nova with a $V$-band decline time as fast as $t_{2}=1.65$\,days (see Table~\ref{t2_table}), the early 2015 spectra capture significantly earlier portions of the eruption than had been seen previously.  Additionally, with the peak $V$-band luminosity occurring 1.01\,days after eruption (see Table~\ref{t2_table}), the first two 2015 spectra were taken while the nova light curve was still rising in the visible -- but notably 0.01\,d and 0.3\,d {\it after} the UV light curve peak.  The final 2015 spectrum captures the eruption 0.3\,days later than any previous spectra.  All of the flux-calibrated spectra of the 2015 eruption are shown in the top portion of Figure~\ref{optical_spec}.

An initial summary following the first spectrum of the 2015 eruption of \novak\ was reported in \citet{2015ATel.7965....1D}.  As in 2012 \citep{2012ATel.4503....1S}, 2013 (\ponek), and 2014 (\otwok), the individual 2015 spectra are dominated by hydrogen Balmer series emission lines (H$\alpha$ through H$\delta$).  Emission lines from He\,{\sc i} (4471, 5015, 5876, 6678, and 7065\,\AA), He\,{\sc ii} (4686\,\AA), N\,{\sc ii} (5679\,\AA), and N\,{\sc iii} (4640\,\AA) are also clearly visible, but appear to fade significantly in the later spectra.  There is a clear detection of continuum emission in each of the spectra.  Despite collecting spectra from much earlier in the eruption process, no clear absorption components (i.e.\ P~Cygni profiles) are seen in any of the spectra, which points to a low mass ejecta.  No Fe\,{\sc ii} or O\,{\sc i} lines, characteristic of `Fe\,{\sc ii} novae', or any Ne lines, are detected in the individual spectra.   As in previous eruptions, the observed spectral lines and velocities (see Sections~\ref{sec:line_morph} and \ref{sec:exp_vel}) are consistent with the eruption of a nova belonging to the He/N taxonomic class \citep{1992AJ....104..725W,2012AJ....144...98W,1994ApJS...90..297W}.

\subsection{Multi-eruption combined visible spectrum}\label{sec:comb_spec}

In the bottom plot of Figure~\ref{optical_spec} we present a combined spectrum using data from the 2012 (HET), 2014 (LT and WHT), and 2015 (LT, LCOGT, and KPNO) eruptions.  Here we have re-sampled all spectra to the wavelength scale of the LT SPRAT data (linear 6.4\,\AA\ per pixel), re-scaled, and median combined the data.  We have excluded the final epoch data from 2014 and 2015 as the signal-to-noise of these spectra were particularly low.  As such, this combined spectrum covers the period from 0.67 to 3.84\,days post-eruption.  When accounting for the relevant exposure time and telescope collecting area, this combined spectrum would be the equivalent of a single 48\,ks spectrum as taken by the LT with SPRAT -- by far the deepest spectrum of an \m31 nova yet obtained.  The combined spectrum is, as expected, very similar to the individual spectra, but a number of fainter features increase in significance.  For example, we note that the He\,{\sc i} (5015\,\AA) line identified in the individual spectra is likely a blend of the He\,{\sc i} 5015 and 5048\,\AA\ lines.  In the combined spectrum, there is still no convincing evidence for the presence of Fe\,{\sc ii}, O\,{\sc i}, or Ne lines.

Newly visible lines at $\sim4200$ and $\sim4542$\,\AA\ are roughly coincident with the H-like He\,{\sc ii} Pickering series \citep[transitions to the $n=4$ state]{1896ApJ.....4..369P}, many stronger Pickering lines are blended with the Balmer series, but the apparent lack of the He\,{\sc ii} (5412\,\AA) line makes these identifications unlikely.  The line at $\sim4200$ may therefore be C\,{\sc iii} (4187\,\AA).  The second line remains unidentified, and we believe it is unlikely to be Fe\,{\sc ii} (4549\,\AA) due to the lack of other visible multiplet 38 lines \citep[see][]{1945CoPri..20....1M}.  Here, we also note that a number of apparently strong lines in the combined spectrum remain unidentified.

One of the most prominent (see Figure~\ref{optical_spec}) features newly resolved in the combined spectrum is the double-peaked line at $\sim6360$\,\AA. If this nova belonged to the Fe\,{\sc ii} taxonomic class, then the most likely identification of this feature would be the Si\,{\sc ii} (6347/6371\,\AA) doublet.  However, despite the greatly improved signal-to-noise of the combined spectrum there remains no convincing, and self-consistent, evidence of any other defining lines of the Fe\,{\sc ii} class (e.g.\ complete sets of Fe\,{\sc ii} multiplets themselves or O\,{\sc i} lines).  As such, we instead tentatively identify this pair of lines as N\,{\sc ii} (6346\,\AA) and the coronal [Fe\,{\sc x}] (6375\,\AA) line.

The combined spectrum also presents tentative evidence of a full series of [Fe\,{\sc vii}] lines, of which there would be nine expected within the observed wavelength range \citep{1982A&A...113...21N}.  Here we address each of the [Fe\,{\sc vii}] line identifications separately.  

The [Fe\,{\sc vii}] (4698\,\AA) line would be blended with the He\,{\sc ii} emission seen at 4686\,\AA\ and therefore, due to its low radiative transition probability, unobservable.

Any [Fe\,{\sc vii}] (4893\,\AA) line would be blended with the strong H$\beta$ line and hence unobservable.  

The  [Fe\,{\sc vii}] line at 4942\,\AA\ may be observed as the small peak just redward of H$\beta$.  However, by virtue of the low transition probability of this line, a more likely identification for this feature would be the red most peak of a double peaked He\,{\sc i} (4922\,\AA) line (such a profile is observed for the other He\,{\sc i} lines) or possibly N\,{\sc v} (4945\,\AA).  However, we note that the corresponding, and similar probability, N\,{\sc v} (4604/4620\,\AA) lines would be mixed with the N\,{\sc iii}/He\,{\sc ii} blend \citep[a strong N\,{\sc v} (1240\,\AA) line is also reported by][in early FUV spectra]{HST2016}.  

There is a possible [Fe\,{\sc vii}] line at 4989\,\AA, but this is within the blue wing of the He\,{\sc i} (5015/5048\,\AA) blend.  

There is a tentative detection of  [Fe\,{\sc vii}] at 5158\,\AA, the only other possible identifications here would be other, less ionized, but still forbidden, Fe lines.  

There is no clear sign of an [Fe\,{\sc vii}] line at 5276\,\AA\ -- a line that could be confused with Fe\,{\sc ii} (5276\,\AA) if there were any other Fe\,{\sc ii} multiplet 49 lines present -- although it could be blended with the line just redward which may be [Fe\,{\sc xiv}] (5303\,\AA), but could in principle be O\,{\sc vi} (5292\,\AA; see later).  

The [Fe\,{\sc vii}] line at 5721\,\AA\ is tentatively detected in the red wing of the N\,{\sc ii} (5679\,\AA) multiplet (\#3), which otherwise appears broader than expected based on expected line strength ratios.  

The eighth [Fe\,{\sc vii}] line is seen at 6086\,\AA, another possible identification would be Fe\,{\sc ii} (6084\,\AA) but no other multiplet 46 lines are observed.  We also note that the [Fe\,{\sc vii}]  5721 and 6086\,\AA\ lines have the largest transition probabilities within this series \citep{1982A&A...113...21N}.

The final [Fe\,{\sc vii}] line at 6601\,\AA\ has the lowest transition probability, but even so, it would be blended with the H$\alpha$ emission and undetectable in such low resolution spectra.

After weighing all the above evidence we believe that it is likely that a series of [Fe\,{\sc vii}] lines, the [Fe\,{\sc x}] (6375\,\AA) line, and [Fe\,{\sc xiv}] (5303\,\AA) line are all visible in the combined spectrum of \novak.  The implication of the presence of these highly ionized forbidden lines is discussed in detail in Section~\ref{sec:decel}.

Finally, we point to the emission feature at $\sim6830$\,\AA.  As noted by \citet{2014A&A...570L...4S}, a possible interpretation of this line is `simply' emission from C\,{\sc i} (6830\,\AA) -- especially in CNe.  However, we also note that other \citep[typically stronger; see][]{NIST_ASD} C\,{\sc i} lines (e.g.\ 6014 and 7115\,\AA) are not seen in the combined spectrum.  \citet{1989A&A...211L..27N} and \citet{1989A&A...211L..31S} were the first to propose that an emission band at $\sim6830$\,\AA\ could be due to \citet{raman} scattering of the O\,{\sc vi} resonance doublet (1032/1038\,\AA) by neutral hydrogen.  As \citet{2014A&A...570L...4S} also points out, such Raman features are unlikely to be formed in the ejecta of MS- or SG-novae, but such features have been observed in the spectra of RG-novae \citep[most notably, RS~Oph;][]{1945ApJ...102..353J,1986PASP...98..875W,2009A&A...505..287I} and are common features of the wider group of symbiotic stars \citep{1980MNRAS.190...75A}.  We note that the weaker Raman band at 7088\,\AA\ would be blended with the strong He\,{\sc i} emission at 7065\,\AA.  Unfortunately, the 6830\,\AA\ Raman band is situated adjacent to the Telluric B-band (6867--6884\,\AA, from O$_2$) and as such the continuum subtraction around the Raman region may be unreliable, and the Raman identification should be treated with some degree of caution. However, we further discuss the potential Raman band emission in Section~\ref{sec:decel}, and we stress the importance of targeted  follow-up spectroscopy for future eruptions.

\subsection{Visible emission-line morphology}\label{sec:line_morph}

\begin{figure*}
\begin{center}
\includegraphics[width=0.95\columnwidth]{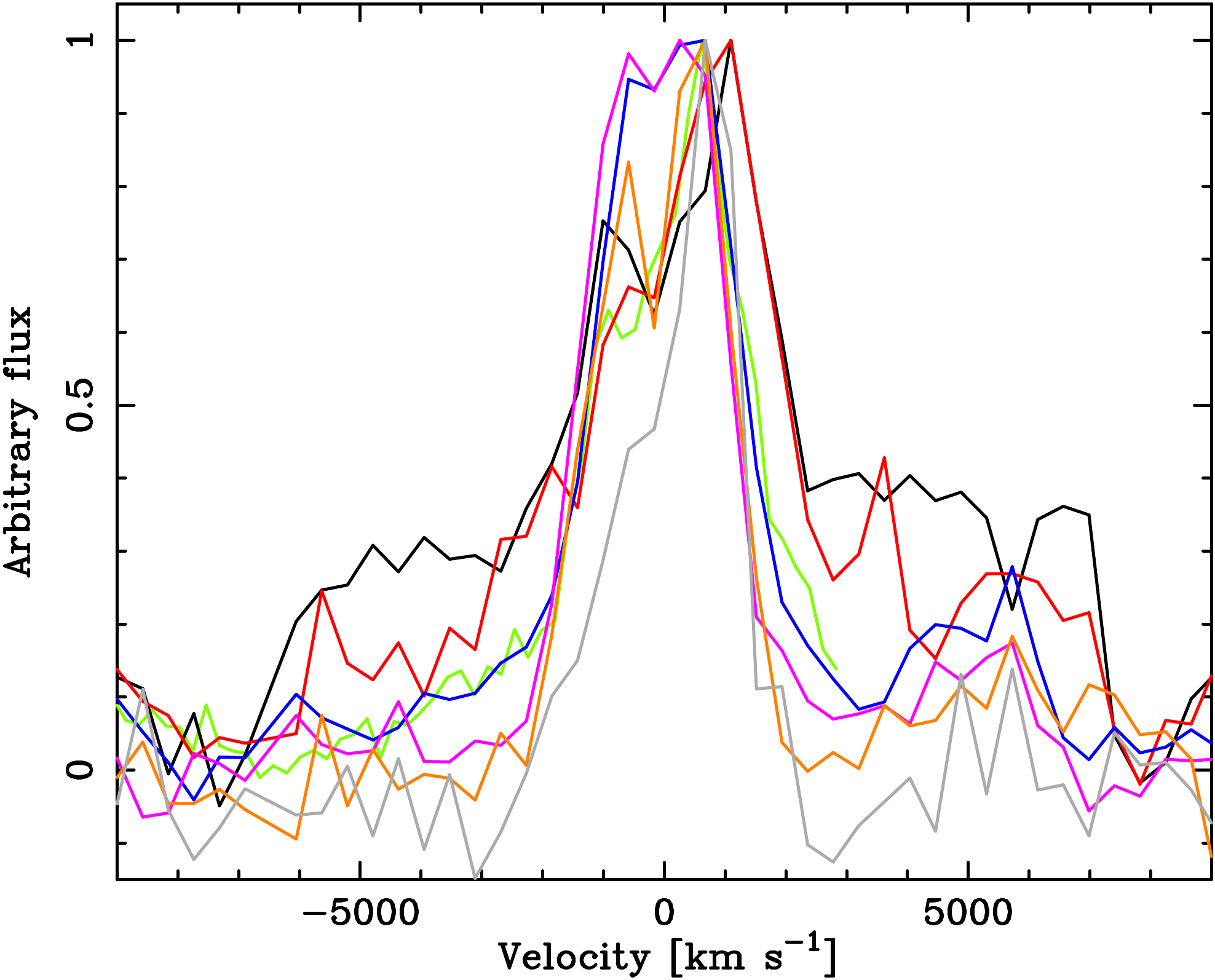}\hfill
\includegraphics[width=0.95\columnwidth]{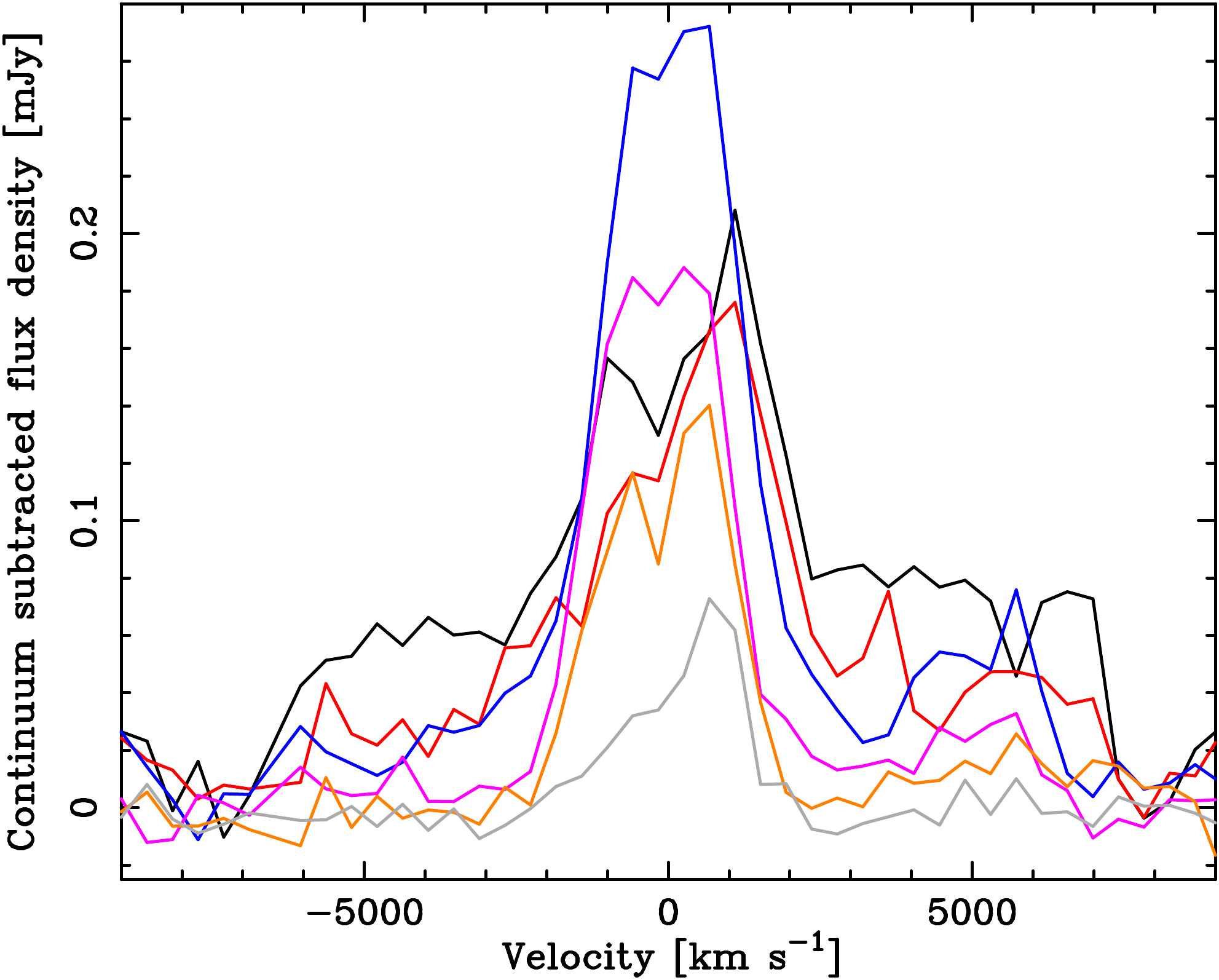}\\
\includegraphics[width=0.95\columnwidth]{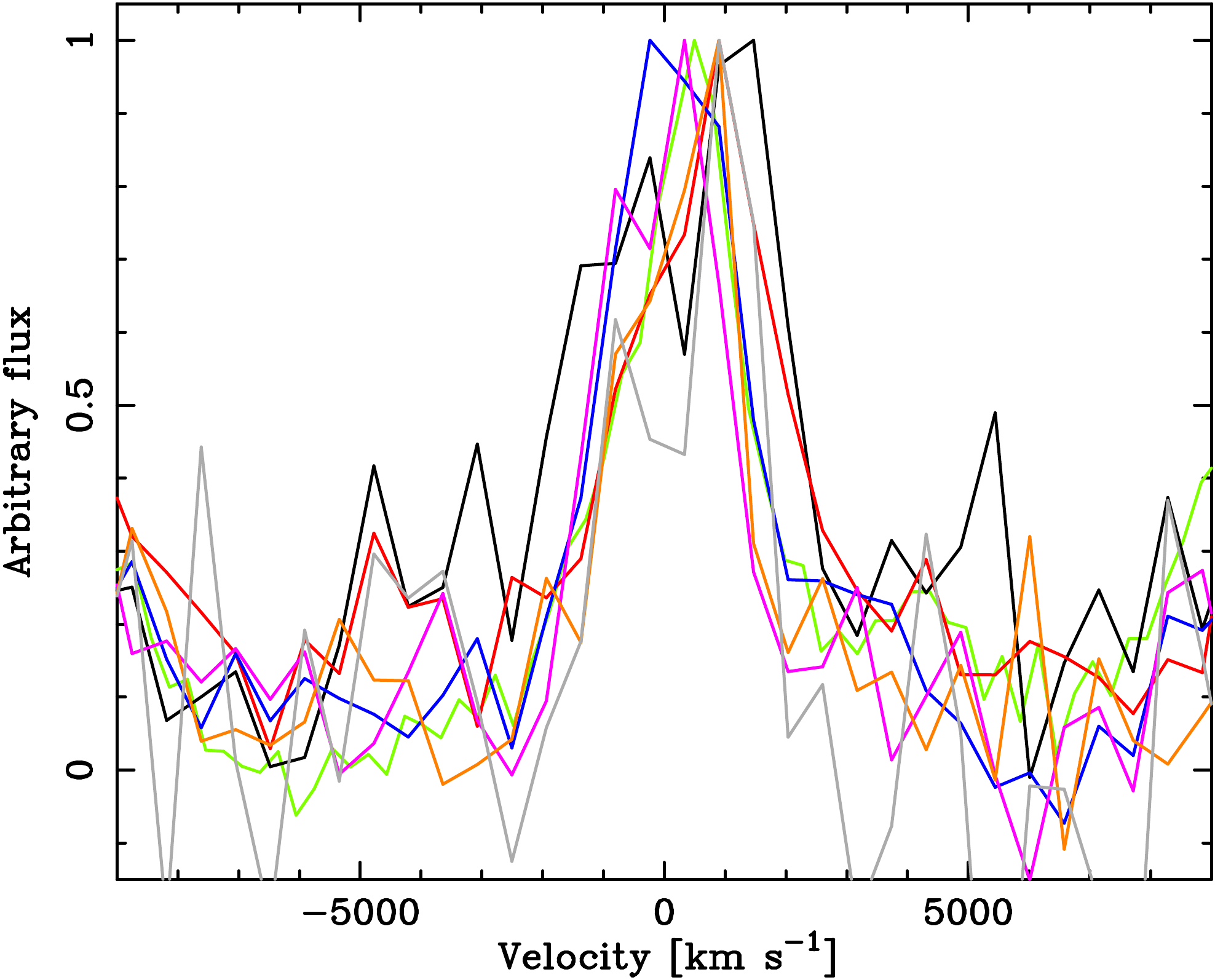}\hfill
\includegraphics[width=0.95\columnwidth]{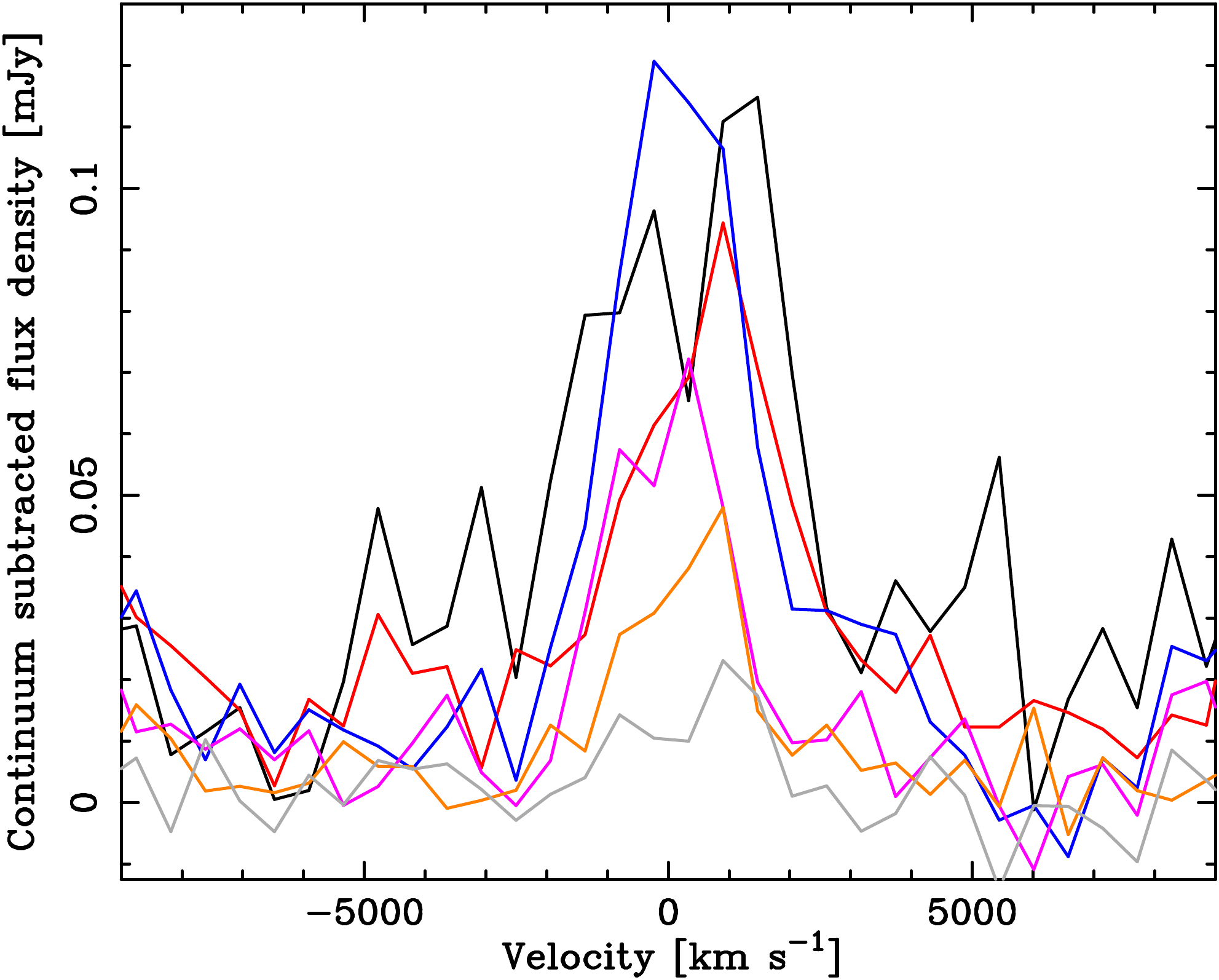}\\
\includegraphics[width=0.95\columnwidth]{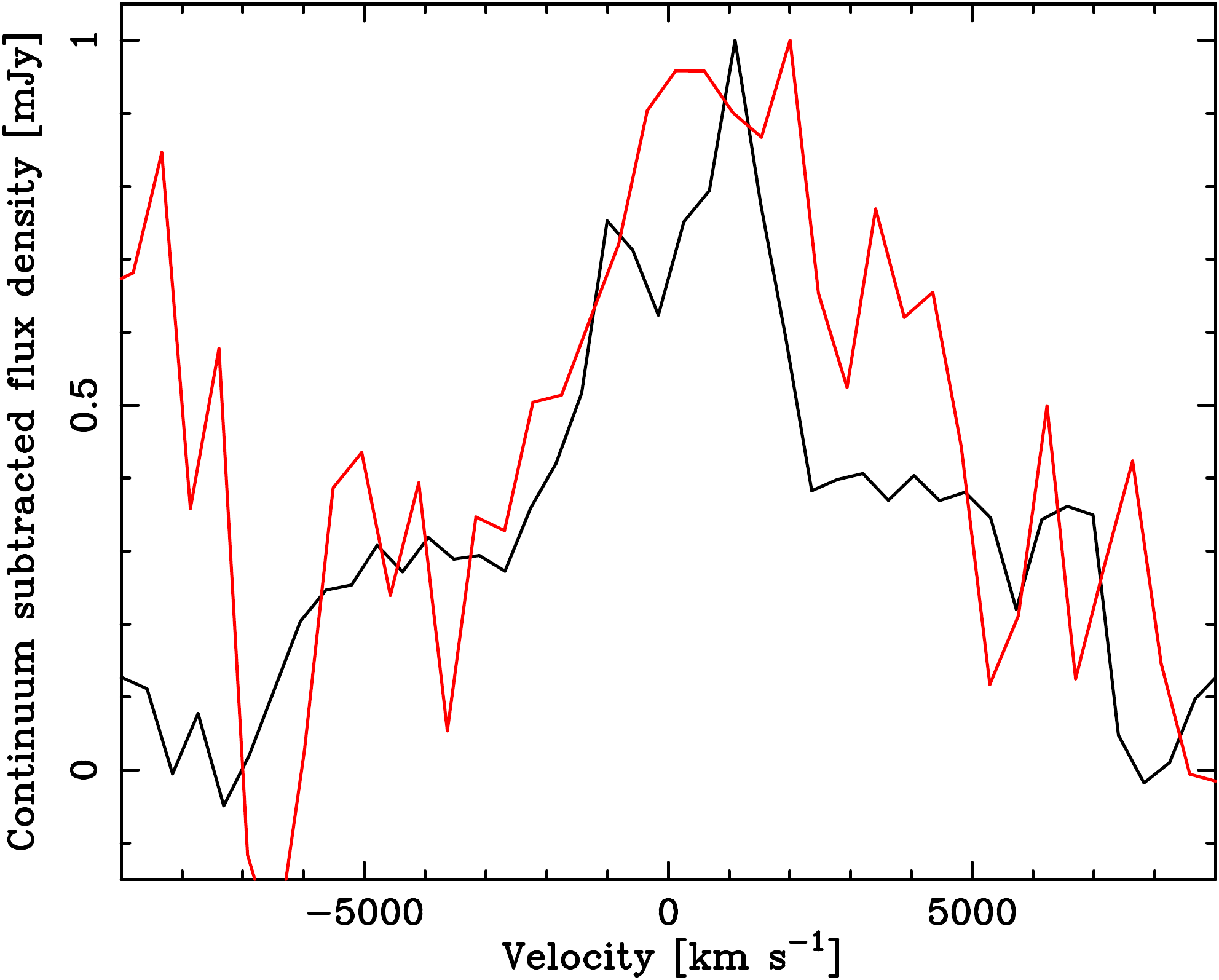}\hfill
\includegraphics[width=0.95\columnwidth]{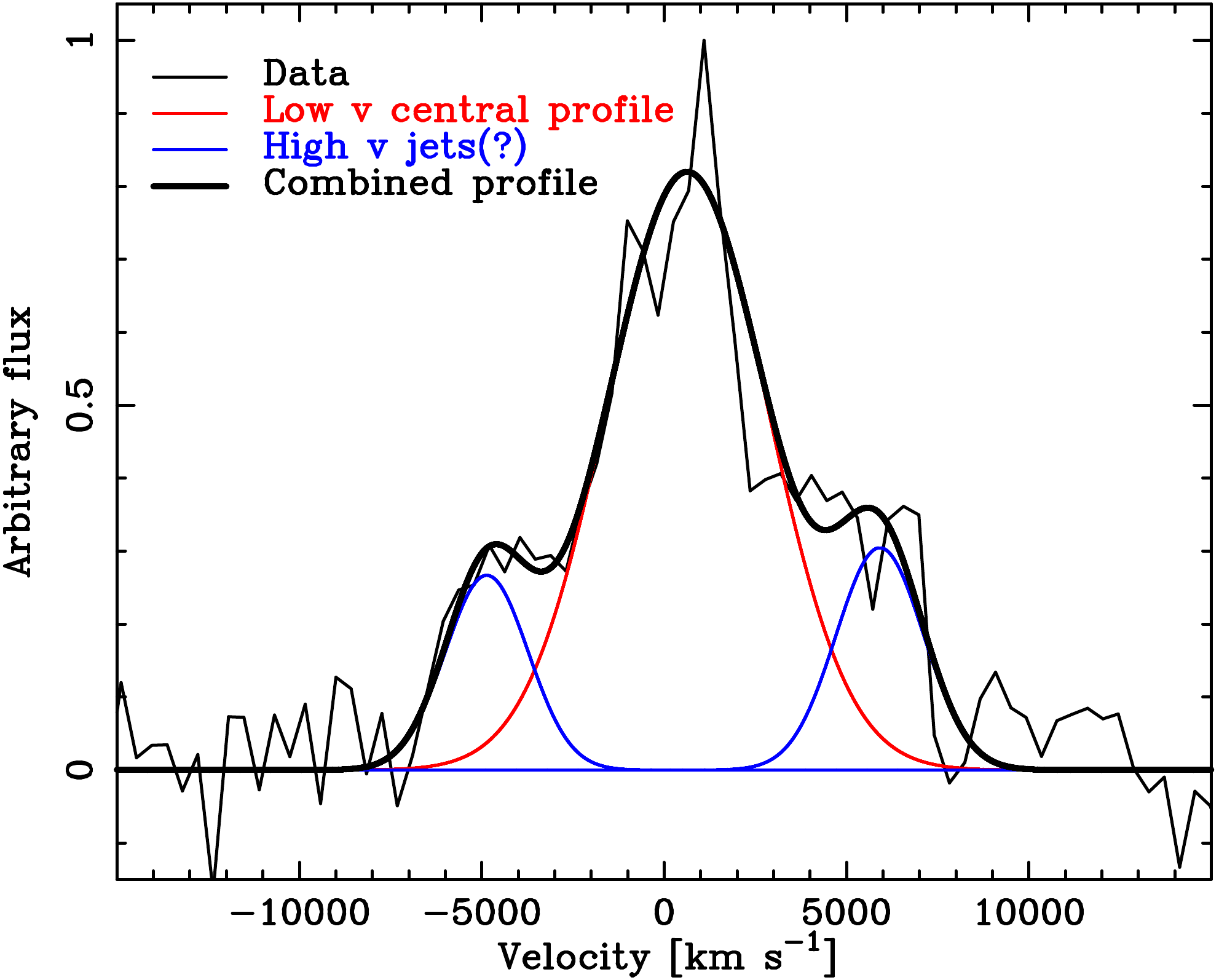}\\
\end{center}
\caption{Evolution of the H$\alpha$ {\bf(top)} and H$\beta$ {\bf(middle)} line profiles following the 2015 eruption of \novak.  {\bf Left:} The peak flux of each $H\alpha$/H$\beta$ line has been normalized to 1 -- to indicate the evolution of the line morphology.  {\bf Right:}  Flux calibrated spectra are shown to illustrate the change in flux of the H$\alpha$/H$\beta$ emission.  {\bf Bottom left:} Comparison between the H$\alpha$ and He\,{\sc i}\,(5876\AA) lines at $\Delta t=0.67$\,days; the line blueward of the He\,{\sc i} line ($\sim-8000$\,km\,s$^{-1}$) is N\,{\sc ii}\,(5679\AA).  {\bf Bottom right:} H$\alpha$ line profile at $t=0.67$\,d , the red and the blue lines show a best-fit model of three Gaussian profiles, the thick black line their combined flux (see Section~\ref{sec:incl}).  Line colors as Figure~\ref{optical_spec} with the additional green data $\Delta t=1.10\pm0.01$\,days (KPNO; rebinned).  The peak redward of H$\alpha$ ($\sim+5300$\,km\,s$^{-1}$) is He\,{\sc i} (6678\,\AA).  The data in all these plots have been continuum subtracted as described in the text.\label{halpha_line}\label{fig:jets}}
\end{figure*}

As was seen following the 2013 eruption (\ponek) and the 2014 eruption (\otwok), and as noted by \ponek, the morphology of the emission lines evolves significantly as the eruption progresses; in particular there is a marked decrease in the width of the line profiles.  Figure~\ref{halpha_line} presents the evolution of the H$\alpha$ (top) and H$\beta$ (middle) lines during the 2015 eruption; the left-hand plots normalize the flux of the continuum and H$\alpha/\mathrm{H}\beta$ peak to highlight the morphological evolution, the right-hand plots show the flux calibrated spectra to illustrate the change in intensity of the lines.  As seen in previous eruptions, the Balmer emission lines have a well-defined central double-peaked profile, and the overall width of the profiles decreases with time.  At all epochs the redward peak of the double-peak is of similar or higher flux than the blueward peak (see Figure~\ref{halpha_line} top left) in the H$\alpha$ line.  Both H$\alpha$ peaks are at approximately equal flux when the H$\alpha$ line has its maximum integrated flux ($t=1.79$\,d; blue line).  The blueward peak appears to wane significantly in later spectra ($t=4.91$\,d; gray line).  Although at lower signal-to-noise, similar behavior appears present for the H$\beta$ line.

In 2014, there was some evidence for higher velocity material beyond the central peak at early times.  In the two earlier 2015 spectra (the earliest spectra yet obtained; see the black and red spectra in Figure~\ref{halpha_line}) we witness evidence of significant emission from very high radial velocity material.

The high velocity material seen in the $t=0.67$\,d spectrum has an approximately `rectangular' profile about the H$\alpha$ central wavelength, with a FWHM of $\simeq13000\,\mathrm{km}\,\mathrm{s}^{-1}$ and $\mathrm{FWZI}\simeq14500\,\mathrm{km}\,\mathrm{s}^{-1}$ (see Figure~\ref{halpha_line}, particularly the bottom plots).  The H$\beta$ equivalent is fainter, and therefore noisier, but still shows a FWZI of $\simeq12000\,\mathrm{km}\,\mathrm{s}^{-1}$.  By $t=0.96$\,d the emission from this high velocity material has begun to fade (note that the emission from central H$\alpha$/H$\beta$ component remains approximately constant during this period), and the measured width of the high-velocity profile has reduced by $\sim1000\,\mathrm{km}\,\mathrm{s}^{-1}$.  On day 1.1, the high-velocity profile has diminished further, becoming indistinguishable from the continuum around H$\beta$; but around H$\alpha$, the appearance of the He\,{\sc i}\,(6678\AA; $\sim+5300\,\mathrm{km}\,\mathrm{s}^{-1}$) emission line additionally complicates the profile.  In all later spectra (including all spectra obtained prior to the 2015 eruption) any emission from such high velocity material is absent or at least indistinguishable from the continuum.

There is also evidence for such high velocity material around the profiles of other (non-H\,{\sc i}) lines in the early spectrum. For example, the He\,{\sc i}\,(5876\AA; see Figure~\ref{halpha_line}, bottom left) and He\,{\sc i}\,(7065\AA) lines both appear to have a similar profile to the Balmer lines, a double-peaked central profile that is bracketed by a high-velocity `rectangular' profile in early spectra.  The similar line profile morphologies and evolution imply that the H\,{\sc i} and He\,{\sc i} emission arise from the same part of the ejecta.

\subsection{Ejecta expansion velocity}\label{sec:exp_vel}

To determine the total flux and FWHM of the spectral lines, a fit to the continuum of each spectrum was made using a third-order polynomial.  Each spectral line was then separately fit using a single Gaussian profile and a background level, all lines were fit in a consistent manner using data within $\pm4000$\,km\,s$^{-1}$ of the line centre\footnote{The two early epoch H$\alpha$ lines were fit between $-7000\le v \le 4000$\,km\,s$^{-1}$ to permit a better fit to the background level, and to avoid contamination from He\,{\sc i} (6678\,\AA; +5260\,km\,s$^{-1}$).}.  The line velocities for the Balmer and He\,{\sc i} lines are shown in Table~\ref{fwhm_tab}, and the corresponding line fluxes in Table~\ref{line_list}.  Generally, a Gaussian profile produced a good fit to the spectral lines however, the early epoch Balmer lines with their high velocity components were not well reproduced with just a single Gaussian (see Figure~\ref{halpha_line}), leading to the larger velocity uncertainties seen in Table~\ref{fwhm_tab}.   For these two early epochs, only the velocity (and line flux) of the central component was calculated using a Gaussian fit.  To fit the He\,{\sc i} (6678\,\AA) line the best-fitting H$\alpha$ profile was first subtracted from the spectrum to aid de-blending of the lines.  The H$\delta$, He\,{\sc i} (5015/5048\,\AA), He\,{\sc ii}, and N lines were not modeled due to a combination of complex profiles, significant blending, or low signal-to-noise.  No data were recorded in Tables~\ref{fwhm_tab} or \ref{line_list} if the fitted flux of a line reported a signal-to-noise ratio $<3$.  The available spectra from 2012 and 2014 were also re-analyzed in a consistent manner to the 2015 spectra and these data, along with those from 2013 (\ponek), are also included in Tables~\ref{fwhm_tab} and \ref{line_list}.  There was no significant evolution observed in the line flux ratios among the H\,{\sc i} or He\,{\sc i} lines, or the overall H\,{\sc i}/He\,{\sc i} ratio.

\begin{table*}
\caption{Evolution of the FWHM of the H$\alpha$ profile.\label{fwhm_tab}}
\begin{center}
\begin{tabular}{lllllllll}
\tableline\tableline
$\Delta t$ (days) & Source & Year & \multicolumn{6}{c}{Best-fit Gaussian FWHM (km\,s$^{-1}$)} \\
 &  &  & H$\alpha$ & H$\beta$ & H$\gamma$ & He\,{\sc i} (7065\,\AA) & He\,{\sc i} (6678\,\AA)$^{\dag}$ & He\,{\sc i} (5876\,\AA)\\
\tableline
$0.67\pm0.02$ & LT$^\ddag$ & 2015 & $4760\pm520$ & $4140\pm670$ & \nodata & \nodata & \nodata & $5900\pm1200$\\ 
$0.96\pm0.02$ & LT$^\ddag$ & 2015 & $3610\pm310$ & $2520\pm260$ & $3270\pm720$ & $4170\pm860$ & \nodata & $2930\pm610$\\ 
$1.10\pm0.01$ & KPNO & 2015 & $2940\pm50$ & $2440\pm80$ & $2150\pm100$ & \nodata & \nodata  & $3570\pm230$\\ 
$1.14\pm0.02$ & LCOGT & 2015 & $3180\pm110$ & $2470\pm200$ & $2020\pm320$ & \nodata & \nodata & \nodata\\ 
$1.27\pm0.01$ & WHT & 2014 & $2740\pm70$ & \nodata & \nodata & $2720\pm220$ & $2620\pm240$ & $2300\pm180$\\
$1.33\pm0.14$ & LT & 2014 & $2800\pm110$ & $2370\pm100$ & $2080\pm290$ & $2160\pm330$ & $1540\pm320$ &  $3420\pm420$ \\
$1.63\pm0.10$ & HET & 2012 & $2520\pm40$ & $2250\pm90$ & $2510\pm110$ & $3220\pm130$ & $1920\pm160$ & $2450\pm110$ \\ 
$1.79\pm0.11$ & LT & 2015 & $2440\pm110$ & $2320\pm170$ & $4960\pm660$ & $5570\pm760$ & $1940\pm460$ & $2490\pm260$ \\ 
$1.8\pm0.2$ & Keck & 2013 & $2600\pm200$ & \nodata & \nodata & \nodata & \nodata & \nodata\\
$2.03\pm0.02$ & LCOGT & 2015 & $2740\pm100$ & $1850\pm220$ & \nodata & \nodata & \nodata & \nodata\\ 
$2.45\pm0.18$ & LT & 2014 & $2340\pm100$ & $2230\pm160$ & $2190\pm360$ & $2040\pm280$ & $1960\pm490$ & $2670\pm250$ \\
$2.84\pm0.11$ & LT & 2015 & $2430\pm160$ & $2140\pm230$  & \nodata & $2620\pm570$ & \nodata & $2300\pm310$\\ 
$3.18\pm0.01$ & LT & 2014 & $2300\pm230$ & \nodata & \nodata & \nodata & \nodata & \nodata\\
$3.84\pm0.02$ & LT & 2015 & $2540\pm290$ & $2070\pm300$  & \nodata & \nodata & \nodata & $1860\pm330$ \\ 
$4.6\pm0.2$ & Keck & 2013 & $1900\pm200$ & \nodata & \nodata & \nodata & \nodata & \nodata\\
$4.91\pm0.02$ & LT & 2015 & $2020\pm290$ & \nodata & \nodata & \nodata & \nodata & \nodata \\ 
\tableline
\end{tabular}
\end{center}
\tablenotetext{\dag}{The He\,{\sc i} (6678\,\AA) line flux was computed by first subtracting the best fitting, and nearby, H$\alpha$ line profile.  As such the values reported here are dependent upon the H$\alpha$ modeling.} 
\tablenotetext{\ddag}{The high velocity material beyond the central profile, seen predominantly in the two early spectra, is not included in the computed line widths.}
\end{table*}

\begin{table*}
\caption{Selected observed emission lines and fluxes from the nine epochs of Liverpool Telescope SPRAT spectra of the 2014 and 2015 eruptions of \novak.\label{line_list}}
\begin{center}
\begin{tabular}{lllllllll}
\tableline\tableline
$\Delta t$ (days) & Source & Year & \multicolumn{6}{c}{Flux ($\times10^{-15}$\,erg\,cm$^{-2}$\,s$^{-1}$)$^\S$} \\
 &  &  & H$\alpha$ & H$\beta$ & H$\gamma$ & He\,{\sc i} (7065\,\AA) & He\,{\sc i} (6678\,\AA)$^{\dag}$ & He\,{\sc i} (5876\,\AA)\\ 
\tableline
$0.67\pm0.02$ & LT$^\ddag$ & 2015 & $10.5\pm1.4$ & $7.4\pm1.5$ & \nodata & \nodata & \nodata & $5.0\pm1.3$ \\
$0.96\pm0.02$ & LT$^\ddag$ & 2015 & $7.7\pm0.8$ & $3.6\pm0.3$ & $3.8\pm1.1$ & $2.4\pm0.7$ & \nodata & $2.1\pm0.5$ \\
$1.33\pm0.14$ & LT & 2014 & $6.8\pm0.4$ & $4.2\pm0.2$ & $3.1\pm0.5$ & $1.8\pm0.4$ & $1.0\pm0.3$ & $2.4\pm0.4$\\
$1.79\pm0.11$ & LT & 2015 & $10.4\pm0.6$ & $5.5\pm0.5$ & $3.6\pm0.3$ & $3.6\pm0.6$ & $1.4\pm0.4$ & $2.7\pm0.4$\\
$2.45\pm0.18$ & LT & 2014 & $5.5\pm0.3$ & $2.7\pm0.2$ & $2.0\pm0.4$ & $1.2\pm0.2$ & $0.8\pm0.3$ & $1.6\pm0.2$\\
$2.84\pm0.11$ & LT & 2015 & $7.8\pm0.6$ & $2.8\pm0.4$ & \nodata & $1.3\pm0.4$ & \nodata & $1.7\pm0.3$\\
$3.18\pm0.01$ & LT & 2014 & $5.1\pm0.6$ & \nodata & \nodata & \nodata & \nodata & \nodata \\
$3.84\pm0.02$ & LT & 2015 & $5.5\pm0.8$ & $1.7\pm0.3$ & \nodata & \nodata & \nodata & $0.9\pm0.2$\\
$4.91\pm0.02$ & LT & 2015 & $2.1\pm0.4$ & \nodata & \nodata & \nodata & \nodata & \nodata \\
\tableline
\end{tabular}
\end{center}
\tablecomments{Line flux is derived from the best-fit Gaussian profile for each emission line and is strongly dependent upon the adopted continuum level.}
\tablenotetext{\dag}{The He\,{\sc i} (6678\,\AA) line flux was computed by first subtracting the best fitting, and nearby, H$\alpha$ line profile.  As such the values reported here are dependent upon the H$\alpha$ modeling.}
\tablenotetext{\ddag}{The high velocity material beyond the central profile, seen predominantly in the two early spectra, is not included in the computed line fluxes.}
\tablenotetext{\S}{Here we note that the flux units reported in \otwok\ (see their Table~3) were incorrect \cite[see][]{erratum}.}
\end{table*}

\begin{figure*}
\includegraphics[width=\columnwidth]{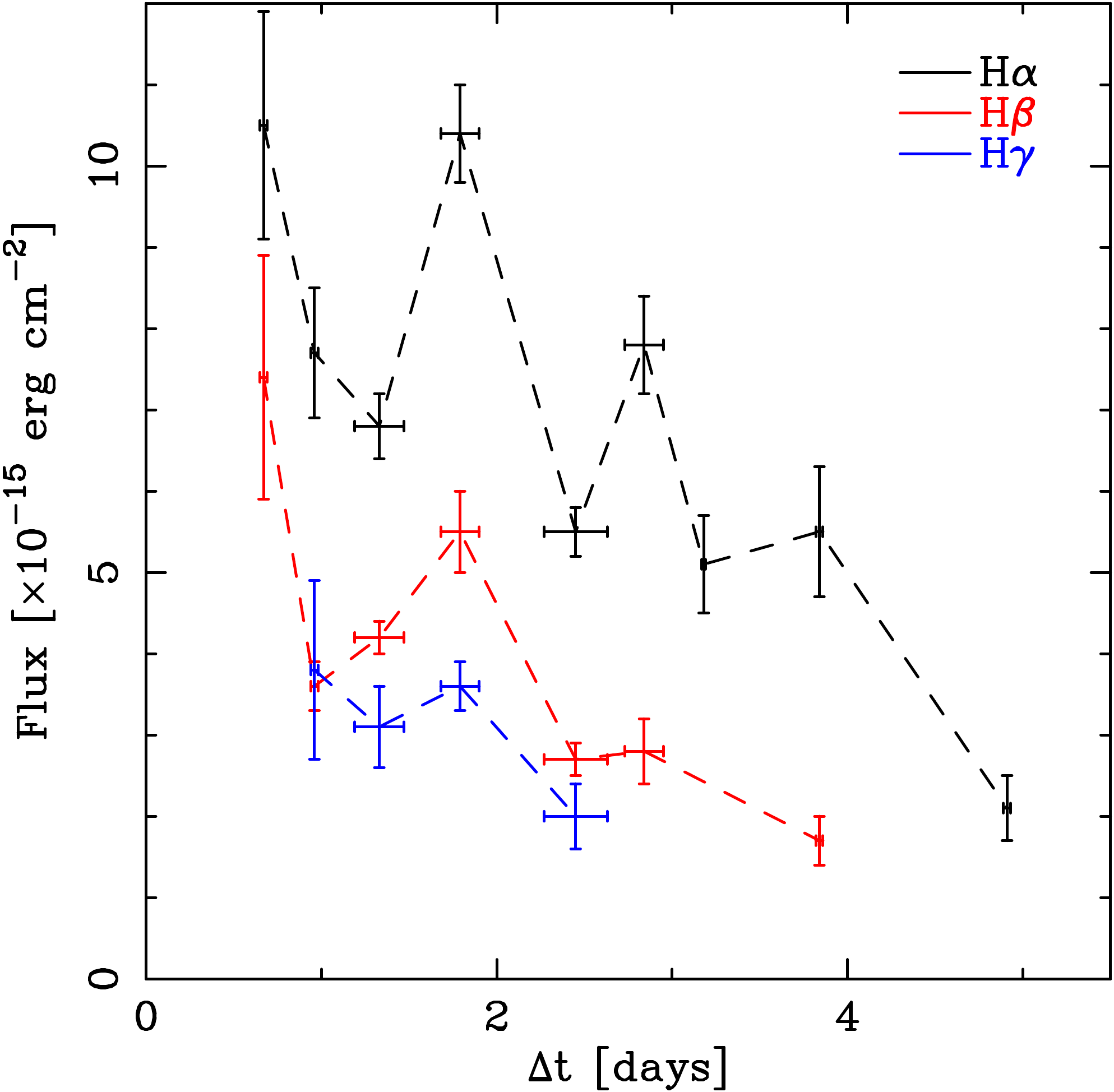}\hfill
\includegraphics[width=\columnwidth]{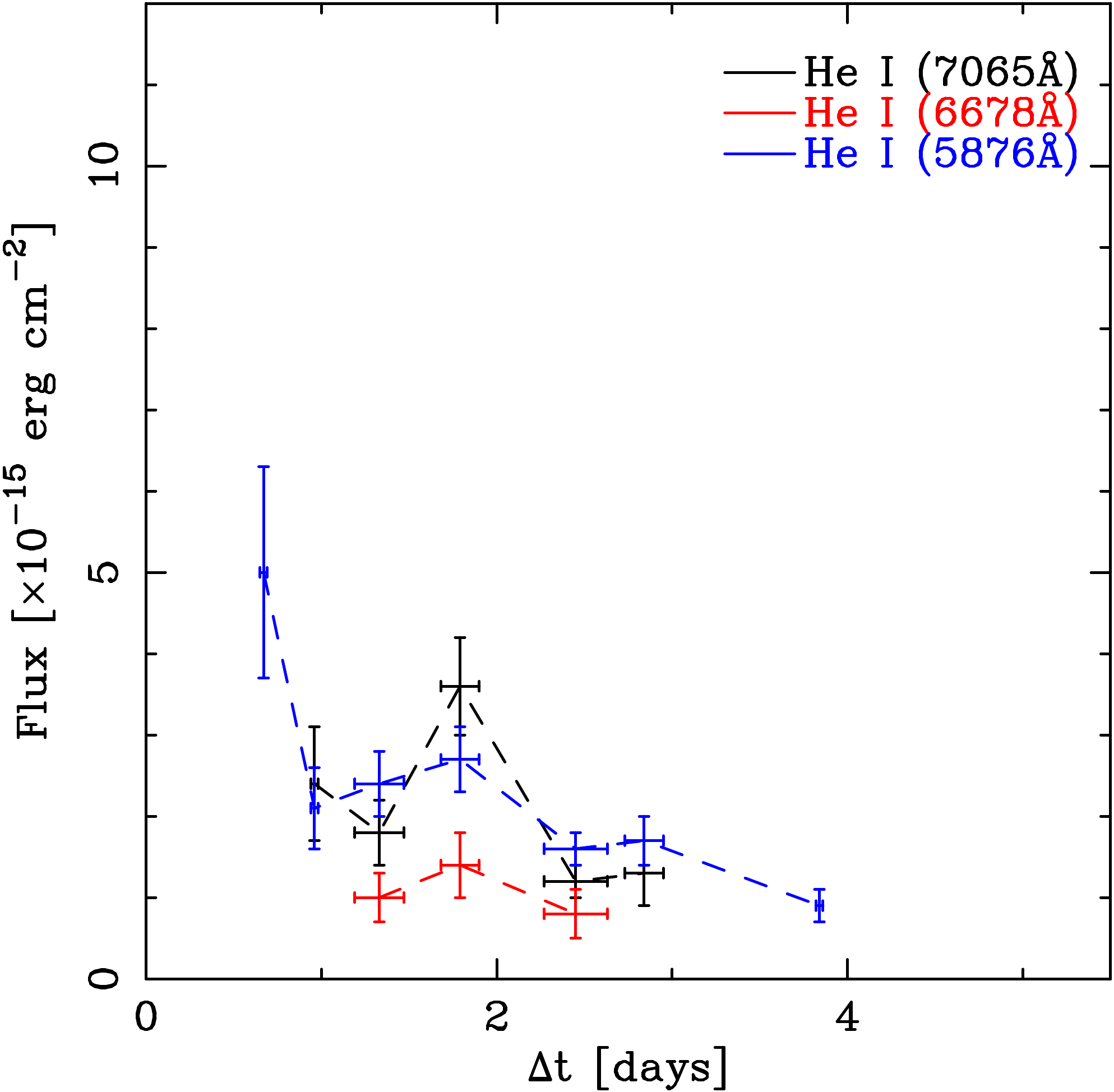}\\
\caption{Evolution of the integrated line fluxes of the Balmer (left) and He\,{\sc i} (right) lines since the onset of the eruption.  Data are from spectra of the 2014, and 2015 eruptions.  The dashed-lines connecting the points are to aid the reader only.  These fluxes were derived by only fitting the central cores of the emission lines, and do not include the higher velocity material seen during the early spectral epochs.\label{lineflux_plot}}
\end{figure*}

In Figure~\ref{lineflux_plot} we present a plot showing the evolution of the H\,{\sc i} (left) and He\,{\sc i} (right) integrated line fluxes with time.  It should be noted that only the flux of the central part of the emission lines was computed, not the contribution from the early higher velocity material.  As was reported by \otwok, the general trend shows a decreasing of line flux with time for the H\,{\sc i} and He\,{\sc i} lines.  All the H\,{\sc i} and He\,{\sc i} lines show a decrease in flux during the final rise phase ($0\le t\le 1$\,d), followed by a brief `recovery' at $t\simeq1.8$\,d, before entering a consistent decline.

\begin{figure*}
\includegraphics[width=\columnwidth]{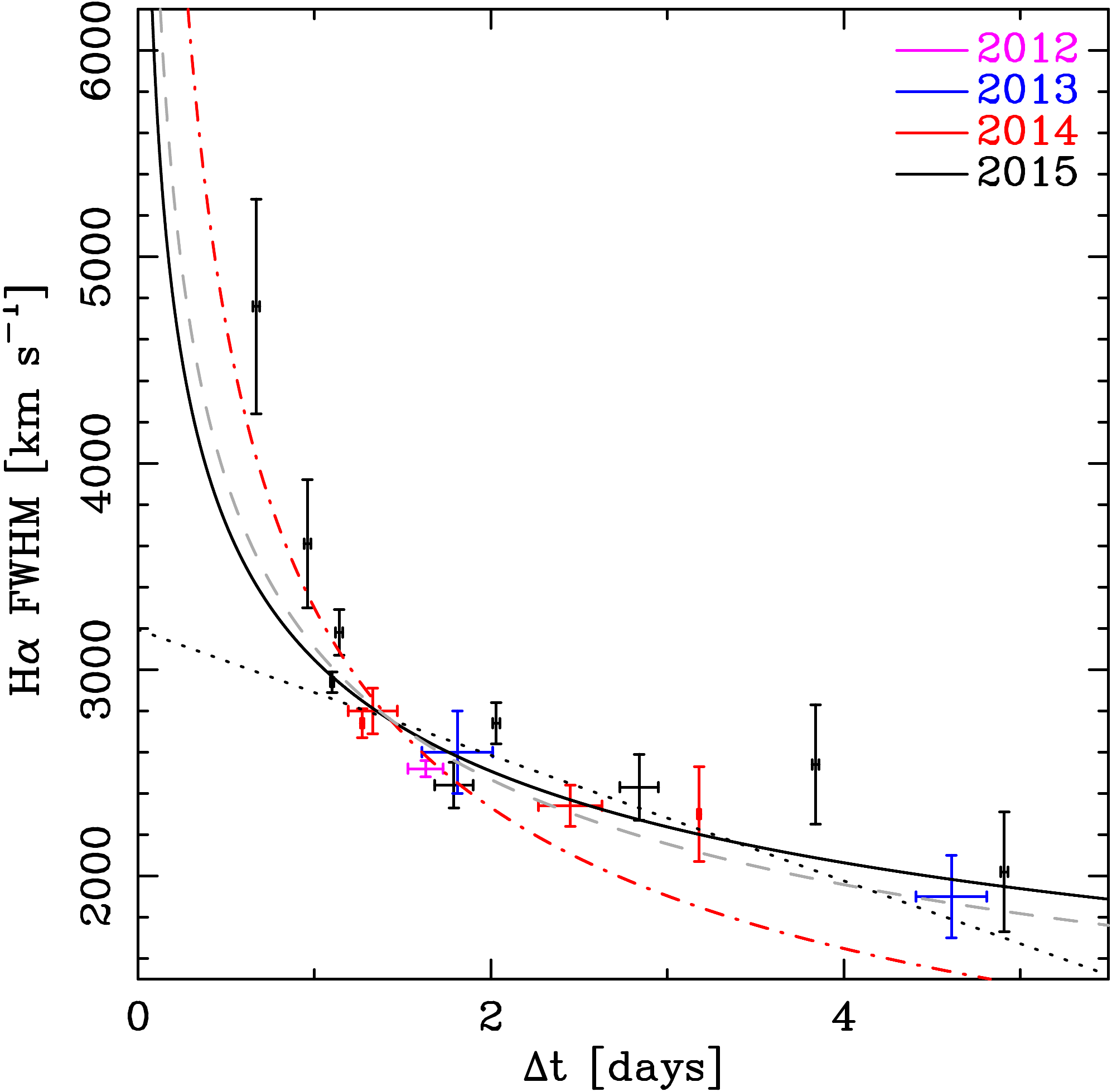}\hfill
\includegraphics[width=\columnwidth]{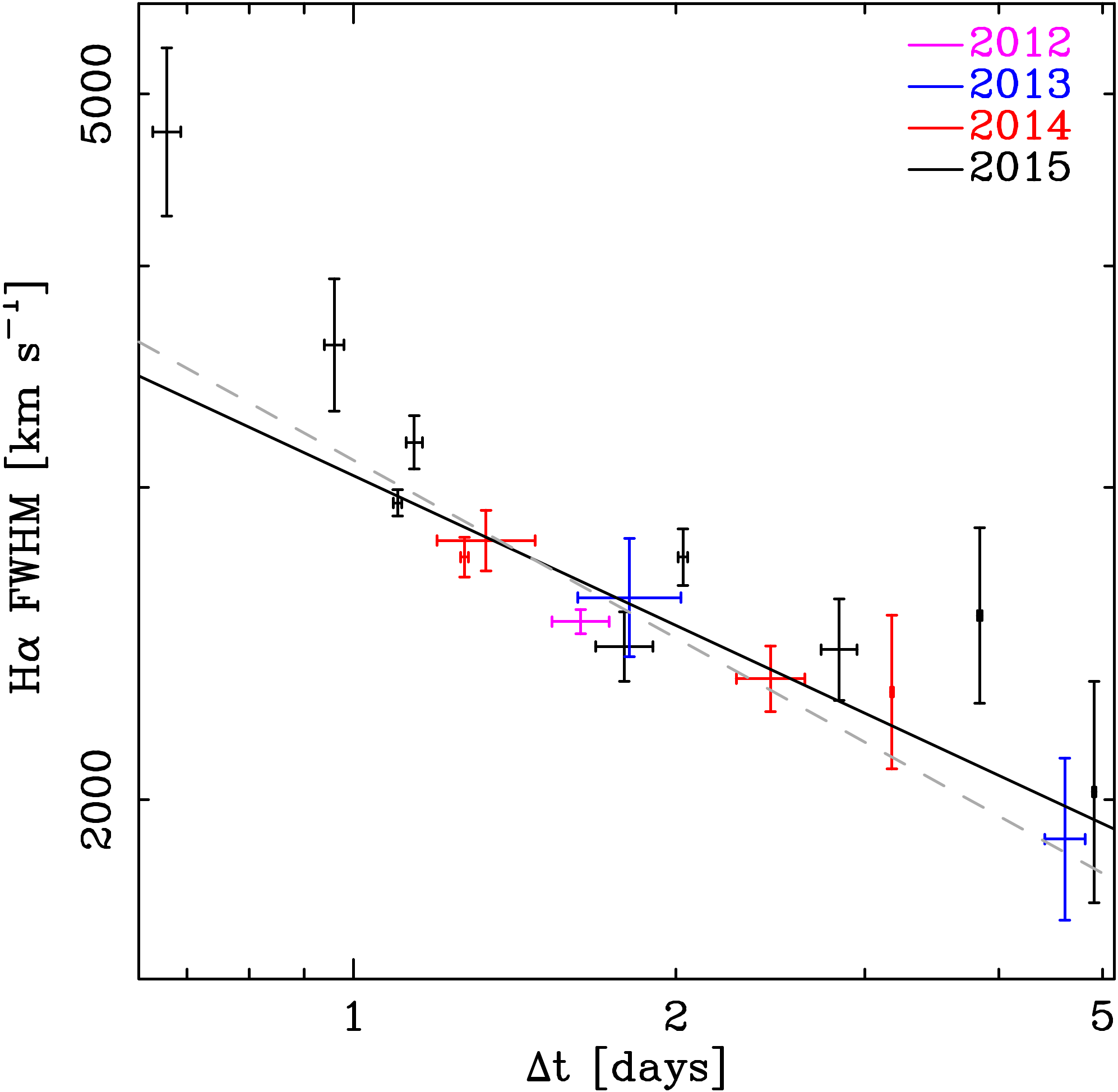}
\caption{Evolution of the FWHM of the H$\alpha$ emission line following the eruption of \novak\ (left: linear axes, right: logarithmic axes). See the key for data point identification. These velocities were derived only from the central cores of the emission lines and do not include the higher velocity material seen in the first two spectral epochs.  The dotted black line shows a simple linear least-squares fit to the 2012--2015 data (gradient = $-300\pm70$\,km\,s$^{-1}$\,day$^{-1}$, $\chi^{2}_{\mathrm{/dof}}=3.9$; if the first two data points are ignored $\chi^{2}_{\mathrm{/dof}}$ reduces to 3.1, the gray dashed line a power-law of index $-1/3$ ($\chi^{2}_{\mathrm{/dof}}=2.7$; Phase II of shocked remnant development), the red dot-dashed line a power-law of index $-1/2$ ($\chi^{2}_{\mathrm{/dof}}=5.8$; Phase III), and the solid black line the best fit power-law with index $-0.28\pm0.05$ ($\chi^{2}_{\mathrm{/dof}}=2.7$), see text for details.  These observations indicate that the ejecta shock pre-existing circumbinary material close to the central system.  The most likely conclusion is that the donor is seeding the local environment via a stellar (red giant) wind.\label{fwhm_plot}}
\end{figure*}

Throughout we use the FWHM of the best-fit Gaussian profile to the emission lines as a proxy for the line-of-sight ejection velocity; the velocities of the Balmer and He\,{\sc i} emission lines for the 2012--2015 eruptions are recorded in Table~\ref{fwhm_tab}.  As previously discussed, the Gaussian profile generally provided a good fit to the lines, at least down to half-maximum flux, notable exceptions being the Balmer emission in the two earliest spectra.  The weighted mean expansion velocity from the H$\alpha$ line from the 2012--2015 eruptions is $2670\pm70$\,km\,s$^{-1}$, consistent with the average found following the 2014 eruption (\otwok).  However, the measured expansion velocities from the two earliest epochs, $t=0.67$ and 0.96\,days, are significantly higher than the mean, and represent velocities not previously seen, or predicted \citep[see, for example,][]{2005ApJ...623..398Y}, from this system.	

In Figure~\ref{fwhm_plot} we show the evolution with time of the H$\alpha$ profile FWHM velocity from the 2012--2015 eruptions.  As a similar analysis following the 2014 eruption indicated (\otwok), there is a clear measurement of a decreasing velocity with time.  A linear least-squares fit to these data reveals a declining gradient of $-300\pm70$\,km\,s$^{-1}$\,day$^{-1}$ ($\chi^{2}_{\mathrm{/dof}}=3.9$).  If the first two, high velocity, data points are excluded the linear fit is essentially unchanged ($\chi^{2}_{\mathrm{/dof}}=3.1$).  Again note that the additional high velocity components seen in the early spectra are not included in these data.

The right-hand plot in Figure~\ref{fwhm_plot} shows a log--log plot of expansion velocity against time; by simple inspection these data appear to be well represented by a power law.  The best fitting power law to these data (of the form $v_{\mathrm{exp}}\propto t^{n}$) has index $n=-0.28\pm0.05$ ($\chi^{2}_{\mathrm{/dof}}=2.7$).  If we choose to fix the power law index at $1/3$ and $1/2$ (see Section~\ref{sec:decel}) then the best fits have $\chi^{2}_{\mathrm{/dof}}=2.7$ and $\chi^{2}_{\mathrm{/dof}}=5.8$, respectively.

\subsection{The X-ray temperature and spectral variability}
\label{sec:xspec}

\begin{figure}
\includegraphics[width=\columnwidth]{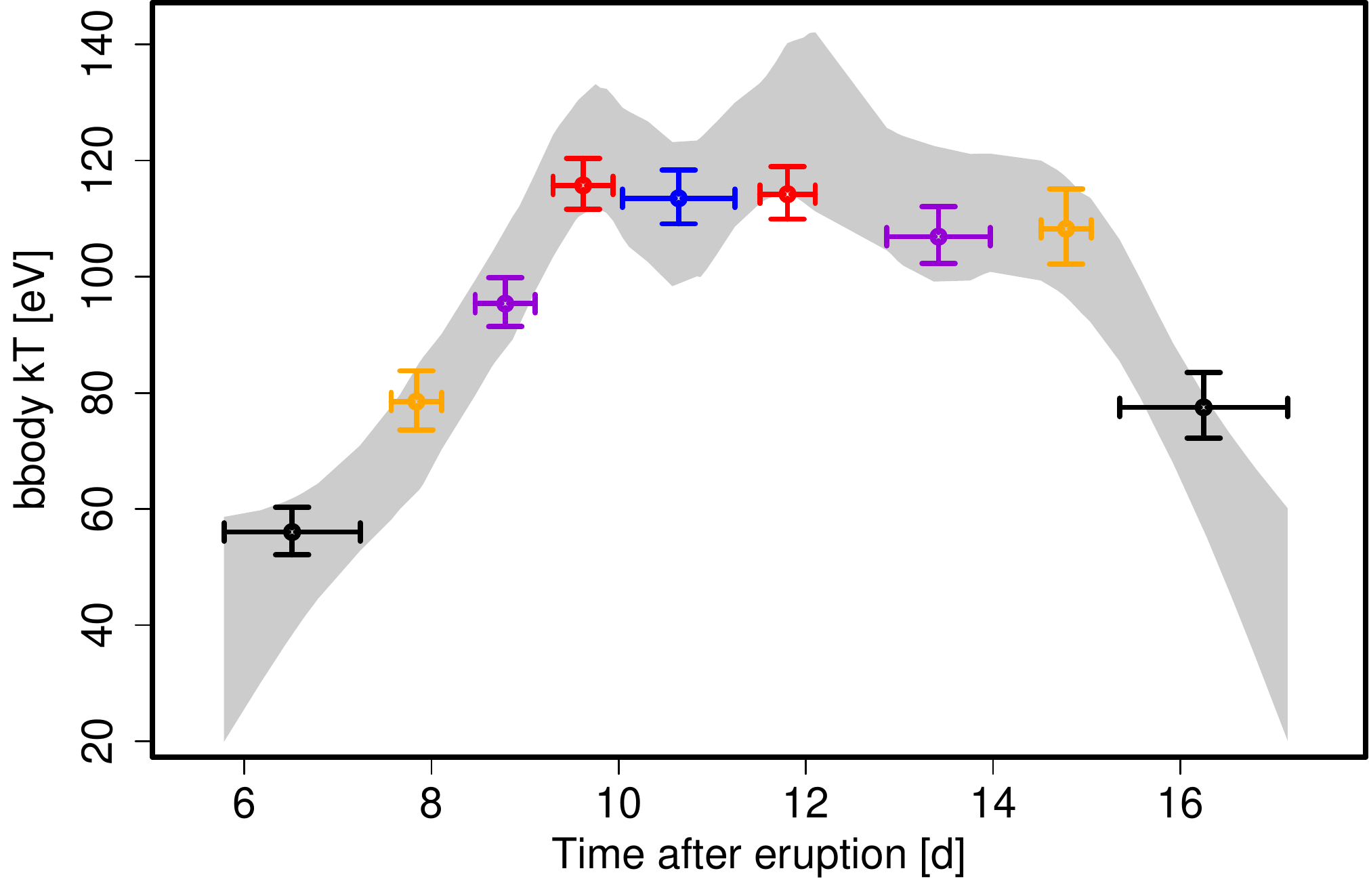}
\caption{The effective black body temperature of \nova depending on the time after eruption. Based on X-ray spectra from the 2015/14/13 eruptions. Sets of spectra with similar temperature (cf.\ Figure~\ref{fig:xray_lc}b) have been fitted simultaneously. Colored data points show the best fit $kT$ and corresponding uncertainty. The colors are only used for quick identification of the eruption stages in Figure \ref{fig:xray_group_spec} and Table \ref{tab:xspec} and carry no specific physical meaning. The error bars in time after eruption extend from the first to the last observation of each group. The gray region shows the 95\% confidence prediction interval derived from smoothing temperature fits based on individual snapshots. For clarity, these individual fits are not shown. A temperature plateau is suggested between days 9--15, with small scale variations possible due to more complex spectral changes (see Figure \ref{fig:xray_group_spec}).\label{fig:xray_temp}}
\end{figure}

The temperature evolution of the SSS phase is shown in Figure~\ref{fig:xray_temp}. This plot is based on simple black body fits to all 2013/14/15 X-ray spectra. Like in Figure~\ref{fig:xray_lc}b the black body parametrization assumes a fixed \nh = $1.4$ \hcm{21}. The spectra have been parametrized individually (see the gray smoothed fit in Figure~\ref{fig:xray_temp}a) and also simultaneously in nine groups similar to those in Figure~\ref{fig:xray_lc}b. Compared to Figure~\ref{fig:xray_lc}b the combined group fits have significantly reduced temperature uncertainties as well as a higher time resolution (9 bins in Figure~\ref{fig:xray_temp} versus 7 bins in Figure~\ref{fig:xray_lc}b).

The individual spectral fits in Figure~\ref{fig:xray_temp} (gray band) tentatively suggest that the observed dip in X-ray flux (cf.\ Figures~\ref{fig:xray_split} and \ref{fig:xray_dip}) is associated with a dip in temperature. While the substructure of the effective temperature evolution is otherwise well represented by the grouped fits, the temperature dip is not visible there. The reasons for this are most likely that the actual SSS spectrum (a) differs strongly from a simple black body continuum and (b) is highly variable. The effective temperature parametrisation in Figure~\ref{fig:xray_temp} represents only a first-order approximation that does not fully capture the actual spectral variations.

\begin{sidewaysfigure*}
\vspace{-10cm}
\includegraphics[width=\textwidth,height=15cm]{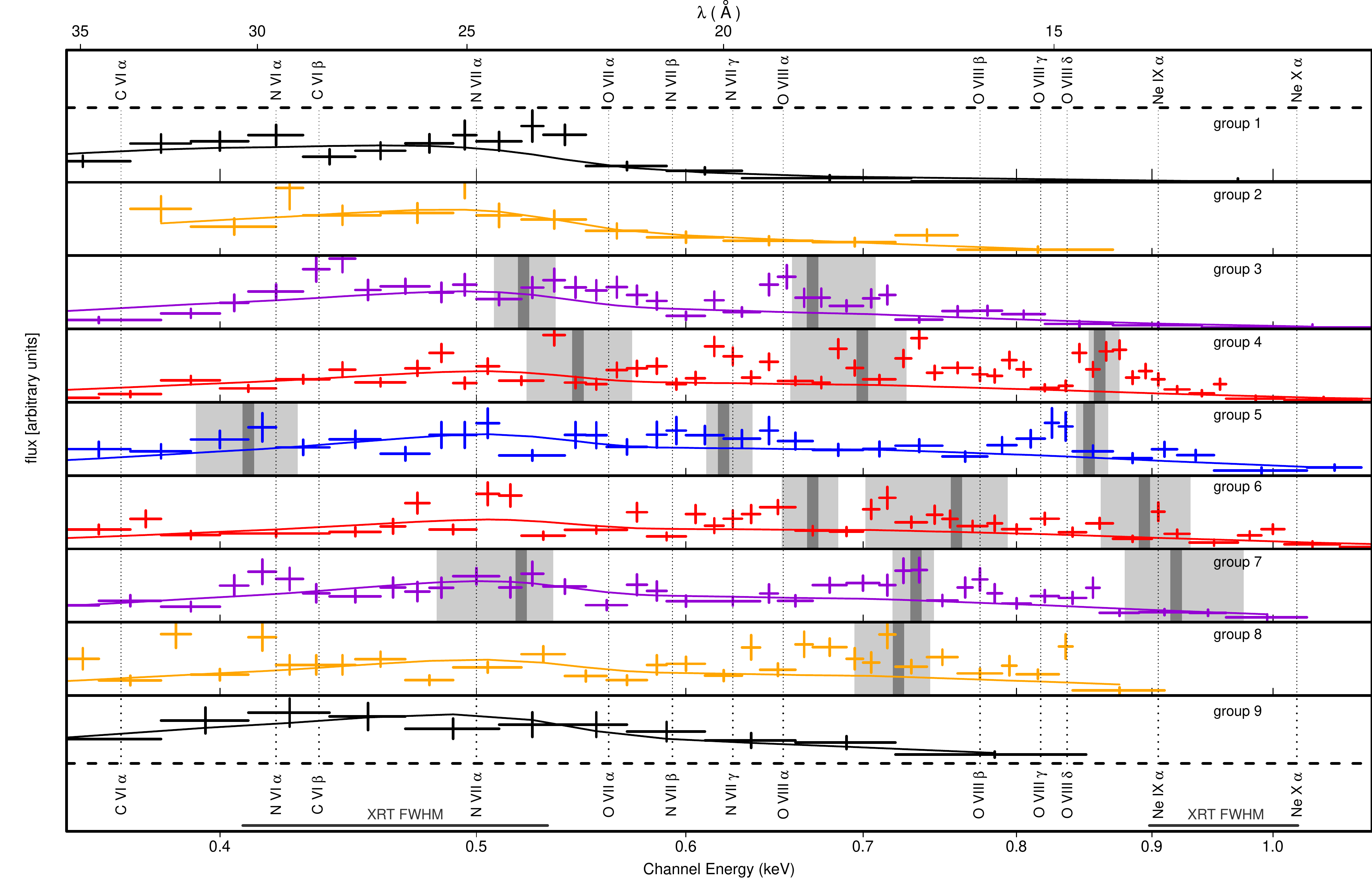}
\caption{Binned combined XRT spectra in arbitrary flux units with black body fits (solid lines). The colors correspond to the eruption stages in Figure~\ref{fig:xray_temp} with time progressing from top to bottom. For groups 3--8 we show the energies (dark gray) and corresponding uncertainties (light gray) of possible emission lines (see Table~\ref{tab:xspec}). Additionally, the energies of known prominent emission lines of H-like and He-like C, N, O, and Ne transitions are indicated by dashed lines (for the He-like triplets only the forbidden line locations are marked). The relevant line identifications are given at the top and bottom of the figure. The horizontal gray lines at the bottom of the plot show the \swift XRT FWHM of $\sim 125$\,eV in this energy range as determined in \texttt{XSPEC} based on the current XRT calibration files (CALDB version 20160121). \label{fig:xray_group_spec}}
\end{sidewaysfigure*}

The shortcomings of the black body model become evident when looking at the merged and binned spectra of the nine spectral groups which are shown in Figure~\ref{fig:xray_group_spec} together with the corresponding black body fits. The early and late low-temperature spectra (groups 1, 2, and 9) can still be reasonably well approximated by a black body continuum based on the residuals in Figure~\ref{fig:xray_group_spec} and the consistent absorption estimates (see Table~\ref{tab:xspec}). However, there is little doubt that around the flux maximum (groups 3-8) the spectra show strong additional features and deviate considerably from a simple black body continuum.  Any further study of the spectral (and flux) variability during the SSS phase has to take into account these features.

For the three groups (1, 2, and 9) that can still be described by black body fits we derive a \nh of $0.7^{+0.5}_{-0.5}$ \hcm{21} from a simultaneous fit. This estimate should be considered as more accurate than the previous value of $1.4$ \hcm{21} which was based on a total spectrum including  possible additional features (\xonek, \xtwok). It does however, still assume a black body continuum. The new value is in excellent agreement with the $E_{B-V} \sim 0.1$ \citep[corresponding to \nh $= 0.69$ \hcm{21} via the relation between optical extinction and hydrogen column density from][]{2009MNRAS.400.2050G} found by \citet{HST2016}.

For the remaining groups (3--8) we attempted to model the X-ray spectra using additional emission components. Here, we first created an approximate model for each merged spectrum shown in Figure~\ref{fig:xray_group_spec} by adding one to three Gaussian emission lines to the black body continuum by eye in \texttt{XSPEC}. We then fitted the line parameters using $\chi^2$ minimization until the residuals showed no strong deviations. In a second step, these models were fitted simultaneously to the (effectively) unbinned individual spectra of each group using Poisson statistics according to \citet{1979ApJ...228..939C}. The model parameters were linked, with only the normalisations free to vary for the single spectra. Not all lines for a given group are detected in all the individual spectra, but almost every line has a flux that is significant at the 95\% confidence level for at least two different spectra. The only exceptions are the line at 0.76\,keV in group 6 and the 0.92\,keV line in group 7 which are only significant in a single spectrum each.

The results of the spectral exploration are summarized in Table~\ref{tab:xspec} where we compare the emission line-enhanced models to pure black body fits. We found that likelihood ratio tests (\texttt{lrt} in \texttt{XSPEC}) preferred the models with emission lines over the pure continuum models, with $>85$~\% better likelihood ratios for all groups and $>95$~\% for most. The table also shows that the addition of line components not only improved the fit statistics but led to considerably more consistent (and realistic) values for the absorption column and the black body temperature throughout. The \nh values are now consistent with the black body fits for groups 1, 2, and 9, as well as the extinction determination by \citet{HST2016}. Also the effective temperatures, albeit with large uncertainties, are in general consistent with the overall trend in Figure~\ref{fig:xray_temp}.

In Figure~\ref{fig:xray_group_spec} we also show the location of known emission lines from H-like and He-like carbon, nitrogen, oxygen, and neon. Table~\ref{tab:xspec} names those lines that are close to the energies of the suggested emission lines. This is not a clear identification of those lines but a first tentative suggestion of those elements that might contribute the additional emission features. 

Note, that the spectral FWHM of the \swift XRT, which is illustrated at the bottom of Figure~\ref{fig:xray_group_spec}, is broader than some of the excess features that only consist of one or two spectral bins and are therefore unlikely to correspond to individual emission lines. For this reason, we fitted the spectral lines first before overlaying the known transitions, so that our analysis would not be biased by associating narrow features with certain laboratory lines.

In terms of Galactic nova low-resolution X-ray spectroscopy, the RNe RS~Oph \citep{2011ApJ...727..124O} and V745~Sco \citep[][see also Section~\ref{sec:v745}]{2015MNRAS.454.3108P} showed significant spectral excesses in high statistical quality XRT spectra.

For practical purposes the \m31 blueshift is negligible ($z \sim -0.001$), well below the binning resolution in Figure~\ref{fig:xray_group_spec}. While additional velocity shifts have been reported for \textit{absorption} lines of Galactic novae \citep[e.g.][]{2010AN....331..175V,2010AN....331..179N}, the \textit{emission} features normally show no such effects.

Hot WD atmosphere models \citep[e.g.][]{2010ApJ...717..363R,2012ApJ...756...43V} resemble black body-like continua cut by sharp absorption features. These features have been observed in X-ray grating spectra of several Galactic novae \citep[e.g.][]{2013A&A...559A..50N}. Therefore, we tested fitting the absorption edge of neutral oxygen at 0.54\,keV \citep[prominently seen in RS~Oph, e.g.][]{2007ApJ...665.1334N}, as well as the N\,{\sc vii} ionization edge at 0.67\,keV and the O\,{\sc viii} edge at 0.87\,keV.  In no case did the inclusion of these features result in a statistically better fit or more consistent values of \nh or $kT$ compared to the pure black body. Therefore, the more complex model atmosphere fits were not attempted. Significantly more exposure time and/or high-resolution \xmm RGS data are required for a solid test of our tentative results.

\section{Discussion}\label{sec:discussion}

\subsection{Ejecta deceleration and eruption environment}\label{sec:decel}

In \otwok\ we presented evidence showing a marked decrease of the inferred velocity of the ejecta based on the line width of the H$\alpha$ emission from the 2012, 2013, and 2014 eruptions.  Similar `decelerations' have been seen in a number of RNe, most notable RS~Oph \citep{1964AnAp...27..555D,2009A&A...505..287I}, and have been linked to the physical deceleration of the ejecta as they interact with significant circumbinary material \citep[in the case of RS~Oph, the red giant wind;][]{1967BAN....19..227P,2006ApJ...652..629B,2007MNRAS.374L...1E}.

\citet{1985MNRAS.217..205B} described the standard three phase model of the interaction of the ejecta with a shocked $1/r^{2}$ density profile stellar wind.  Phase~I is the early stage of the interaction and the ejecta are still imparting energy to the shocked wind and the reverse shock, running into the ejecta, remains important.  In Phase~II there is a period of adiabatic forward shock expansion until the shock temperature decreases and the shocked gas becomes well cooled with the momentum-conserving Phase~III of development established.  The expected behavior of the observed shock velocities during Phase~II and III are well represented by the power laws $v\propto t^{-1/3}$ and $v\propto t^{-1/2}$, respectively.

However, for the 2006 eruption of RS~Oph, the X-ray emission has revealed that after an ejecta-dominated, free expansion stage (Phase~I) lasting $\sim6$ days \citep{2006ApJ...652..629B}, the remnant rapidly evolved to display behavior characteristic of a shock experiencing significant radiative cooling (Phase~III).   The duration of an adiabatic, Sedov-Taylor phase (Phase II) was rather shorter than predicted by the remnant evolution model developed by \citet{1985MNRAS.217..205B,1987MNRAS.228..277O,1992MNRAS.255..683O} after the 1985 eruption of RS~Oph.  This was in part due to not appreciating at that time the nature of the SSS phase in RS~Oph but also because of particle acceleration in the shock \citep{2007ApJ...663L.101T}.

Line narrowing has also been witnessed in a number of CNe and RNe that are not expected to have erupted into dense circumbinary environments.  \citet{1996ApJ...456..717S} presents an alternative interpretation, that simply higher velocity material has always traveled the furthest distance, so its emissivity decreases at a greater rate than that of slower moving material, causing the emission lines to narrow.  In systems where the ejecta interact with significant circumbinary material one would expect a combination of both effects.

The H$\alpha$ velocity evolution, as presented in \otwok, was best described by a power law of the form $v\propto t^{-0.12\pm0.05}$, clearly incompatible with the expectations of Phase~II or III.  As such, \otwok\ interpreted the velocity evolution between $1.27\le t\le 4.6$ as Phase~I of shock evolution, and based their conclusions upon this interpretation.  By direct comparison to the RS~Oph system, and using the observed velocities from the 2014 eruption, \otwok\ determined that Phase~I following an eruption of \novak\ should therefore last for $\ga3.6$\,days after maximum visible light (or $\ga4.6$\,days after the onset of the eruption).  The comparison led to an inferred ejected mass from \novak\ of $\ga3\times10^{-8}\,M_\odot$.

However, the addition of the 2015 eruption data significantly alters the picture and subsequent interpretation.  The \otwok\ investigation tied the velocities to the time of maximum visible light, which would make any power-law relation appear too shallow; here we relate the velocities to the estimated time of the eruption (i.e.\ the start of mass ejection; around one day prior to maximum visible light; see Table~\ref{t2_table}).  The early epoch spectral observations of the 2015 eruption contain complex line morphologies showing evidence of very high velocity material.  The addition of the early- and late-time 2015 data, and the shifting of the time-axis (see Table~\ref{fwhm_tab} and Figure~\ref{fwhm_plot}) have the effect of steepening the best-fit power-law to $v\propto t^{-0.28\pm0.05}$.  These updated data are now entirely consistent with the expected deceleration from Phase~II shock behavior as the \novak\ ejecta interact with surrounding pre-existing material.  We also note that the time-scale of Phase~II would therefore run from $t\simeq1$\,d until $\ga4.9$\,d post-eruption -- the time of the final spectrum.  The end of Phase~II is poorly constrained by the lack of later-time spectra, but it appears that Phase~II is consistent with the linear early-decline of the NIR--UV light curve (see Section~\ref{sec:vis_lc}).

With knowledge of high, early time, ejecta velocities (FWHM $\sim13000$\,km\,s$^{-1}$) we can update the Phase~I time-scale estimate as presented in \otwok.  \citet{1985MNRAS.217..205B} show that the time-scale of Phase~I is given by $t\propto M_{\mathrm{e}}u/\dot{M}v_{\mathrm{e}}$, where $M_{\mathrm{e}}$ and $v_{\mathrm{e}}$ are the ejecta mass and initial velocity, respectively, and $\dot{M}$ and $u$ are the donor mass loss rate and wind velocity, respectively; which we will again assume are similar to those seen in RS\,Oph ($v_\mathrm{e}=5100$\,km\,s$^{-1}$; \citealt{2009ApJ...703.1955R}, and $M_\mathrm{e}=2\times10^{-7}\,M_\odot$; \citealt{2006Natur.442..279O}; \citealt{2009A&A...493.1049O}).  For RS\,Oph, \citet{2006ApJ...652..629B} derived a Phase~I timescale of $\sim6$\,days, whereas the updated early high-velocities for \novak\ ($v_e=6500$\,km\,s$^{-1}$; taken as the HWHM of the `rectangular' emission line profile) give a time scale of $0.9\pm0.2$\, days (post-eruption) for an ejected hydrogen mass of $M_{\mathrm{e,H}}=\left(2.6\pm0.4\right)\times10^{-8}\,M_{\odot}$\footnote{Here, we have assumed Solar abundances in the ejecta.  We note that \citet{XrayFlash} assume X=0.53 for the ejecta, if we therefore assume such a higher mass ejecta the Phase~I timescale increases slightly to $1.2\pm0.2$\,days.} (\xtwok), consistent with the earliest ($t<1$\,d) spectra of the 2015 eruption where very high velocity material is seen.  Based on Figure~\ref{fwhm_plot}, it appears that Phase~II begins at around $t\simeq1$\,d, consistent with this timescale estimate for Phase~I.

These observations, now spanning four consecutively detected eruptions, clearly indicate that the ejecta interact with, and shocks, significant pre-existing circumbinary material close to the central system.  With eruptions occurring perhaps as frequently as every six months, the local environment will need to be regularly replenished as we expect RNe to be long-lasting phenomena.  Therefore, it seems likely that the donor star is seeding the circumbinary environment, not just the WD, via a high mass loss rate -- a stellar wind.

If we assume that material is lost from the donor with a velocity of $33\,\mathrm{km}\,\mathrm{s}^{-1}$ \citep[akin to the red giant wind velocity of RS\,Oph;][]{2009A&A...505..287I} then the maximum extent of such material at time of eruption would be 6.6\,AU (or 3.3\,AU for $P_\mathrm{rec}=174$\,d; see Section~\ref{sec:prec}).  Assuming the ejecta initially expand with a (HWHM) velocity of $6500$\,km\,s$^{-1}$, this wind could {\it begin} to be cleared from as early as 1.8\,d (or 0.9\,d for a six month recurrence period).  With the bulk of the ejected material presumably travelling with the mean (but decelerating) velocity of $2670\pm70$\,km\,s$^{-1}$ the ejecta would {\it begin} to run-off the wind at around 4.3\,d (or 2.2\,d for $P_\mathrm{rec}\simeq6$\,months).  As the spectra imply that Phase~II continues until at least day 4, this suggests that the recurrence period may indeed be $\sim1$\,yr; although we note that the donor wind velocity may be different to that of RS~Oph. 

In such a scenario, the high temperatures developed as the ejecta shock any surrounding material is expected to give rise to the so-called `coronal' lines of, for example, [Fe\,{\sc vii}], [Fe\,{\sc x}], and [Fe\,{\sc xiv}], as are observed around 30 days after the eruptions of RS~Oph \citep{1987rorn.conf...27R,2009A&A...505..287I}.  \otwok\ reported that they had observed no evidence of such lines in the {\it individual} spectra from the 2012 and 2014 eruptions.  Such a non-detection was not inconsistent with the shock timescales derived in \otwok.  However, with the much accelerated timescale, as derived above, one might expect to see such high-ionization coronal lines in the early time spectra.  By rough extrapolation, day 30 in RS~Oph is approximately equivalent to day 4--5 in \novak\footnote{But also see \citet{XrayFlash} who predict an even earlier onset of the eruption in \novak.}.  As described in Section~\ref{sec:vis_spec}, there are tentative detections of [Fe\,{\sc vii}], [Fe\,{\sc x}], and [Fe\,{\sc xiv}] emission lines in the combined 2012, 2014, and 2015 spectrum.  \citet{2009A&A...505..287I} reported the appearance of such coronal lines between day 29 and 35 after the 2006 eruption of RS~Oph, these lines then strengthened significantly in later spectra.  Therefore, the weak coronal lines detected before day 4 in \novak\ are roughly consistent with this timescale, and we may expect them to strengthen in later-time spectra.  As discussed in \otwok, any hard X-ray emission from such shocks, as was seen by \swift\ from RS~Oph \citep{2006ApJ...652..629B}, would be undetectable at the distance of \m31. The same conclusion is found when scaling from the more similar nova V745~Sco (see Section~\ref{sec:v745}). Building an argument based on each of these coronal lines individually would be folly; but with five of such lines possibly detected and the `missing' lines easily accounted for, the evidence is quite compelling. 

The tentative identification of Raman scattered O\,{\sc vi} emission at $\sim6830$\,\AA\ in the combined spectrum (as described in Section~\ref{sec:vis_spec}) potentially provides another independent line of evidence pointing directly at the donor star in the system.  Although we cannot completely rule out a C\,{\sc i} origin for this line, the lack of other C\,{\sc i} lines in the spectrum is somewhat telling.  Here we again point to the additional caveats discussed in Section~\ref{sec:comb_spec}.  Such Raman emission is not seen in classical novae (MS- or SG-novae), but is readily observed in symbiotic stars and RG-novae (nova eruptions within symbiotic systems, e.g.\ RS~Oph).

Taken together, the color--magnitude evolution (see Section~\ref{col:mag}), the $v\propto t^{-1/3}$ power-law ejecta deceleration, the coronal lines, and the possible Raman emission band provide strong evidence describing the environment of the nova.  The simplest coherent picture being a WD accreting from the extensive stellar wind of a red giant, with the subsequent nova eruptions then interacting with, and shocking, the extended wind.  By virtue of the low ejected mass and high ejection velocity of \novak\ (both at the extremes of the ranges observed in novae), the early ejecta evolution occurs on timescales significantly shorter than seen in any other novae -- days rather than weeks.

Therefore, we conclude that the mass donor in \novak\ is a red giant.  As such, the companion itself will be accessible to NIR photometric \citep[see][for a detailed discussion of the existing photometry]{HST2016} and possibly even spectroscopic observations.  As a natural consequence, the orbital period of the system must be long (of the order of hundreds of days), and will therefore -- again uniquely -- be similar to the recurrence period.  With no strict requirement for the orbits in a long-orbital period nova to have been completely circularized, the inter eruption periods for \novak\ may be, intriguingly, sensitive to the orbital phase of the system. 

\subsection{Spectral Energy Distribution}\label{sed_sec}

\begin{figure*}
\begin{center}
\includegraphics[width=0.8\textwidth]{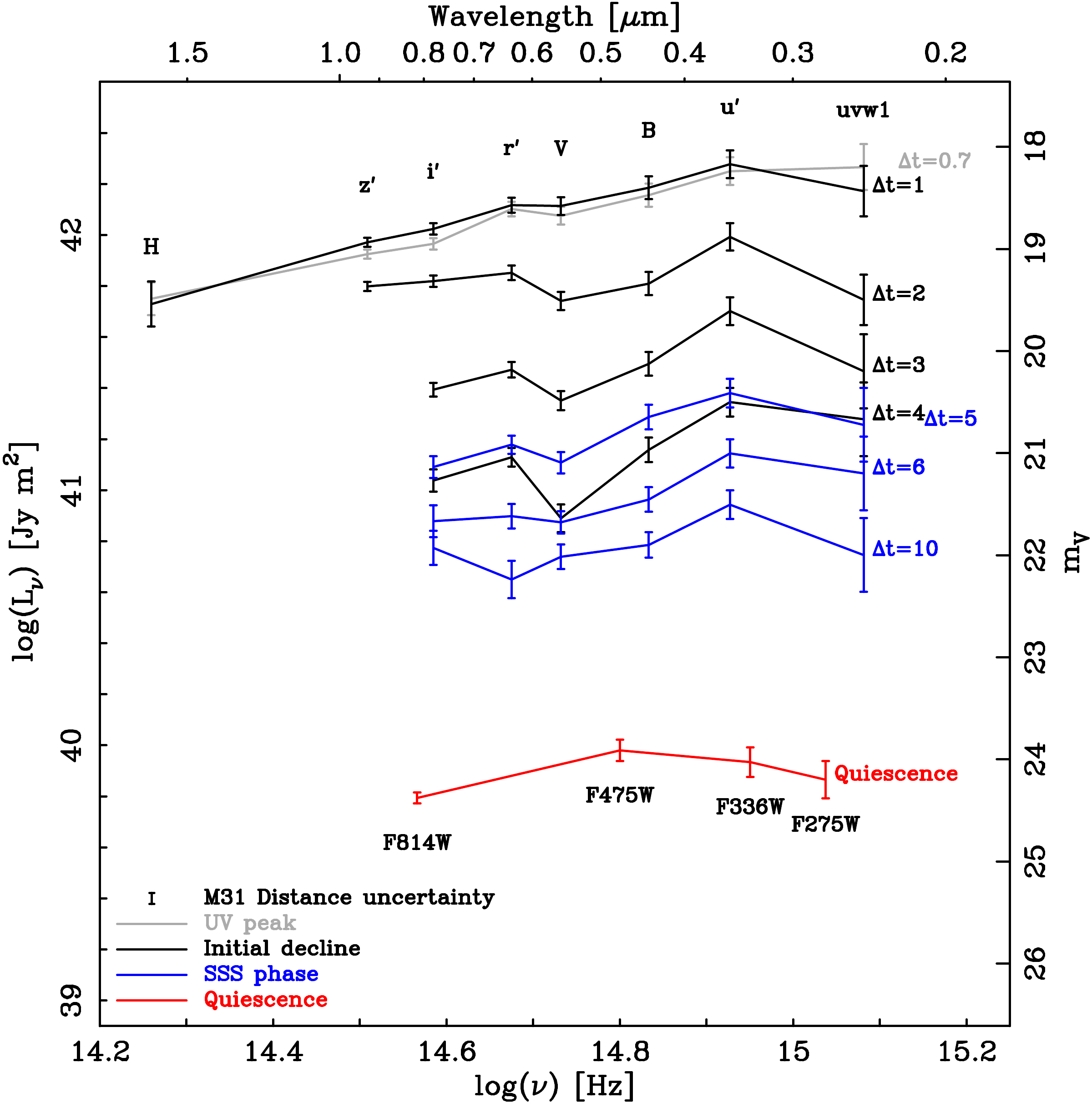}
\end{center}
\caption{Distance and extinction corrected SED plots showing the evolution of the SED of the 2015 eruption ($t=0.7$\,d corresponds to the UV peak, 1--4\,d the initial decline, 5--6 and 10\,d the SSS phase). Units chosen for consistency with similar plots in \citet[see their Figure~71]{2010ApJS..187..275S} and \oonek\ (see their Figure~4). The central wavelength locations of the Johnson-Cousins, Sloan, {\it HST}, and \swift filters are shown to assist the reader.  Here the extinction is treated as just the line-of-sight (Galactic) extinction towards \m31 \citep[$E_{B-V}^{\mathrm{Galactic}}=0.1$;][]{1992ApJS...79...77S}; see the detailed discussion in \citet{HST2016}.  The error bars include contributions from the photometric and extinction uncertainties, the single black point above the key indicates the systematic uncertainty based on the distance of \m31.  A $V$-band apparent magnitude scale (not corrected for extinction) is shown on the right-hand y-axis to aid the reader.\label{sed_evo}}
\end{figure*}

In Figure~\ref{sed_evo} we illustrate the spectral evolution of the 2015 eruption of \novak\ from $\Delta t\simeq0.7$\,days after the eruption (gray points), then from $\Delta t\simeq1$\,days at one day intervals up to and including $\Delta t\simeq6$\,days, at $\Delta t\simeq10$\,days, and at quiescence (red data points).  In this plot the black data points indicate epochs before the SSS turn-on (all during the linear early-decline of the NIR--UV light curve; see Section~\ref{sec:vis_lc}), with blue data points showing the evolution during the SSS phase.  The quiescent data are taken from archival {\it HST} observations (see \oonek).  We have utilized visible and \swift UV absolute calibrations from \citet{1979PASP...91..589B} and \citet{2010Bre}, respectively.  Here we have assumed a distance to \m31 of $770\pm19$\,kpc \citep{1990ApJ...365..186F} and line-of-sight reddening of $E_{B-V}=0.096\pm0.026$ \citep[all of a foreground Galactic origin, see][for a complete reddening analysis and discussion]{HST2016}.  We have utilized the analytical Galactic extinction law of \citet[assuming $R_{V}=3.1$]{1989ApJ...345..245C} to determine the extinction values suitable for the \swift UV filters (calculated at the central wavelength of each filter).	

Compared to the SED of the 2014 eruption presented in \otwok\ (see their Figure~11) the SED coverage from the 2015 is more extensive, following the evolution for over three times as long and over a much broader wavelength regime. In 2014 there was a relatively large time offset between the ground-based (LT) data and the \swift UV data which, given the rapid nature of the evolution, was not ideal.  The temporal matching between the ground and \swift for the 2015 eruption is much improved. 

When nova ejecta are still optically thick, the continuum emission can be well described by a black body with the wavelength of peak emission closely related to the radius of the pseudo-photosphere \citep[see, for example,][]{1976ApJ...204L..35G}.  As initially discussed in \otwok, these SEDs, even at $\Delta t=0.7$\,d are not consistent with black body emission, not even with the Rayleigh-Jeans tail of a `hot', low-photospheric radius, ejecta, as expected from systems with very low ejected mass.

Based on the visible spectra, we know that the emission up to at least $t\simeq5$\,d is a combination of continuum and  emission line flux, and we have no reason to expect that this will change significantly at later epochs.  As the line flux contribution from later times is unknown, we chose not to correct any of the SEDs for lines.  With this caveat in mind, we fit the SEDs with simple power-laws.  The spectra up to $t\simeq1$\,d are continuum dominated (see Figure~\ref{optical_spec}) and, as indicated by Figure~\ref{lineflux_plot}, the integrated line fluxes decrease until after day~1 as the high-velocity component wanes.  Fits to the $t=0.7$ and $t=1$\,d SEDs show power-laws with indices $0.67\pm0.13_\mathrm{random}\pm0.09_\mathrm{systematic}$ and $0.66\pm0.07_\mathrm{random}\pm0.09_\mathrm{systematic}$, respectively (where the systematic error arises from the reddening determination, and persists for all power-laws fit to these data).  Therefore, at this time, the SED is completely consistent with optically-thick free-free emission \citep[$f_\nu\propto\nu^{2/3}$;][]{1975MNRAS.170...41W}.  

From $t\simeq1.5$\,d, the line fluxes initially increase (see Figure~\ref{lineflux_plot}) against a continuum which decays until $t\simeq4$\,d (the end of linear early-decline phase) -- any power-law fits to the SEDs therefore may become increasingly confused by the emission line flux.  At  $t=2$\,d the slope of the SED has decreased to $0.27\pm0.23$, which could indicate a move toward optically-thin free-free emission \citep[$f_\nu\propto\nu^{-0.1}$;][]{1975MNRAS.170...41W}.   However, from $t=3$\,d onward (see Table~\ref{SED_pl}) the general form of the SED is relatively stable and is again consistent with optically {\it thick} free-free emission.  During that period we just see an overall decrease in flux, although between $t=4$\,d and $t=5$\,d the SEDs are essentially unchanged in flux (see below) as the nova enters the quasi-plateau phase.  The mean SED slope across all epochs is $0.69\pm0.06$, consistent with optically thick free-free emission.  Without spectra to further constrain the emission, it is difficult to speculate whether the slope change at $t=2$\,d is a genuine transition to optically thin free-free emission, with the later SEDs becoming increasingly line dominated, or simply a statistical outlier.

\begin{table}
\caption{Indices of power-laws fit to the evolving SED of \novak.\label{SED_pl}}
\begin{center}
\begin{tabular}{ll}
\hline\hline
$\Delta t$ (days) & SED power-law index$^{\dag}$\\
\hline
0.7 & $0.67\pm0.13$ \\
1 & $0.66\pm0.07$ \\
2 & $0.27\pm0.23$ \\
3 & $0.92\pm0.29$ \\
4 & $0.91\pm0.33$ \\
5 & $0.86\pm0.20$ \\
6 & $1.06\pm0.25$ \\
10 & $0.82\pm0.29$ \\
\hline
\end{tabular}
\end{center}
\tablenotetext{\dag}{Quoted uncertainties are based on random photometric errors, an addition systematic error of 0.09 due to the extinction uncertainty should also be applied to all indices.}
\end{table}

It is interesting to note the behavior of the $r'$ and $V$-band relative flux as the SED evolves.  At $t=1$\,d, when the SED is continuum dominated these points follow the general optically-thick free-free trend, with a hint of an $r'$-band excess due to the strong H$\alpha$ emission.  As the evolution continues, the $r'$-band excess strengthens against $V$ as the continuum drops and the H$\alpha$ flux strengthens (as mentioned broadly above) -- continuing throughout the linear early-decline (up to $t\simeq4$\,d).  Then as the nova enters the quasi-plateau phase, and the SSS turns on ($t\simeq5$\,d), this trend starts to reverse.  Between day 4 and 5, the only significant change in the SEDs is an increase in $V$-band flux.  By $t\simeq10$\,d, the $V$-band flux is stronger than the $r'$-band.  While late-time spectra are required to confirm what is causing such a trend, it is likely to be line driven.  Nebular lines typically begin to appear in nova spectra once the UV becomes optically thin \citep[see, for example][which is also related to the unveiling of the SSS]{2001AJ....121.1636M,2002A&A...390..155D}.  The strongest nebular lines in novae are usually the [O\,{\sc iii}]\,(particularly 4959/5007\AA) lines -- located within the $V$-band.  Therefore we predict that \novak\ enters the nebular phase during its quasi-plateau phase, probably between days 4 and 5.

As described by \oonek\ and \ponek, the quiescent SED (indicated by the red data in Figure~\ref{sed_evo}) is consistent with being dominated by a luminous accretion disk; the SED during the late decline-phase and the nature of the quiescent system is discussed in detail in the companion paper by \citet{HST2016}.

We note that the visible peak (for \novak\ at $t\simeq1$\,d) for a `typical' nova corresponds to the maximum extent of the pseudo-photosphere and the minimal effective temperature ($T_\mathrm{eff}\simeq8000$\,K) -- therefore, such a nova is expected to have a black body-like spectrum, which peaks in the visible, at the time of the peak in the visible light curve.  As stated in \otwok, we again find that the SED at visible peak does not correspond to a black body peaking in the visible.  But now that the extinction is constrained \citep{HST2016}, we have also confirmed that the SED at the visible peak is {\it not} the Rayleigh-Jeans tail ($f_\nu\propto\nu^2$) of black body emission from a `hotter' source.  Even at such an early stage ($t=1$\,d), the visible emission has already evolved to an (optically thick) free-free form.  During the linear initial-decline phase, the NUV \swift uvw1 data points are systematically lower than the $u'$-band data -- once the SSS turns on this `discrepancy' may disappear.  Although a single data point alone cannot confirm this, it may be evidence of a transition to an optically thick regime at bluer wavelengths.  Together, these SEDs confirm that the emission peak from \novak\ never moves as red as the visible, and probably not even into the NUV or FUV, and that it may always be constrained to the EUV (before shifting back into the X-ray as the SSS is unveiled).  Therefore, as the optical depth of the ejecta is so low, we can conclude that the ejected mass in an eruption of \novak\ must be significantly lower than in a `typical' CN, or even all other observed RNe.  This is in agreement with the theoretical estimates obtained through hydrodynamic simulations when very high values of the WD mass and accretion rate are adopted \citep[see, for example,][]{2008NewAR..52..386H}

\subsection{X-ray spectral variability}
\label{sec:disc_xvar}

The spectral models summarized in Table~\ref{tab:xspec} suggest that the SSS phase emission of \nova can be consistently described using emission lines superimposed on an absorbed black body continuum. The statistical significance of these detections is modest, and has not be subjected to the rigorous examination described by \citet{2008ApJ...679..587H}, which is beyond the scope of this work. Nevertheless, we consider discussion of the potential origin of these features to be of interest.

Even though no fitted line appears in all groups (compare also Figure~\ref{fig:xray_group_spec}), there are certain features that several groups have in common. Many of the putative X-ray emission lines fall near known transitions of N\,{\sc vii} (0.50\,keV), O\,{\sc vii} (0.56\,keV), or O\,{\sc viii} (0.65\,keV). In the highest temperature spectra (i.e.\ groups 4--7) there is weak evidence for Ne emission features (Ne\,{\sc ix}\,$\alpha$ and Ne\,{\sc x}\,$\alpha$) around 0.9 or 1\,keV (group 6 only). However, not all fitted features have obvious laboratory counterparts (see Figure~\ref{fig:xray_group_spec}); and one must keep in mind, that due to the broad XRT spectral response width, narrow observed features in the count spectrum are likely a noisy representation of the underlying spectrum at best.

The identification and classification of possible emission lines is complicated further by the strong SSS variability of \nova up to day 13 (see Figure~\ref{fig:xray_split}). All emission components are potentially variable and some lines will not necessarily be present in all snapshot spectra of a certain group.  The visible differences between the group spectra in Figure~\ref{fig:xray_group_spec} already suggest a more complex spectral variability. Incidentally, the group spectra after the end of the early variability phase (i.e.\ groups 7, 8, and 9, see Table\,\ref{tab:xspec}) appeared to be somewhat more homogeneous and easier to fit than the earlier spectra. These tentative first results need to be tested robustly with high-resolution (\xmm RGS) X-ray spectra to enable a confident interpretation of the underlying physics.

For now, the X-ray spectral models suggested for \novak, i.e.\ a hot photospheric continuum with superimposed emission lines of highly ionized nitrogen, oxygen, and possibly neon, is reminiscent of the high-resolution spectra of the Galactic RN U~Sco as discussed by \citet{2012ApJ...745...43N}. Their Figure~8 shows strong oxygen features (and weak Ne lines) to appear as the continuum temperature increased. \citet{2012ApJ...745...43N} suggest that the strongest emission lines appearing at the peak of the continuum flux indicates that those lines were photoexcitated and that therefore the plasma that produces them should have been close to the central SSS. Note, however, that \citet{2012ApJ...745...43N} suggest that the Ne lines, together with potential Mg lines, originate more likely in a collisional plasma. In our case, there seem to be no detectable lines beyond the Wien tail of the black body model.

Recently, \citet{2013A&A...559A..50N} introduced a phenomenological classification of SSSs, according to their high-resolution spectra, into those exhibiting clear emission lines (SSe) or absorption lines (SSa) in addition to a continuum component. They note that SSe objects have on average higher inclination angles. \citet{2013A&A...559A..50N} suggest that SSe spectra indicate an obscuration of the central WD, with residual continuum emission being observable because of Thomson scattering. In this picture, emission lines are photoexcitated and arise from resonant line scattering. In this model interpretation, \nova would be classified as a SSe and we discuss the possible implications on its inclination angle in Section~\ref{sec:incl}.

We also tried to model the separate high- and low-flux spectra in Figure~\ref{fig:xray_split}c using the same approach as for the group spectra. The results are included in Table~\ref{tab:xspec}. We found that a black body continuum plus emission lines again leads to statistically improved and physically more consistent values than the pure black body. The estimated column densities have relatively large uncertainties but are consistent with the best group fit of \nh $= 0.7^{+0.5}_{-0.5}$ \hcm{21}. The continuum temperatures are not well constrained either. However, they suggest that the high flux spectra might have a higher black body temperature ($\sim120$\,eV) than the low-flux spectra ($\sim90$\,eV). Taken at face value, this difference would translate (via the Stefan-Boltzmann law) to a factor of $\sim 3$ larger flux for the high-state bins, which is consistent with an average factor of $2.6$ flux difference between high and low-flux snapshots (see Section\,\ref{sec:lc_xvar}), without the need for a change in radius.

For the identified emission lines we found no obvious overlap in energies between the high- and low-flux spectra. The two higher-energy lines in Table~\ref{tab:xspec} overlap within their $2\sigma$ uncertainties. However, given the current spectral resolution and relatively low number of counts it is not possible to study the emission line variability with any confidence.

\begin{figure}
\includegraphics[width=\columnwidth]{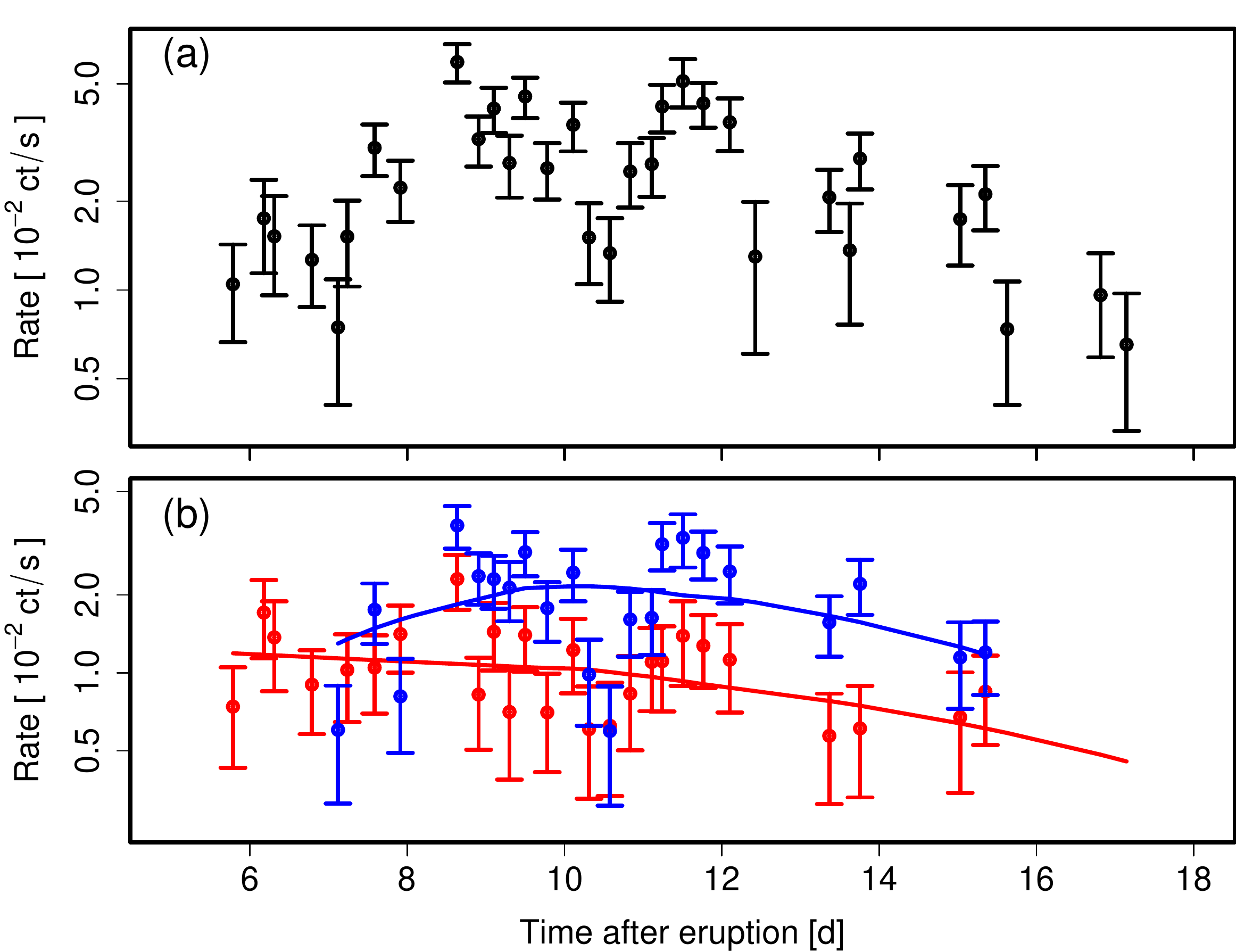}
\caption{\textit{Panel a}: Same as Figure~\ref{fig:xray_lc}a for the 2015 detections only. \textit{Panel b}: Light curves for the energy bands of 0.3-0.5\,keV (red) and 0.5-1.5\,keV (blue) with corresponding smoothed fits based on local regression (solid lines).\label{fig:xray_hr15}}
\end{figure}

Additionally, based on the group spectra shown in Figure~\ref{fig:xray_group_spec} we examined the SSS variability in two different energy bands: 0.3--0.5\,keV (soft) and 0.5--1.5\,keV (hard). Almost all of the potential emission features were found in the hard band. The resulting light curves for the 2015 eruption are shown in Figure~\ref{fig:xray_hr15}. The apparent dip around day 11 appears to be confined to the hard band. Statistical tests confirm that there is significantly (F-test: beyond $3\sigma$ level; p-value = $4.154\times10^{-6}$) less variability in the soft band (standard deviation = 0.36) compared to the hard band ($\sigma = 0.88$). This is further evidence that the early SSS variability is connected to spectral variations.

High-resolution X-ray spectroscopy data will be obtained for a future eruption, using a dedicated \xmm RGS observation, in order to study the putative spectral features and their possible variability with much more confidence.

\subsection{Geometry, inclination, and jets?}\label{sec:incl}

The inclination of the \novak\ system to the line-of-sight is unknown, but it is one of the key missing ingredients to fully understanding the observations of the eruptions.

Working under the assumption that the surrounding nebulosity was related to past eruptions of \novak\ and with the morphology of the H$\alpha$ emission line during the linear early-decline phase, \otwok\ employed morpho-kinematical modeling \cite[see, for example,][]{2009ApJ...703.1955R,2011MNRAS.412.1701R,2013MNRAS.433.1991R} to derive an inclination estimate of $i=46^{+8}_{-38}$\,degrees (where $i=90$ corresponds to an edge on, eclipsing, system).  Here, again, we strongly reinforce the caveats placed on this result by \otwok, for example, this result is likely not unique and is strongly dependent on the assumed connection between the nova and the nebula, and that the nebula has not been significantly re-shaped post-eruption.

Here though, the high velocity material present in the early spectra may provide some useful constraints on the system inclination.  For simplification, in Section~\ref{sec:decel}, we have worked under the assumption that the ejecta and wind are spherically symmetric.  The fleeting nature of this high velocity emission (up to approximately 1\,day post-eruption), surrounding the longer-lived and slower central emission component (see Figure~\ref{halpha_line}), is strongly suggestive -- as proposed in many novae -- of highly asymmetrical ejecta.  For example, the observed line profiles are inconsistent with those expected from either a filled or shell-like spherical system.  Given the high velocities initially observed, higher than any velocities previously recorded in novae, we must assume that this material is essentially traveling along, or close to, the line of sight.  With the expected geometry of the pre-eruption system, the WD, donor, accretion disk, and the bulk of any circumbinary material or stellar wind all lying in the orbital plane, it seems likely that this high velocity material must have been ejected in the polar direction where it can expand relatively unimpeded \citep[see, for example, the ejecta geometry of V959~Monocerotis, as described by][]{2014Natur.514..339C}.  With the emissivity of this essentially free-expanding material diminishing rapidly, the spectral evidence is similarly short-lived.  With the high velocities seen here already approaching those seen in SNe, we must then infer that the orbital plane of \novak\ has to be close to being face-on, and that the central emission component is due to equatorial expansion of the ejecta (aligned close to the plane of the sky).  Finally, we note that the inclination derived in \otwok\ is not inconsistent with such a geometry.

From the X-ray point of view, there appears to be evidence for additional spectral components beyond a simple (black body) continuum model (see Section~\ref{sec:xspec}). The combined spectra in Figure~\ref{fig:xray_group_spec} seem to be consistent with the presence of emission lines. The Galactic study of \citet{2013A&A...559A..50N} discussed a possible link between the presence of strong emission lines in SSS high-resolution X-ray spectra (their SSe class) and the inclination angle of the system. The SSe were interpreted as obscured WDs and the majority of them had high inclinations. If \nova were a SSe with a high inclination then this would be somewhat at odds with our conclusions drawn from the visible spectra.

However, the sample of \citet{2013A&A...559A..50N} was still small and these authors argued for a careful interpretation of the apparent correlation. With only tentative hints at X-ray emission lines in \novak, and insufficient evidence on the impact of the inclination angle, more data and a larger Galactic sample are needed to explore and harness the predictive power of X-ray spectral classifications on the binary geometry.

Following radio observations of the 1985 and 2006 eruptions of RS~Oph, the presence of a jet or jet-like structure was reported \citep[respectively]{1989MNRAS.237...81T,2008ApJ...688..559R}.  \citet{2008ApJ...685L.137S} proposed that the jets in RS~Oph are driven by highly collimated outflows, rather than, for example, inherently asymmetric explosions, or interaction with the circumbinary medium.  The H$\alpha$ profiles in the new early \novak\ spectra (see the bottom right plot within Figure~\ref{halpha_line}) are not dissimilar to the H$\alpha$ lines of RS~Oph at day 12 and 15 after the 2006 eruption \citep[see][their Figure~2; although the velocities in \novak\ are significantly greater]{2008ASPC..401..227S}.  Figure~\ref{fig:jets} (bottom right) shows the H$\alpha$ line profile 0.67\,d after the 2015 eruption of \novak, here the high-velocity emission has been isolated as a pair of Gaussian profiles (blue lines) around the central profile (red line).  These high-velocity Gaussians have consistent fluxes and widths (mean $\mathrm{FWHM}=2800\pm100\,\mathrm{km}\,\mathrm{s}^{-1}$).  The blue and red shifted Gaussians are offset from the rest wavelength by $-4860\pm200\,\mathrm{km}\,\mathrm{s}^{-1}$ and $5920\pm200\,\mathrm{km}\,\mathrm{s}^{-1}$.

With such a high mass accretion rate and accretion disk luminosity (see \oonek) and a proposed red giant donor (hence large orbital separation) it seems a reasonable assumption that the \novak\ accretion disk is particularly massive.  As such, it is a reasonable step to further propose that the accretion disk may survive each eruption.  Such a short recurrence period will therefore require accretion to begin soon after each eruption.  With disk formation timescales being related to the orbital period \cite[see the detailed discussion in][and references therein]{2011ApJ...742..113S}, the long orbital period required by a giant donor would not permit a destroyed or heavily disrupted disk to reform in such a short time-scale.  Therefore the proposal of a long orbital period, short recurrence period, system with a giant donor seems to require that the accretion disk persists post-eruption.  A surviving accretion disk may be able to provide the collimation mechanism required to drive any jets.  This proposal of a surviving accretion disk is further explored in \citet{HST2016}.

\subsection{Recurrence Period}\label{sec:prec}

\begin{figure}
\includegraphics[width=\columnwidth]{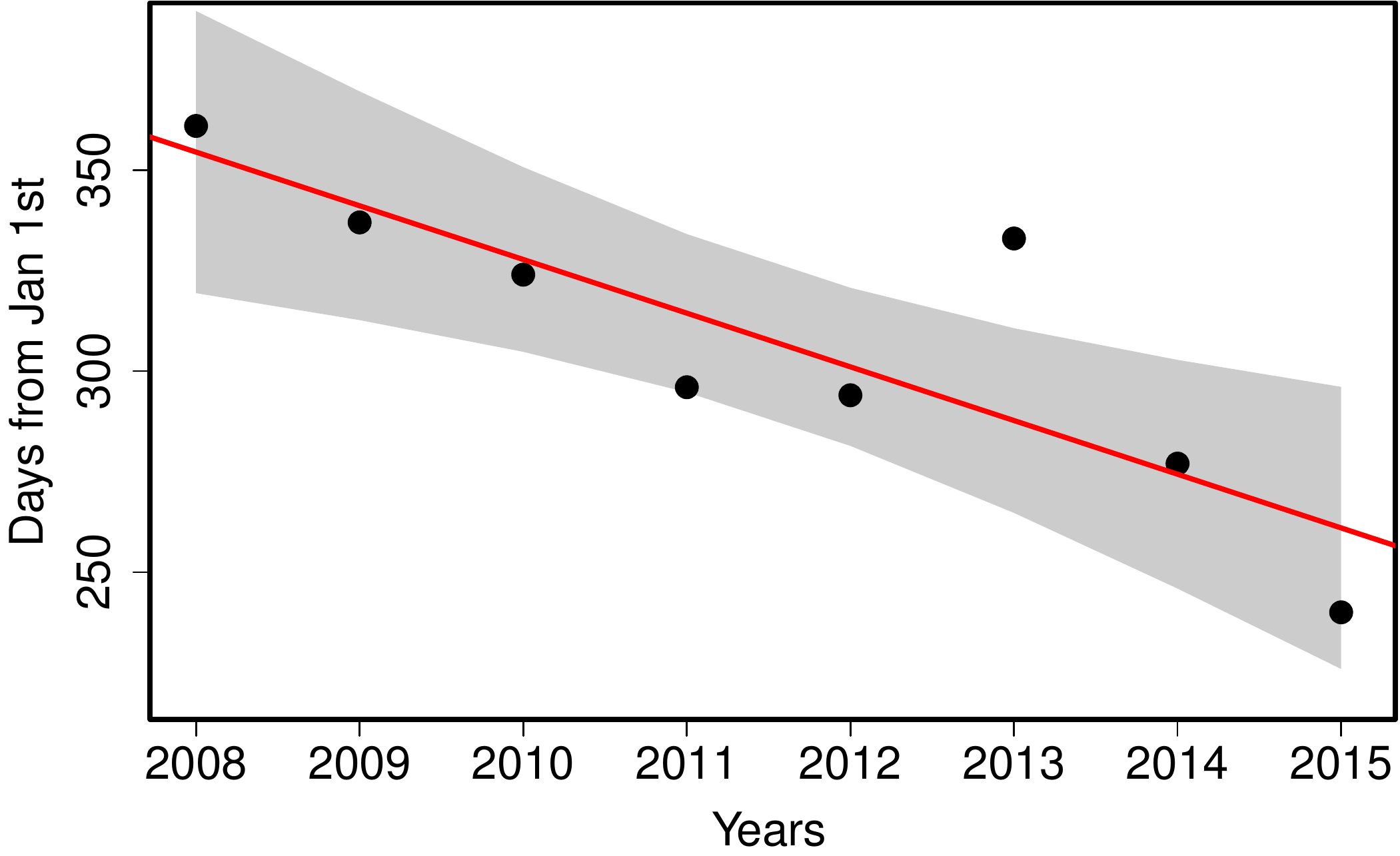}
\caption{Distribution of eruption dates (in days of the year) over time since 2008. Based on Table~\ref{eruption_history}. The red line is the best fit. The gray area is the corresponding 95\% confidence region. The uncertainties of the estimated eruption dates are smaller than the symbol size. Here we show the updated recurrence time fit and scatter, which are the basis for the eruption date predictions in Table \ref{tab:future}. \label{fig:rec}}
\end{figure}

In Figure~\ref{fig:rec} we show the eruption dates, in days of the year, of every visible detection of an eruption during 2008--2015. The plot and corresponding linear fit show that successive eruptions tend to occur slightly earlier in the year. This trend is significant. Therefore, the observed recurrence period appears to be slightly shorter than one year. The Figure~\ref{fig:rec} is based on \halfk, where an apparent period of $351\pm11$\,d was estimated. Including the 2015 eruption here we find a value of $347\pm10$\,d ($0.950\pm0.027$\,yr).

The first long-term analysis of the recurrence period of \novak\ was also presented in \halfk, who proposed a $\sim6$\,month eruption cycle, rather than the approximately yearly one currently observed. This scenario is based on the historical eruption dates inferred from archival X-ray detections in 1992, 1993 (with ROSAT), and 2001 (with \chandrak). The dates of these eruptions are only consistent with the trend shown in Figure~\ref{fig:rec} if a shift of $\sim0.5$\,yr is applied. The simplest explanation for this behavior is that \nova has two eruptions per year. Hereafter we define the ``a'' and ``b'' eruptions as the first and second eruption in a given calendar year, respectively\footnote{Assuming the shorter recurrence period, there will be a short period repeating every $\sim21$\,years in which a third ``c'' eruption may occur each year.}. In this picture, the missing ``a'' eruptions during 2008-15 have occurred during the time of Mar-May while \m31 was in Solar conjunction.

The corresponding recurrence period is $174\pm10$\,d or $0.476\pm0.027$\,yr. As for the above estimate of the apparent recurrence period the given uncertainty is the standard error of the mean. Individual eruptions appear to deviate from the mean by about $\pm1$ month ($1\sigma$; cf.\ Figure~\ref{fig:rec}).  We note that this $1\sigma$ prediction window is $\sim12$ times shorter than that for the Galactic RN U~Sco \citep{2005ApJ...621L..53S,2010AJ....140..925S}.

So far, only one eruption per year has been detected. However, as the eruptions of the established ``b'' sequence tend to occur earlier each year, the predicted ``a'' eruptions (in the first half of the year) are expected slowly to leave the Sun constraint. In Table~\ref{tab:future} we list the predicted future eruption dates, together with their $1\sigma$ prediction uncertainties, based on all known eruptions from Table~\ref{eruption_history}. These estimates assume a 6-month period which we expect to confirm in the coming years.

\begin{table}
\caption{Predicted future eruption dates and $1\sigma$ prediction uncertainty ranges of \nova assuming a $\sim6$ month recurrence period The ``a'' and ``b'' labels refer to the first and second eruptions in a given year, respectively.\label{tab:future}}
\begin{center}
\begin{tabular}{lllll}
\tableline\tableline
ID & MJD & Date & Lower limit & Upper limit\\
\tableline
2016-b & 57647 & 2016-09-16 & 2016-08-21 & 2016-10-13\\
2017-a & 57826 & 2017-03-14 & 2017-02-15 & 2017-04-10\\
2017-b & 58003 & 2017-09-07 & 2017-08-11 & 2017-10-04\\
2018-a & 58179 & 2018-03-02 & 2018-02-03 & 2018-03-29\\
2018-b & 58356 & 2018-08-26 & 2018-07-30 & 2018-09-22\\
2019-a & 58533 & 2019-02-19 & 2019-01-23 & 2019-03-18\\
2019-b & 58710 & 2019-08-15 & 2019-07-19 & 2019-09-11\\
2020-a & 58886 & 2020-02-07 & 2020-01-11 & 2020-03-06\\
2020-b & 59063 & 2020-08-02 & 2020-07-06 & 2020-08-30\\
\tableline
\end{tabular}
\end{center}
\end{table}

The data shown in Table~\ref{tab:future} and Figure~\ref{fig:rec} will be updated after each future eruption, which may also allow us to improve the prediction accuracy. A comprehensive search in various archives for historical eruptions is in progress and the results will be published in the near future.

\subsection{Comparison to V745 Scorpii}\label{sec:v745}

The Galactic RN V745~Sco can be considered as the closest cousin of \novak. Assumed to be hosting the most massive WD in the Galaxy and being fueled by wind accretion from an RG companion, this nova shares many of the extreme observational characteristics of our object. Here we discuss the main similarities and differences between these two promising SN~Ia progenitor candidates.

V745~Sco belongs to the RG-nova class \citep{2012ApJ...746...61D} and has undergone detected eruptions in 1937, 1989, and most recently in 2014 from which a period of $P_\mathrm{rec}\sim25$\,yrs is inferred.  However, it should be noted that with the peak luminosity reaching (only) 10$^\mathrm{th}$ magnitude, some eruptions may have been missed, rendering the inferred recurrence period an upper limit \citep[see][and references therein, for a general Galactic nova completeness discussion]{2016arXiv160602358S}.

\citet{2015MNRAS.454.3108P} presented a comprehensive X-ray and UV analysis of the 2014 eruption of V745~Sco. They found that photospheric X-ray emission was detected only 4--5~d after eruption, taking into account the uncertainty of the eruption date, thereby narrowly surpassing the turn-on time scale of \novak. However, in contrast to \novak, the SSS phase only lasted until about day 10 (see their Figure 11), instead of day 18, giving V745~Sco the fastest SSS turn-off and the shortest SSS phase observed in any nova, so far.

The fact that these time scales are even faster than in \nova suggests that less matter was ejected and burned during the SSS phase. This would indicate that a smaller amount of hydrogen was necessary to trigger the eruption, and that therefore the WD in V745~Sco would be more massive than in our case \citep[see the models of][]{2006ApJS..167...59H}. A lower hydrogen content of the accreted material would also lead to a shortening of these time scales, but they react more sensitively to changes in the WD mass \citep[cf.][]{2006ApJS..167...59H}. In either case, the much longer recurrence time of V745~Sco (25\,yr vs.\ 0.5\,/\,1\,yr for \novak) suggests that its accretion rate is lower than the 1.6\tpower{-7}\,$M_{\odot}$\,yr$^{-1}$ estimated for \nova based on the theoretical models of \citet[who assume a 1~yr recurrence period]{2015ApJ...808...52K}. Additionally, while for \novak\ we speculate, in Section~\ref{sec:incl}, that the short recurrence time may require the accretion disk to stay intact, it may not survive the eruption in V745~Sco, thus delaying the next eruption by at least the time it takes for accretion to resume again. 

\citet{2015MNRAS.454.3108P} reported that V745~Sco showed no variability during the early SSS phase. However, the observed smooth rise to peak flux was exceptionally rapid and essentially covered by only 6 observations, which might not have been sufficient to capture variability. Interestingly, there appeared to be a dip in effective temperature at maximum (their Figure 5) which looks similar to our Figure\,\ref{fig:xray_temp}.

V745~Sco is a symbiotic system with the WD accreting from a RG companion with a possible orbital period of $\sim 500$~d \citep[see][and references therein for the controversy on this period]{2015MNRAS.454.3108P}. \citet{2016ApJ...825...95D} reported on \chandra spectra of the post-SSS phase (day 16) that showed a shock-heated circumstellar medium. They suggest an orbital inclination close to face-on, similar to the visible evidence for \novak. For both novae, the potential presence of strong emission lines on top of the SSS continuum appears to be somewhat at odds with a low inclination angle (see Section \ref{sec:incl}).

In agreement with NIR studies by \citet{2014ApJ...785L..11B}, \citet{2016ApJ...825...95D} interpreted the observational characteristics of the V745~Sco eruption as a high-velocity blast wave interacting with a RG wind. This is consistent with early Fermi-LAT $\gamma$-ray detections \citep{2014ATel.5879....1C}. \citet{2014ApJ...785L..11B} showed the narrowing of the Pa\,$\beta$ FWHM, suggesting that the shock was propagating into a wind that was not spherically symmetric. \citet{2016ApJ...825...95D} inferred a collimation of the blast wave by an equatorial density enhancement. They also concluded that the WD in V745~Sco is likely gaining mass and is another good SN~Ia progenitor candidate.

The early hard X-ray emission in V745~Sco, indicative of the shock-heated plasma, was observed with \swiftk/XRT at count rates which were a factor of $\sim100$ fainter than the maximum during the SSS phase \citep[see Figure 1 in][]{2015MNRAS.454.3108P}. Below we discuss that \nova had a hotter SSS maximum than V745~Sco, which would increase the contrast between the maximum count rate and the early hard emission for our nova by a factor of $\sim3$ based on temperature only. Additionally, the ejected mass of V745~Sco was estimated by \citet{2014ApJ...785L..11B}, \citet{2015MNRAS.454.3108P}, and \citet{2016ApJ...825...95D} consistently as $\sim$~\power{-7}~$M_\sun$, which is more than a factor of two higher than for our object (\xtwok,\otwok).

Scaling from V745~Sco, this suggests that the luminosity of the early hard X-ray emission in \nova would be significantly more than two orders of magnitude below its SSS maximum. Combining all \swift observations between the visible detection and the SSS turn-on from this year and 2014 (\xtwok) results in an upper limit of 6.5 \cts{-4} (for 47.4~ks of total exposure). This is nearly a factor of 100 below the detected SSS peak (5--6 \cts{-2}, see Figure~\ref{fig:xray_lc} and Table~\ref{tab:obs_swift}). Therefore, the non-detection of hard X-rays is expected and significantly more exposure would be needed to observe them.

The X-ray spectral evolution of the SSS phase in V745~Sco reached black body temperatures of about 90~eV, significantly cooler than \nova \citep[see Figure 11 in][and note the artificial shift in temperature for V745~Sco]{2015MNRAS.454.3108P}. This is slightly at odds with the (\m31) nova population correlation models of \citet{2014A&A...563A...2H} which suggest that shorter SSS time scales are linked to higher black body temperatures, possibly via the WD mass. However, \citet{2015MNRAS.454.3108P} discuss the possibility that the hydrogen burning in V745~Sco had ceased before the SSS could reach its potential maximum temperature, as evidenced by an almost negligible temperature plateau of only about 2 days (see their Figure 11). In any case, the effective temperature of V745~Sco is very high and qualitatively consistent with its fast SSS evolution.

By virtue of the higher count rates for their Galactic object, \citet{2015MNRAS.454.3108P} could analyze their \swift XRT spectra in much more detail. They fitted the SSS emission with a black body continuum plus 5 emission lines and two absorption edges (neutral and H-like oxygen at 0.54 and 0.87\,keV). Fixing the line energies to the H-like and He-like transitions of oxygen and neon, and He-like magnesium (1.35\,keV) they reported significantly improved fit statistics. The neon lines were strongest at the SSS peak, while H-like oxygen and He-like magnesium were significant throughout; the He-like oxygen line only occasionally reaching significance.

Apart from the Mg line, which is not evident in our XRT spectra, these emission features are similar to those tentatively suggested in \nova (cf.\ Figure~\ref{fig:xray_group_spec} and Table~\ref{tab:xspec}). The suggested strengthening of the Ne lines with increasing continuum temperature, here and in V745~Sco, indicates photoexcitation. In contrast to \citet{2015MNRAS.454.3108P}, we did not find absorption edges to have a significant impact at any stage of the spectral evolution. Our spectra, so far, did not have sufficient counts to model both emission and absorption features at the same time.

In essence, V745~Sco and \nova are two extreme RNe that share several observational characteristics. In both objects, low-mass ejecta appear to interact strongly with the stellar wind from a companion, slowing significantly in the process and producing high-temperature shocks. Their SSS spectra extend to high temperatures and appear to feature strong, variable emission lines. While the WD in V745~Sco may be more massive, \nova appears to have the higher accretion rate, providing the unique opportunity to observe at least one eruption per year. Assuming a $\sim25$\,year cycle, the next eruption of V745~Sco is expected around 2039, by which time we will have studied \nova to a sufficient extent to provide detailed predictions on the variations on the eruption properties of V745~Sco. 

\section{Summary \& Conclusions}\label{sec:conclusions}

The 2015 eruption of \novak\ was discovered independently by dedicated monitoring programs utilizing the \swift orbiting observatory and the LCOGT 2\,m (Hawaii) on 2015 Aug 28.41 UT and 28.425, respectively, with pre-eruption non-detections constraining the time of the eruption to 2015 Aug $28.28\pm0.12$ UT.  Following detection, a pre-planned pan-chromatic follow-up campaign was initiated which involved ten ground-based telescopes around the globe, but was spearheaded by \swiftk, the LT, and the LCOGT.

The eruption light curves spanning the electromagnetic-spectrum from the super-soft X-rays to the $I/i'$-band show remarkable similarity between the 2013, 2014, and 2015 eruptions.  The combined visible spectrum from the 2012, 2014, and 2015 eruptions shows tentative evidence for high-excitation coronal lines of [Fe\,{\sc vii}], [Fe\,{\sc x}], and [Fe\,{\sc xiv}], often observed during high temperature shocks, and also hints at the presence of Raman scattered O\,{\sc vi} emission, as seen in spectra of symbiotic stars and novae with red giant companions.  The visible spectra from the 2012--2015 eruptions show a consistent decrease in line width.  Between days 1 and 4, post eruption, this deceleration is consistent with a power-law decline of the ejection velocity ($v\propto t^{-1/3}$).  This deceleration is consistent with consistent with the adiabatic Phase~II shock development as the ejecta interact with significant preexisting circumbinary material.  These observations, backed-up by the color--magnitude behavior, point to the donor being a red giant in a long orbital period system.

Below we summarize a number of our conclusions.

\begin{enumerate}

\item The color--magnitude evolution in the visible appears more consistent with the behavior of RS~Oph and V745 Sco (both harboring red giant donors) than that of the sub-giant (e.g.\ U~Sco) or main sequence (T~Pyx) donor RNe.

\item There is no evidence at visible wavelengths for optically thick photospheric emission during the early evolution of the eruption. This points to a large {\it minimum} temperature of the expanding photosphere, with photospheric emission therefore peaking in the FUV or EUV.

\item The evolving SED of the eruption points to optically thick free-free emission being the dominant process (in the NIR--NUV) throughout the evolution from $t=0.7$\,d to $t=10$\,d.  Although significant contribution to the SED from emission lines cannot be ruled beyond day four.

\item The $V$- and $r'$-band trends in the SED leads to a prediction of the nebular phase beginning as early as day 5 post-eruption.

\item Emission from extremely high velocity ($\mathrm{FWHM}\simeq13000\,\mathrm{km}\,\mathrm{s}^{-1}$) material seen only in the early spectra ($t\la1$\,d) is indicative of outflows along the polar direction -- possible highly collimated outflows or jets.

\item We obtained an unprecedentedly detailed UV light curve with \swift UVOT, observing for the first time the rise to maximum and fast decline with subsequent plateaus. The UV peak clearly precedes the visible peak.

\item The X-ray light curve of the 2015 eruption was consistent with the last two years in its time scales, $\eton = 5.6\pm0.7$~d and $\etoff = 18.6\pm0.7$~d, as well as in the properties of the early SSS variability and its cessation around day 13.

\item The 2015 X-ray light curve also showed evidence of a peculiar dip around day 11 which might have been present in the 2013 light curve as well.

\item Merged X-ray spectra tentatively suggest the presence of high-ionization emission lines superimposed on a photospheric continuum that reaches black body temperatures of around 120\,eV.

\end{enumerate}

The next eruption of \novak\ is predicted for mid Sep.\ 2016 with a $1\sigma$ uncertainty of about 1 month. This prediction holds for both the 1 year and the 6 month recurrence scenarios. In the case of the 6 month period, we expect the subsequent eruption in Feb.\ -- Apr.\ 2017.

\acknowledgements
The Liverpool Telescope is operated on the island of La Palma by Liverpool John Moores University (LJMU) in the Spanish Observatorio del Roque de los Muchachos of the Instituto de Astrof\'{i}sica de Canarias with financial support from STFC.  This work makes use of observations from the LCOGT network.  The Pirka telescope is operated by Graduate School of Science, Hokkaido University, and it also participates in the Optical \& Near-Infrared Astronomy Inter-University Cooperation Program,  supported by the MEXT of Japan.  This research made use of data supplied by the UK Swift Science Data Centre at the University of Leicester.  This publication makes use of data products from the Two Micron All Sky Survey, which is a joint project of the University of Massachusetts and the Infrared Processing and Analysis Center/California Institute of Technology, funded by the National Aeronautics and Space Administration and the National Science Foundation.  This research has made use of NASA's Astrophysics Data System Bibliographic Services.  

We thank Brian W.\ Shafter (PHHS) for assistance with the MLO data reduction.  The authors would like to thank K.\ Ho\v{n}kov\'a for sharing her observing time on the Ond\v{r}ejov 0.65m telescope, M.\ Eracleous for his help at the Apache Point Observatory, D.\ Hatzidimitriou for assistance with the 2015 campaign, D.\ Willmarth for obtaining the KPNO spectrum, and P.\ James for advice regarding the figures.  We are, as always, grateful to the \swift Team for making the ToO observations possible, in particular the duty scientists, as well as the science planners. 

E.A.B., A.F.V., and V.P.G.\ acknowledge the grant RFBR No.\ 16-02-00758 for support. J.F., J.J., and G.S.\ acknowledge support from the Spanish MINECO grant AYA2014-59084-P, the E.U.\ FEDER funds, and from the AGAUR/Generalitat de Catalunya grant SGR0038/2014.  S.F.\ acknowledges support from the Russian Scientific Foundation (grant N\,14-50-00043) and the Russian Government Program of Competitive Growth of Kazan Federal University.  M.\ Henze acknowledges the support of the Spanish Ministry of Economy and Competitiveness (MINECO) under the grant FDPI-2013-16933. M.\ Hernanz acknowledges MINECO support under the grant ESP2014-56003-R. K.H.\ was supported by the project RVO:67985815.  J.P.O.\ and K.L.P.\ acknowledge funding from the UK Space Agency.  V.A.R.M.R.\ acknowledges financial support from the Radboud Excellence Initiative.  S.C.W.\ acknowledges a Visiting Research Fellowship at LJMU.  This work has been supported in part by NSF grant AST-1009566 and NASA grant HST-Go-14125.012. 

Finally, the authors would like to express their gratitude to the anonymous referee for their timely and constructive comments.

\facilities{Liverpool:2m, FTN, OO:0.65, MLO:1m, BAT, OAO:0.5m, \swift, Mayall}

\software{IRAF \citep[v2.16.1,][]{1993ASPC...52..173T}, Starlink \citep[v2015B,][]{1982QJRAS..23..485D}, APHOT \citep{1994ExA.....5..375P}, HEASOFT (v6.16), XIMAGE (v4.5.1), XSPEC \citep[v12.8.2;][]{1996ASPC..101...17A}, XSELECT (v2.4c), R \citep{R_manual}.}

\bibliographystyle{aasjournal} 
\bibliography{refs} 

\begin{thebibliography}{}
\expandafter\ifx\csname natexlab\endcsname\relax\def\natexlab#1{#1}\fi

\bibitem[{{Afanasiev} \& {Moiseev}(2005)}]{2005AstL...31..194A}
{Afanasiev}, V.~L., \& {Moiseev}, A.~V. 2005, Astronomy Letters, 31, 194

\bibitem[{{Allen}(1980)}]{1980MNRAS.190...75A}
{Allen}, D.~A. 1980, \mnras, 190, 75

\bibitem[{{Arai} {et~al.}(2015){Arai}, {Naito}, {Watanabe}, {Nakajima}, {Sano},
  {Watanabe}, {Kiyota}, {Maehara}, {Kuroda}, {Kawai}, {Henze}, {Darnley},
  {Shafter}, \& {Kato}}]{2015ATel.7979....1A}
{Arai}, A., {Naito}, H., {Watanabe}, F., {et~al.} 2015, The Astronomer's
  Telegram, 7979, 1

\bibitem[{{Arnaud}(1996)}]{1996ASPC..101...17A}
{Arnaud}, K.~A. 1996, in Astronomical Society of the Pacific Conference Series,
  Vol. 101, Astronomical Data Analysis Software and Systems V, ed.
  {G.~H.~Jacoby \& J.~Barnes}, 17--+

\bibitem[{{Balucinska-Church} \& {McCammon}(1992)}]{1992ApJ...400..699B}
{Balucinska-Church}, M., \& {McCammon}, D. 1992, \apj, 400, 699

\bibitem[{{Banerjee} {et~al.}(2014){Banerjee}, {Joshi}, {Venkataraman},
  {Ashok}, {Marion}, {Hsiao}, \& {Raj}}]{2014ApJ...785L..11B}
{Banerjee}, D.~P.~K., {Joshi}, V., {Venkataraman}, V., {et~al.} 2014, \apjl,
  785, L11

\bibitem[{{Barnsley} {et~al.}(2016){Barnsley}, {Jermak}, {Steele}, {Smith},
  {Bates}, \& {Mottram}}]{2016JATIS...2a5002B}
{Barnsley}, R.~M., {Jermak}, H.~E., {Steele}, I.~A., {et~al.} 2016, Journal of
  Astronomical Telescopes, Instruments, and Systems, 2, 015002

\bibitem[{{Barsukova} {et~al.}(2011){Barsukova}, {Fabrika}, {Hornoch},
  {Fatkhullin}, {Sholukhova}, \& {Pietsch}}]{2011ATel.3725....1B}
{Barsukova}, E., {Fabrika}, S., {Hornoch}, K., {et~al.} 2011, The Astronomer's
  Telegram, 3725, 1

\bibitem[{{Bateson} \& {Hull}(1979)}]{1979IAUC.3341....1B}
{Bateson}, F.~M., \& {Hull}, O. 1979, \iaucirc, 3341

\bibitem[{{Bessell}(1979)}]{1979PASP...91..589B}
{Bessell}, M.~S. 1979, \pasp, 91, 589

\bibitem[{{Bessell}(1990)}]{1990PASP..102.1181B}
---. 1990, \pasp, 102, 1181

\bibitem[{{Bode}(2010)}]{2010AN....331..160B}
{Bode}, M.~F. 2010, Astronomische Nachrichten, 331, 160

\bibitem[{{Bode} {et~al.}(2009){Bode}, {Darnley}, {Shafter}, {Page},
  {Smirnova}, {Anupama}, \& {Hilton}}]{2009ApJ...705.1056B}
{Bode}, M.~F., {Darnley}, M.~J., {Shafter}, A.~W., {et~al.} 2009, \apj, 705,
  1056

\bibitem[{{Bode} \& {Evans}(2008)}]{2008clno.book.....B}
{Bode}, M.~F., \& {Evans}, A., eds. 2008, Cambridge Astrophysics Series,
  Vol.~43, {Classical Novae, 2nd Edition} (Cambridge: Cambridge University
  Press)

\bibitem[{{Bode} \& {Kahn}(1985)}]{1985MNRAS.217..205B}
{Bode}, M.~F., \& {Kahn}, F.~D. 1985, \mnras, 217, 205

\bibitem[{{Bode} {et~al.}(2006){Bode}, {O'Brien}, {Osborne}, {Page},
  {Senziani}, {Skinner}, {Starrfield}, {Ness}, {Drake}, {Schwarz}, {Beardmore},
  {Darnley}, {Eyres}, {Evans}, {Gehrels}, {Goad}, {Jean}, {Krautter}, \&
  {Novara}}]{2006ApJ...652..629B}
{Bode}, M.~F., {O'Brien}, T.~J., {Osborne}, J.~P., {et~al.} 2006, \apj, 652,
  629

\bibitem[{{Bode} {et~al.}(2016){Bode}, {Darnley}, {Beardmore}, {Osborne},
  {Page}, {Walter}, {Krautter}, {Melandri}, {Ness}, {O'Brien}, {Orio},
  {Schwarz}, {Shara}, \& {Starrfield}}]{2016ApJ...818..145B}
{Bode}, M.~F., {Darnley}, M.~J., {Beardmore}, A.~P., {et~al.} 2016, \apj, 818,
  145

\bibitem[{{Breeveld}(2010)}]{2010Bre}
{Breeveld}, A. 2010, {SWIFT-UVOT-CALDB-16-R01},  {Swift Calibration Database},
  \url{http://heasarc.gsfc.nasa.gov/docs/heasarc/caldb/swift/docs/uvot/uvot_caldb_AB_10wa.pdf}

\bibitem[{{Brown} {et~al.}(2013){Brown}, {Baliber}, {Bianco}, {Bowman},
  {Burleson}, {Conway}, {Crellin}, {Depagne}, {De Vera}, {Dilday}, {Dragomir},
  {Dubberley}, {Eastman}, {Elphick}, {Falarski}, {Foale}, {Ford}, {Fulton},
  {Garza}, {Gomez}, {Graham}, {Greene}, {Haldeman}, {Hawkins}, {Haworth},
  {Haynes}, {Hidas}, {Hjelstrom}, {Howell}, {Hygelund}, {Lister}, {Lobdill},
  {Martinez}, {Mullins}, {Norbury}, {Parrent}, {Paulson}, {Petry}, {Pickles},
  {Posner}, {Rosing}, {Ross}, {Sand}, {Saunders}, {Shobbrook}, {Shporer},
  {Street}, {Thomas}, {Tsapras}, {Tufts}, {Valenti}, {Vander Horst}, {Walker},
  {White}, \& {Willis}}]{2013PASP..125.1031B}
{Brown}, T.~M., {Baliber}, N., {Bianco}, F.~B., {et~al.} 2013, \pasp, 125, 1031

\bibitem[{{Burrows} {et~al.}(2005){Burrows}, {Hill}, {Nousek}, {Kennea},
  {Wells}, {Osborne}, {Abbey}, {Beardmore}, {Mukerjee}, {Short}, {Chincarini},
  {Campana}, {Citterio}, {Moretti}, {Pagani}, {Tagliaferri}, {Giommi},
  {Capalbi}, {Tamburelli}, {Angelini}, {Cusumano}, {Br{\"a}uninger}, {Burkert},
  \& {Hartner}}]{2005SSRv..120..165B}
{Burrows}, D.~N., {Hill}, J.~E., {Nousek}, J.~A., {et~al.} 2005, \ssr, 120, 165

\bibitem[{{Cameron}(1959)}]{1959ApJ...130..916C}
{Cameron}, A.~G.~W. 1959, \apj, 130, 916

\bibitem[{{Cardelli} {et~al.}(1989){Cardelli}, {Clayton}, \&
  {Mathis}}]{1989ApJ...345..245C}
{Cardelli}, J.~A., {Clayton}, G.~C., \& {Mathis}, J.~S. 1989, \apj, 345, 245

\bibitem[{{Cash}(1979)}]{1979ApJ...228..939C}
{Cash}, W. 1979, \apj, 228, 939

\bibitem[{{Chandrasekhar}(1931)}]{1931ApJ....74...81C}
{Chandrasekhar}, S. 1931, \apj, 74, 81

\bibitem[{{Chen} {et~al.}(2016){Chen}, {Woods}, {Yungelson}, {Gilfanov}, \&
  {Han}}]{2016MNRAS.458.2916C}
{Chen}, H.-L., {Woods}, T.~E., {Yungelson}, L.~R., {Gilfanov}, M., \& {Han}, Z.
  2016, \mnras, 458, 2916

\bibitem[{{Chen} {et~al.}(2015){Chen}, {Lau}, \& {Xu}}]{2015ATel.7974....1C}
{Chen}, T., {Lau}, K.~M., \& {Xu}, Z. 2015, The Astronomer's Telegram, 7974, 1

\bibitem[{{Cheung} {et~al.}(2014){Cheung}, {Jean}, \&
  {Shore}}]{2014ATel.5879....1C}
{Cheung}, C.~C., {Jean}, P., \& {Shore}, S.~N. 2014, The Astronomer's Telegram,
  5879

\bibitem[{{Chomiuk} {et~al.}(2014){Chomiuk}, {Linford}, {Yang}, {O'Brien},
  {Paragi}, {Mioduszewski}, {Beswick}, {Cheung}, {Mukai}, {Nelson}, {Ribeiro},
  {Rupen}, {Sokoloski}, {Weston}, {Zheng}, {Bode}, {Eyres}, {Roy}, \&
  {Taylor}}]{2014Natur.514..339C}
{Chomiuk}, L., {Linford}, J.~D., {Yang}, J., {et~al.} 2014, \nat, 514, 339

\bibitem[{{Ciardullo} {et~al.}(1987){Ciardullo}, {Ford}, {Neill}, {Jacoby}, \&
  {Shafter}}]{1987ApJ...318..520C}
{Ciardullo}, R., {Ford}, H.~C., {Neill}, J.~D., {Jacoby}, G.~H., \& {Shafter},
  A.~W. 1987, \apj, 318, 520

\bibitem[{{Ciardullo} {et~al.}(1990{\natexlab{a}}){Ciardullo}, {Shafter},
  {Ford}, {Neill}, {Shara}, \& {Tomaney}}]{1990ApJ...356..472C}
{Ciardullo}, R., {Shafter}, A.~W., {Ford}, H.~C., {et~al.} 1990{\natexlab{a}},
  \apj, 356, 472

\bibitem[{{Ciardullo} {et~al.}(1990{\natexlab{b}}){Ciardullo}, {Tamblyn},
  {Jacoby}, {Ford}, \& {Williams}}]{1990AJ.....99.1079C}
{Ciardullo}, R., {Tamblyn}, P., {Jacoby}, G.~H., {Ford}, H.~C., \& {Williams},
  R.~E. 1990{\natexlab{b}}, \aj, 99, 1079

\bibitem[{{Cohen}(1988)}]{1988igbo.conf..448C}
{Cohen}, J.~G. 1988, in Instrumentation for Ground-Based Optical Astronomy, ed.
  L.~B. {Robinson}, 448

\bibitem[{{Connelley} \& {Sandage}(1958)}]{1958PASP...70..600C}
{Connelley}, M., \& {Sandage}, A. 1958, \pasp, 70, 600

\bibitem[{{Curtin} {et~al.}(2015){Curtin}, {Shafter}, {Pritchet}, {Neill},
  {Kundu}, \& {Maccarone}}]{2015ApJ...811...34C}
{Curtin}, C., {Shafter}, A.~W., {Pritchet}, C.~J., {et~al.} 2015, \apj, 811, 34

\bibitem[{{Darnley} {et~al.}(2015{\natexlab{a}}){Darnley}, {Henze}, {Shafter},
  \& {Kato}}]{2015ATel.8029....1D}
{Darnley}, M.~J., {Henze}, M., {Shafter}, A.~W., \& {Kato}, M.
  2015{\natexlab{a}}, The Astronomer's Telegram, 8029, 1

\bibitem[{{Darnley} {et~al.}(2015{\natexlab{b}}){Darnley}, {Henze}, {Shafter},
  \& {Kato}}]{2015ATel.7964....1D}
---. 2015{\natexlab{b}}, The Astronomer's Telegram, 7964, 1

\bibitem[{{Darnley} {et~al.}(2015{\natexlab{c}}){Darnley}, {Henze}, {Shafter},
  \& {Kato}}]{2015ATel.7980....1D}
---. 2015{\natexlab{c}}, The Astronomer's Telegram, 7980, 1

\bibitem[{{Darnley} {et~al.}(2015{\natexlab{d}}){Darnley}, {Henze}, {Shafter},
  \& {Kato}}]{2015ATel.7965....1D}
---. 2015{\natexlab{d}}, The Astronomer's Telegram, 7965, 1

\bibitem[{{Darnley} {et~al.}(2008){Darnley}, {Hounsell}, \&
  {Bode}}]{2008ASPC..401..203D}
{Darnley}, M.~J., {Hounsell}, R.~A., \& {Bode}, M.~F. 2008, in Astronomical
  Society of the Pacific Conference Series, Vol. 401, RS Ophiuchi (2006) and
  the Recurrent Nova Phenomenon, ed. A.~{Evans}, M.~F. {Bode}, T.~J. {O'Brien},
  \& M.~J. {Darnley}, 203

\bibitem[{{Darnley} {et~al.}(2016{\natexlab{a}}){Darnley}, {Kuin}, {Page},
  {Osborne}, {Schwarz}, {Shore}, {Starrfield}, \&
  {Williams}}]{2016ATel.8587....1D}
{Darnley}, M.~J., {Kuin}, N.~P.~M., {Page}, K.~L., {et~al.} 2016{\natexlab{a}},
  The Astronomer's Telegram, 8587

\bibitem[{{Darnley} {et~al.}(2012){Darnley}, {Ribeiro}, {Bode}, {Hounsell}, \&
  {Williams}}]{2012ApJ...746...61D}
{Darnley}, M.~J., {Ribeiro}, V.~A.~R.~M., {Bode}, M.~F., {Hounsell}, R.~A., \&
  {Williams}, R.~P. 2012, \apj, 746, 61

\bibitem[{{Darnley} {et~al.}(2011){Darnley}, {Ribeiro}, {Bode}, \&
  {Munari}}]{2011A&A...530A..70D}
{Darnley}, M.~J., {Ribeiro}, V.~A.~R.~M., {Bode}, M.~F., \& {Munari}, U. 2011,
  \aap, 530, A70

\bibitem[{{Darnley} {et~al.}(2014{\natexlab{a}}){Darnley}, {Williams}, {Bode},
  {Henze}, {Ness}, {Shafter}, {Hornoch}, \& {Votruba}}]{2014A&A...563L...9D}
{Darnley}, M.~J., {Williams}, S.~C., {Bode}, M.~F., {et~al.}
  2014{\natexlab{a}}, \aap, 563, L9

\bibitem[{{Darnley} {et~al.}(2014{\natexlab{b}}){Darnley}, {Williams}, {Bode},
  {Shafter}, {Henze}, {Ness}, \& {Hounsell}}]{2014ATel.6527....1D}
---. 2014{\natexlab{b}}, The Astronomer's Telegram, 6527, 1

\bibitem[{{Darnley} {et~al.}(2004){Darnley}, {Bode}, {Kerins}, {Newsam}, {An},
  {Baillon}, {Novati}, {Carr}, {Cr{\'e}z{\'e}}, {Evans}, {Giraud-H{\'e}raud},
  {Gould}, {Hewett}, {Jetzer}, {Kaplan}, {Paulin-Henriksson}, {Smartt},
  {Stalin}, \& {Tsapras}}]{2004MNRAS.353..571D}
{Darnley}, M.~J., {Bode}, M.~F., {Kerins}, E., {et~al.} 2004, \mnras, 353, 571

\bibitem[{{Darnley} {et~al.}(2006){Darnley}, {Bode}, {Kerins}, {Newsam}, {An},
  {Baillon}, {Belokurov}, {Calchi Novati}, {Carr}, {Cr{\'e}z{\'e}}, {Evans},
  {Giraud-H{\'e}raud}, {Gould}, {Hewett}, {Jetzer}, {Kaplan},
  {Paulin-Henriksson}, {Smartt}, {Tsapras}, \& {Weston}}]{2006MNRAS.369..257D}
---. 2006, \mnras, 369, 257

\bibitem[{{Darnley} {et~al.}(2007){Darnley}, {Kerins}, {Newsam}, {Duke},
  {Gould}, {Han}, {Ibrahimov}, {Im}, {Jeon}, {Karimov}, {Lee}, \&
  {Park}}]{2007ApJ...661L..45D}
{Darnley}, M.~J., {Kerins}, E., {Newsam}, A., {et~al.} 2007, \apjl, 661, L45

\bibitem[{{Darnley} {et~al.}(2014{\natexlab{c}}){Darnley}, {Bode}, {Harman},
  {Hounsell}, {Munari}, {Ribeiro}, {Surina}, {Williams}, \&
  {Williams}}]{2014ASPC..490...49D}
{Darnley}, M.~J., {Bode}, M.~F., {Harman}, D.~J., {et~al.} 2014{\natexlab{c}},
  in Astronomical Society of the Pacific Conference Series, Vol. 490, Stellar
  Novae: Past and Future Decades, ed. P.~A. {Woudt} \& V.~A.~R.~M. {Ribeiro},
  49

\bibitem[{{Darnley} {et~al.}(2015{\natexlab{e}}){Darnley}, {Henze}, {Steele},
  {Bode}, {Ribeiro}, {Rodr{\'{\i}}guez-Gil}, {Shafter}, {Williams}, {Baer},
  {Hachisu}, {Hernanz}, {Hornoch}, {Hounsell}, {Kato}, {Kiyota}, {Ku{\v
  c}{\'a}kov{\'a}}, {Maehara}, {Ness}, {Piascik}, {Sala}, {Skillen}, {Smith},
  \& {Wolf}}]{2015A&A...580A..45D}
{Darnley}, M.~J., {Henze}, M., {Steele}, I.~A., {et~al.} 2015{\natexlab{e}},
  \aap, 580, A45

\bibitem[{{Darnley} {et~al.}(2016{\natexlab{b}}){Darnley}, {Henze}, {Steele},
  {Bode}, {Ribeiro}, {Rodr{\'{\i}}guez-Gil}, {Shafter}, {Williams}, {Baer},
  {Hachisu}, {Hernanz}, {Hornoch}, {Hounsell}, {Kato}, {Kiyota}, {Ku{\v
  c}{\'a}kov{\'a}}, {Maehara}, {Ness}, {Piascik}, {Sala}, {Skillen}, {Smith},
  \& {Wolf}}]{erratum}
---. 2016{\natexlab{b}}, \aap, doi:10.1051/0004-6361/201526027e

\bibitem[{{Darnley} {et~al.}(2016{\natexlab{c}}){Darnley}, {Hounsell}, {Bode},
  {Harman}, {Henze}, {Hornoch}, {Ness}, {Ribeiro}, {Shafter}, {Shara}, \&
  {Williams}}]{HST2016}
{Darnley}, M.~J., {Hounsell}, R., {Bode}, M.~F., {et~al.} 2016{\natexlab{c}},
  in preparation, for submission to ApJL

\bibitem[{{Della Valle} {et~al.}(2002){Della Valle}, {Pasquini}, {Daou}, \&
  {Williams}}]{2002A&A...390..155D}
{Della Valle}, M., {Pasquini}, L., {Daou}, D., \& {Williams}, R.~E. 2002, \aap,
  390, 155

\bibitem[{{Dilday} {et~al.}(2012){Dilday}, {Howell}, {Cenko}, {Silverman},
  {Nugent}, {Sullivan}, {Ben-Ami}, {Bildsten}, {Bolte}, {Endl}, {Filippenko},
  {Gnat}, {Horesh}, {Hsiao}, {Kasliwal}, {Kirkman}, {Maguire}, {Marcy},
  {Moore}, {Pan}, {Parrent}, {Podsiadlowski}, {Quimby}, {Sternberg}, {Suzuki},
  {Tytler}, {Xu}, {Bloom}, {Gal-Yam}, {Hook}, {Kulkarni}, {Law}, {Ofek},
  {Polishook}, \& {Poznanski}}]{2012Sci...337..942D}
{Dilday}, B., {Howell}, D.~A., {Cenko}, S.~B., {et~al.} 2012, Science, 337, 942

\bibitem[{{Disney} \& {Wallace}(1982)}]{1982QJRAS..23..485D}
{Disney}, M.~J., \& {Wallace}, P.~T. 1982, \qjras, 23, 485

\bibitem[{{Drake} {et~al.}(2016){Drake}, {Delgado}, {Laming}, {Starrfield},
  {Kashyap}, {Orlando}, {Page}, {Hernanz}, {Ness}, {Gehrz}, {van Rossum}, \&
  {Woodward}}]{2016ApJ...825...95D}
{Drake}, J.~J., {Delgado}, L., {Laming}, J.~M., {et~al.} 2016, \apj, 825, 95

\bibitem[{{Dufay} {et~al.}(1964){Dufay}, {Bloch}, {Bertaud}, \&
  {Dufay}}]{1964AnAp...27..555D}
{Dufay}, J., {Bloch}, M., {Bertaud}, C., \& {Dufay}, M. 1964, Annales
  d'Astrophysique, 27, 555

\bibitem[{{Evans} {et~al.}(2008){Evans}, {Bode}, {O'Brien}, \&
  {Darnley}}]{2008ASPC..401.....E}
{Evans}, A., {Bode}, M.~F., {O'Brien}, T.~J., \& {Darnley}, M.~J., eds. 2008,
  Astronomical Society of the Pacific Conference Series, Vol. 401, {RS Ophiuchi
  (2006) and the Recurrent Nova Phenomenon} (San Francisco: Astronomical
  Society of the Pacific)

\bibitem[{{Evans} {et~al.}(2007){Evans}, {Kerr}, {Yang}, {Matsuoka}, {Tsuzuki},
  {Bode}, {Eyres}, {Geballe}, {Woodward}, {Gehrz}, {Lynch}, {Rudy}, {Russell},
  {O'Brien}, {Starrfield}, {Davis}, {Ness}, {Drake}, {Osborne}, {Page},
  {Adamson}, {Schwarz}, \& {Krautter}}]{2007MNRAS.374L...1E}
{Evans}, A., {Kerr}, T., {Yang}, B., {et~al.} 2007, \mnras, 374, L1

\bibitem[{{Fabrika} {et~al.}(2015){Fabrika}, {Barsukova}, {Valeev},
  {Vinokurov}, {Sholukhova}, {Goranskij}, {Hornoch}, {Henze}, {Hernanz},
  {Sala}, {Jose}, {Figueira}, \& {Shafter}}]{2015ATel.8033....1F}
{Fabrika}, S., {Barsukova}, E.~A., {Valeev}, A.~F., {et~al.} 2015, The
  Astronomer's Telegram, 8033

\bibitem[{{Freedman} \& {Madore}(1990)}]{1990ApJ...365..186F}
{Freedman}, W.~L., \& {Madore}, B.~F. 1990, \apj, 365, 186

\bibitem[{{Gallagher} \& {Ney}(1976)}]{1976ApJ...204L..35G}
{Gallagher}, J.~S., \& {Ney}, E.~P. 1976, \apjl, 204, L35

\bibitem[{{Gurevitch} \& {Lebedinsky}(1957)}]{1957IAUS....3...77G}
{Gurevitch}, L.~E., \& {Lebedinsky}, A.~I. 1957, in IAU Symposium, Vol.~3,
  Non-stable stars, ed. G.~H. {Herbig} (Cambridge Univ. Press, Cambridge, UK),
  77

\bibitem[{{Gutierrez} {et~al.}(1996){Gutierrez}, {Garcia-Berro}, {Iben},
  {Isern}, {Labay}, \& {Canal}}]{1996ApJ...459..701G}
{Gutierrez}, J., {Garcia-Berro}, E., {Iben}, Jr., I., {et~al.} 1996, \apj, 459,
  701

\bibitem[{{G{\"u}ver} \& {{\"O}zel}(2009)}]{2009MNRAS.400.2050G}
{G{\"u}ver}, T., \& {{\"O}zel}, F. 2009, \mnras, 400, 2050

\bibitem[{{Hachisu} \& {Kato}(2006)}]{2006ApJS..167...59H}
{Hachisu}, I., \& {Kato}, M. 2006, \apjs, 167, 59

\bibitem[{{Hachisu} \& {Kato}(2007)}]{2007ApJ...662..552H}
---. 2007, \apj, 662, 552

\bibitem[{{Hachisu} \& {Kato}(2010)}]{2010ApJ...709..680H}
---. 2010, \apj, 709, 680

\bibitem[{{Hachisu} \& {Kato}(2014)}]{2014ApJ...785...97H}
---. 2014, \apj, 785, 97

\bibitem[{{Hachisu} \& {Kato}(2016)}]{2016ApJS..223...21H}
---. 2016, \apjs, 223, 21

\bibitem[{{Hachisu} {et~al.}(1999{\natexlab{a}}){Hachisu}, {Kato}, \&
  {Nomoto}}]{1999ApJ...522..487H}
{Hachisu}, I., {Kato}, M., \& {Nomoto}, K. 1999{\natexlab{a}}, \apj, 522, 487

\bibitem[{{Hachisu} {et~al.}(1999{\natexlab{b}}){Hachisu}, {Kato}, {Nomoto}, \&
  {Umeda}}]{1999ApJ...519..314H}
{Hachisu}, I., {Kato}, M., {Nomoto}, K., \& {Umeda}, H. 1999{\natexlab{b}},
  \apj, 519, 314

\bibitem[{{Hachisu} {et~al.}(2016){Hachisu}, {Saio}, \&
  {Kato}}]{2016ApJ...824...22H}
{Hachisu}, I., {Saio}, H., \& {Kato}, M. 2016, \apj, 824, 22

\bibitem[{{Hachisu} {et~al.}(2008){Hachisu}, {Kato}, {Kiyota}, {Kubotera},
  {Maehara}, {Nakajima}, {Ishii}, {Kamada}, {Mizoguchi}, {Nishiyama},
  {Sumitomo}, {Tanaka}, {Yamanaka}, \& {Sadakane}}]{2008ASPC..401..206H}
{Hachisu}, I., {Kato}, M., {Kiyota}, S., {et~al.} 2008, in Astronomical Society
  of the Pacific Conference Series, Vol. 401, RS Ophiuchi (2006) and the
  Recurrent Nova Phenomenon, ed. A.~{Evans}, M.~F. {Bode}, T.~J. {O'Brien}, \&
  M.~J. {Darnley}, 206

\bibitem[{{Henze} {et~al.}(2015{\natexlab{a}}){Henze}, {Darnley}, {Kabashima},
  {Nishiyama}, {Itagaki}, \& {Gao}}]{2015A&A...582L...8H}
{Henze}, M., {Darnley}, M.~J., {Kabashima}, F., {et~al.} 2015{\natexlab{a}},
  \aap, 582, L8

\bibitem[{{Henze} {et~al.}(2008){Henze}, {Meusinger}, \&
  {Pietsch}}]{2008A&A...477...67H}
{Henze}, M., {Meusinger}, H., \& {Pietsch}, W. 2008, \aap, 477, 67

\bibitem[{{Henze} {et~al.}(2014{\natexlab{a}}){Henze}, {Ness}, {Darnley},
  {Bode}, {Williams}, {Shafter}, {Kato}, \& {Hachisu}}]{2014A&A...563L...8H}
{Henze}, M., {Ness}, J.-U., {Darnley}, M.~J., {et~al.} 2014{\natexlab{a}},
  \aap, 563, L8

\bibitem[{{Henze} {et~al.}(2010){Henze}, {Pietsch}, {Haberl}, {Hernanz},
  {Sala}, {Della Valle}, {Hatzidimitriou}, {Rau}, {Hartmann}, {Greiner},
  {Burwitz}, \& {Fliri}}]{2010A&A...523A..89H}
{Henze}, M., {Pietsch}, W., {Haberl}, F., {et~al.} 2010, \aap, 523, A89

\bibitem[{{Henze} {et~al.}(2011){Henze}, {Pietsch}, {Haberl}, {Hernanz},
  {Sala}, {Hatzidimitriou}, {Della Valle}, {Rau}, {Hartmann}, \&
  {Burwitz}}]{2011A&A...533A..52H}
---. 2011, \aap, 533, A52

\bibitem[{{Henze} {et~al.}(2014{\natexlab{b}}){Henze}, {Pietsch}, {Haberl},
  {Della Valle}, {Sala}, {Hatzidimitriou}, {Hofmann}, {Hernanz}, {Hartmann}, \&
  {Greiner}}]{2014A&A...563A...2H}
---. 2014{\natexlab{b}}, \aap, 563, A2

\bibitem[{{Henze} {et~al.}(2015{\natexlab{b}}){Henze}, {Ness}, {Darnley},
  {Bode}, {Williams}, {Shafter}, {Sala}, {Kato}, {Hachisu}, \&
  {Hernanz}}]{2015A&A...580A..46H}
{Henze}, M., {Ness}, J.-U., {Darnley}, M.~J., {et~al.} 2015{\natexlab{b}},
  \aap, 580, A46

\bibitem[{{Henze} {et~al.}(2015{\natexlab{c}}){Henze}, {Darnley}, {Shafter},
  {Kato}, {Hachisu}, {Bode}, {Ness}, {Osborne}, {Kennea}, {Gehrels}, \& {et
  al.}}]{2015ATel.8062....1H}
{Henze}, M., {Darnley}, M.~J., {Shafter}, A.~W., {et~al.} 2015{\natexlab{c}},
  The Astronomer's Telegram, 8062, 1

\bibitem[{{Henze} {et~al.}(2015{\natexlab{d}}){Henze}, {Darnley}, {Shafter},
  {Kato}, {Hachisu}, {Bode}, {Ness}, {Osborne}, {Kennea}, \&
  {Gehrels}}]{2015ATel.7984....1H}
---. 2015{\natexlab{d}}, The Astronomer's Telegram, 7984, 1

\bibitem[{{Hernanz} \& {Jos{\'e}}(2008)}]{2008NewAR..52..386H}
{Hernanz}, M., \& {Jos{\'e}}, J. 2008, \nar, 52, 386

\bibitem[{{Hillebrandt} \& {Niemeyer}(2000)}]{2000ARA&A..38..191H}
{Hillebrandt}, W., \& {Niemeyer}, J.~C. 2000, \araa, 38, 191

\bibitem[{{Hillman} {et~al.}(2015){Hillman}, {Prialnik}, {Kovetz}, \&
  {Shara}}]{2015MNRAS.446.1924H}
{Hillman}, Y., {Prialnik}, D., {Kovetz}, A., \& {Shara}, M.~M. 2015, \mnras,
  446, 1924

\bibitem[{{Hillman} {et~al.}(2016){Hillman}, {Prialnik}, {Kovetz}, \&
  {Shara}}]{2016ApJ...819..168H}
---. 2016, \apj, 819, 168

\bibitem[{{Hornoch} \& {Fabrika}(2015)}]{2015ATel.8038....1H}
{Hornoch}, K., \& {Fabrika}, S. 2015, The Astronomer's Telegram, 8038

\bibitem[{{Hornoch} {et~al.}(2015{\natexlab{a}}){Hornoch}, {Henze}, {Darnley},
  {Shafter}, \& {Kato}}]{2015ATel.7969....1H}
{Hornoch}, K., {Henze}, M., {Darnley}, M.~J., {Shafter}, A.~W., \& {Kato}, M.
  2015{\natexlab{a}}, The Astronomer's Telegram, 7969, 1

\bibitem[{{Hornoch} {et~al.}(2015{\natexlab{b}}){Hornoch}, {Kucakova},
  {Quimby}, {Shafter}, {Henze}, {Darnley}, \& {Kato}}]{2015ATel.7976....1H}
{Hornoch}, K., {Kucakova}, H., {Quimby}, R., {et~al.} 2015{\natexlab{b}}, The
  Astronomer's Telegram, 7976, 1

\bibitem[{{Hounsell} {et~al.}(2010){Hounsell}, {Bode}, {Hick}, {Buffington},
  {Jackson}, {Clover}, {Shafter}, {Darnley}, {Mawson}, {Steele}, {Evans},
  {Eyres}, \& {O'Brien}}]{2010ApJ...724..480H}
{Hounsell}, R., {Bode}, M.~F., {Hick}, P.~P., {et~al.} 2010, \apj, 724, 480

\bibitem[{{Hounsell} {et~al.}(2016){Hounsell}, {Darnley}, {Bode}, {Harman},
  {Surina}, {Starrfield}, {Holdsworth}, {Bewsher}, {Hick}, {Jackson},
  {Buffington}, {Clover}, \& {Shafter}}]{2016ApJ...820..104H}
{Hounsell}, R., {Darnley}, M.~J., {Bode}, M.~F., {et~al.} 2016, \apj, 820, 104

\bibitem[{{Hubble}(1929)}]{1929ApJ....69..103H}
{Hubble}, E.~P. 1929, \apj, 69, 103

\bibitem[{{Hurkett} {et~al.}(2008){Hurkett}, {Vaughan}, {Osborne}, {O'Brien},
  {Page}, {Beardmore}, {Godet}, {Burrows}, {Capalbi}, {Evans}, {Gehrels},
  {Goad}, {Hill}, {Kennea}, {Mineo}, {Perri}, \&
  {Starling}}]{2008ApJ...679..587H}
{Hurkett}, C.~P., {Vaughan}, S., {Osborne}, J.~P., {et~al.} 2008, \apj, 679,
  587

\bibitem[{{Iijima}(2009)}]{2009A&A...505..287I}
{Iijima}, T. 2009, \aap, 505, 287

\bibitem[{{Jester} {et~al.}(2005){Jester}, {Schneider}, {Richards}, {Green},
  {Schmidt}, {Hall}, {Strauss}, {Vanden Berk}, {Stoughton}, {Gunn},
  {Brinkmann}, {Kent}, {Smith}, {Tucker}, \& {Yanny}}]{2005AJ....130..873J}
{Jester}, S., {Schneider}, D.~P., {Richards}, G.~T., {et~al.} 2005, \aj, 130,
  873

\bibitem[{{Jos\'e}(2016)}]{Jos16}
{Jos\'e}, J. 2016, {Stellar Explosions: Hydrodynamics and Nucleosynthesis}
  (CRC/Taylor and Francis, Boca Raton, FL, USA), doi:10.1201/b19165

\bibitem[{{Jos\'e} \& {Shore}(2008)}]{JS08}
{Jos\'e}, J., \& {Shore}, S.~N. 2008, in {Classical Novae, 2nd Edition}, ed.
  M.~F. {Bode} \& A.~{Evans} (Cambridge: Cambridge University Press), 121--140

\bibitem[{{Joy}(1954)}]{1954ApJ...120..377J}
{Joy}, A.~H. 1954, \apj, 120, 377

\bibitem[{{Joy} \& {Swings}(1945)}]{1945ApJ...102..353J}
{Joy}, A.~H., \& {Swings}, P. 1945, \apj, 102, 353

\bibitem[{{Kato} {et~al.}(2015){Kato}, {Saio}, \&
  {Hachisu}}]{2015ApJ...808...52K}
{Kato}, M., {Saio}, H., \& {Hachisu}, I. 2015, \apj, 808, 52

\bibitem[{{Kato} {et~al.}(2014){Kato}, {Saio}, {Hachisu}, \&
  {Nomoto}}]{2014ApJ...793..136K}
{Kato}, M., {Saio}, H., {Hachisu}, I., \& {Nomoto}, K. 2014, \apj, 793, 136

\bibitem[{{Kato} {et~al.}(2016){Kato}, {Saio}, {Henze}, {Ness}, {Osborne},
  {Page}, {Darnley}, {Bode}, {Shafter}, {Hernanz}, {Gehrels}, {Kennea}, \&
  {Hachisu}}]{XrayFlash}
{Kato}, M., {Saio}, H., {Henze}, M., {et~al.} 2016, \apj, in press

\bibitem[{{Kerins} {et~al.}(2010){Kerins}, {Darnley}, {Duke}, {Gould}, {Han},
  {Newsam}, {Park}, \& {Street}}]{2010MNRAS.409..247K}
{Kerins}, E., {Darnley}, M.~J., {Duke}, J.~P., {et~al.} 2010, \mnras, 409, 247

\bibitem[{{Korotkiy} \& {Elenin}(2011)}]{2011Kor}
{Korotkiy}, S., \& {Elenin}, L. 2011, {CBAT},  {IAU},
  \url{http://www.cbat.eps.harvard.edu/unconf/followups/J00452885+4154094.html}

\bibitem[{{Kotani} {et~al.}(2005){Kotani}, {Kawai}, {Yanagisawa}, {Watanabe},
  {Arimoto}, {Fukushima}, {Hattori}, {Inata}, {Izumiura}, {Kataoka}, {Koyano},
  {Kubota}, {Kuroda}, {Mori}, {Nagayama}, {Ohta}, {Okada}, {Okita}, {Sato},
  {Serino}, {Shimizu}, {Shimokawabe}, {Suzuki}, {Toda}, {Ushiyama}, {Yatsu},
  {Yoshida}, \& {Yoshida}}]{2005NCimC..28..755K}
{Kotani}, T., {Kawai}, N., {Yanagisawa}, K., {et~al.} 2005, Nuovo Cimento C
  Geophysics Space Physics C, 28, 755

\bibitem[{{Kovetz} \& {Prialnik}(1994)}]{1994ApJ...424..319K}
{Kovetz}, A., \& {Prialnik}, D. 1994, \apj, 424, 319

\bibitem[{{Kraft}(1964)}]{1964ApJ...139..457K}
{Kraft}, R.~P. 1964, \apj, 139, 457

\bibitem[{Kramida {et~al.}(2015)Kramida, {Yu.~Ralchenko}, Reader, \& {and NIST
  ASD Team}}]{NIST_ASD}
Kramida, A., {Yu.~Ralchenko}, Reader, J., \& {and NIST ASD Team}. 2015, {NIST
  Atomic Spectra Database (ver. 5.3), [Online]. Available:
  {\url{http://physics.nist.gov/asd}} [2016, July 7].},  National Institute of
  Standards and Technology, Gaithersburg, MD.

\bibitem[{{Krautter}(2008)}]{2008ASPC..401..139K}
{Krautter}, J. 2008, in Astronomical Society of the Pacific Conference Series,
  Vol. 401, RS Ophiuchi (2006) and the Recurrent Nova Phenomenon, ed.
  A.~{Evans}, M.~F. {Bode}, T.~J. {O'Brien}, \& M.~J. {Darnley}, 139

\bibitem[{{Litvinchova} {et~al.}(2011){Litvinchova}, {Pavlenko}, \&
  {Shugarov}}]{2011Ap.....54...36L}
{Litvinchova}, A.~A., {Pavlenko}, E.~P., \& {Shugarov}, S.~Y. 2011,
  Astrophysics, 54, 36

\bibitem[{{Lomb}(1976)}]{1976Ap&SS..39..447L}
{Lomb}, N.~R. 1976, \apss, 39, 447

\bibitem[{{Maguire} {et~al.}(2016){Maguire}, {Taubenberger}, {Sullivan}, \&
  {Mazzali}}]{2016MNRAS.457.3254M}
{Maguire}, K., {Taubenberger}, S., {Sullivan}, M., \& {Mazzali}, P.~A. 2016,
  \mnras, 457, 3254

\bibitem[{{Maoz} {et~al.}(2014){Maoz}, {Mannucci}, \&
  {Nelemans}}]{2014ARA&A..52..107M}
{Maoz}, D., {Mannucci}, F., \& {Nelemans}, G. 2014, \araa, 52, 107

\bibitem[{{Massey} {et~al.}(2016){Massey}, {Neugent}, \&
  {Smart}}]{2016arXiv160400112M}
{Massey}, P., {Neugent}, K.~F., \& {Smart}, B.~M. 2016, ArXiv e-prints,
  arXiv:1604.00112

\bibitem[{{Massey} {et~al.}(2011){Massey}, {Olsen}, {Hodge}, {Jacoby},
  {McNeill}, {Smith}, \& {Strong}}]{2011AJ....141...28M}
{Massey}, P., {Olsen}, K.~A.~G., {Hodge}, P.~W., {et~al.} 2011, \aj, 141, 28

\bibitem[{{Massey} {et~al.}(2006){Massey}, {Olsen}, {Hodge}, {Strong},
  {Jacoby}, {Schlingman}, \& {Smith}}]{2006AJ....131.2478M}
---. 2006, \aj, 131, 2478

\bibitem[{{Mingo} {et~al.}(2016){Mingo}, {Pagani}, {Beardmore}, {Abbey},
  {Perri}, \& {Capalbi}}]{2016Min}
{Mingo}, B., {Pagani}, C., {Beardmore}, A., {et~al.} 2016,
  {SWIFT-XRT-CALDB-04},  {Swift Calibration Database},
  \url{http://heasarc.gsfc.nasa.gov/docs/heasarc/caldb/swift/docs/xrt/xrt_gain_CALDB-04_v16.pdf}

\bibitem[{{Moore}(1945)}]{1945CoPri..20....1M}
{Moore}, C.~E. 1945, Contributions from the Princeton University Observatory,
  20, 1

\bibitem[{{Moro-Mart{\'{\i}}n} {et~al.}(2001){Moro-Mart{\'{\i}}n}, {Garnavich},
  \& {Noriega-Crespo}}]{2001AJ....121.1636M}
{Moro-Mart{\'{\i}}n}, A., {Garnavich}, P.~M., \& {Noriega-Crespo}, A. 2001,
  \aj, 121, 1636

\bibitem[{{Mr{\'o}z} {et~al.}(2016){Mr{\'o}z}, {Udalski}, {Poleski},
  {Soszy{\'n}ski}, {Szyma{\'n}ski}, {Pietrzy{\'n}ski}, {Wyrzykowski},
  {Ulaczyk}, {Koz{\l}owski}, {Pietrukowicz}, \&
  {Skowron}}]{2016ApJS..222....9M}
{Mr{\'o}z}, P., {Udalski}, A., {Poleski}, R., {et~al.} 2016, \apjs, 222, 9

\bibitem[{{Munari} {et~al.}(2011){Munari}, {Ribeiro}, {Bode}, \&
  {Saguner}}]{2011MNRAS.410..525M}
{Munari}, U., {Ribeiro}, V.~A.~R.~M., {Bode}, M.~F., \& {Saguner}, T. 2011,
  \mnras, 410, 525

\bibitem[{{Ness} {et~al.}(2007){Ness}, {Starrfield}, {Beardmore}, {Bode},
  {Drake}, {Evans}, {Gehrz}, {Goad}, {Gonzalez-Riestra}, {Hauschildt},
  {Krautter}, {O'Brien}, {Osborne}, {Page}, {Sch{\"o}nrich}, \&
  {Woodward}}]{2007ApJ...665.1334N}
{Ness}, J., {Starrfield}, S., {Beardmore}, A.~P., {et~al.} 2007, \apj, 665,
  1334

\bibitem[{{Ness}(2010)}]{2010AN....331..179N}
{Ness}, J.-U. 2010, Astronomische Nachrichten, 331, 179

\bibitem[{{Ness} {et~al.}(2012){Ness}, {Schaefer}, {Dobrotka}, {Sadowski},
  {Drake}, {Barnard}, {Talavera}, {Gonzalez-Riestra}, {Page}, {Hernanz},
  {Sala}, \& {Starrfield}}]{2012ApJ...745...43N}
{Ness}, J.-U., {Schaefer}, B.~E., {Dobrotka}, A., {et~al.} 2012, \apj, 745, 43

\bibitem[{{Ness} {et~al.}(2013){Ness}, {Osborne}, {Henze}, {Dobrotka}, {Drake},
  {Ribeiro}, {Starrfield}, {Kuulkers}, {Behar}, {Hernanz}, {Schwarz}, {Page},
  {Beardmore}, \& {Bode}}]{2013A&A...559A..50N}
{Ness}, J.-U., {Osborne}, J.~P., {Henze}, M., {et~al.} 2013, \aap, 559, A50

\bibitem[{{Ness} {et~al.}(2015){Ness}, {Beardmore}, {Osborne}, {Kuulkers},
  {Henze}, {Piro}, {Drake}, {Dobrotka}, {Schwarz}, {Starrfield}, {Kretschmar},
  {Hirsch}, \& {Wilms}}]{2015A&A...578A..39N}
{Ness}, J.-U., {Beardmore}, A.~P., {Osborne}, J.~P., {et~al.} 2015, \aap, 578,
  A39

\bibitem[{{Nishiyama} \& {Kabashima}(2008)}]{2008Nis}
{Nishiyama}, K., \& {Kabashima}, F. 2008, {CBAT},  {IAU},
  \url{http://www.cbat.eps.harvard.edu/iau/CBAT\_M31.html\#2008-12a}

\bibitem[{{Nishiyama} \& {Kabashima}(2012)}]{2012Nis}
---. 2012, {CBAT},  {IAU},
  \url{http://www.cbat.eps.harvard.edu/unconf/followups/J00452884+4154095.html}

\bibitem[{{Norton} {et~al.}(2004){Norton}, {Wynn}, \&
  {Somerscales}}]{2004ApJ...614..349N}
{Norton}, A.~J., {Wynn}, G.~A., \& {Somerscales}, R.~V. 2004, \apj, 614, 349

\bibitem[{{Nussbaumer} {et~al.}(1989){Nussbaumer}, {Schmid}, \&
  {Vogel}}]{1989A&A...211L..27N}
{Nussbaumer}, H., {Schmid}, H.~M., \& {Vogel}, M. 1989, \aap, 211, L27

\bibitem[{{Nussbaumer} {et~al.}(1982){Nussbaumer}, {Storey}, \&
  {Storey}}]{1982A&A...113...21N}
{Nussbaumer}, H., {Storey}, P.~J., \& {Storey}, P.~J. 1982, \aap, 113, 21

\bibitem[{{O'Brien} {et~al.}(1992){O'Brien}, {Bode}, \&
  {Kahn}}]{1992MNRAS.255..683O}
{O'Brien}, T.~J., {Bode}, M.~F., \& {Kahn}, F.~D. 1992, \mnras, 255, 683

\bibitem[{{O'Brien} \& {Kahn}(1987)}]{1987MNRAS.228..277O}
{O'Brien}, T.~J., \& {Kahn}, F.~D. 1987, \mnras, 228, 277

\bibitem[{{O'Brien} {et~al.}(2006){O'Brien}, {Bode}, {Porcas}, {Muxlow},
  {Eyres}, {Beswick}, {Garrington}, {Davis}, \& {Evans}}]{2006Natur.442..279O}
{O'Brien}, T.~J., {Bode}, M.~F., {Porcas}, R.~W., {et~al.} 2006, \nat, 442, 279

\bibitem[{{Orio} {et~al.}(2015){Orio}, {Rana}, {Page}, {Sokoloski}, \&
  {Harrison}}]{2015MNRAS.448L..35O}
{Orio}, M., {Rana}, V., {Page}, K.~L., {Sokoloski}, J., \& {Harrison}, F. 2015,
  \mnras, 448, L35

\bibitem[{{Orlando} {et~al.}(2009){Orlando}, {Drake}, \&
  {Laming}}]{2009A&A...493.1049O}
{Orlando}, S., {Drake}, J.~J., \& {Laming}, J.~M. 2009, \aap, 493, 1049

\bibitem[{{Osborne}(2015)}]{2015JHEAp...7..117O}
{Osborne}, J.~P. 2015, Journal of High Energy Astrophysics, 7, 117

\bibitem[{{Osborne} {et~al.}(2011){Osborne}, {Page}, {Beardmore}, {Bode},
  {Goad}, {O'Brien}, {Starrfield}, {Rauch}, {Ness}, {Krautter}, {Schwarz},
  {Burrows}, {Gehrels}, {Drake}, {Evans}, \& {Eyres}}]{2011ApJ...727..124O}
{Osborne}, J.~P., {Page}, K.~L., {Beardmore}, A.~P., {et~al.} 2011, \apj, 727,
  124

\bibitem[{{Overbeek} {et~al.}(1987){Overbeek}, {McNaught}, {Whitelock},
  {Cragg}, \& {Verdenet}}]{1987IAUC.4395....1O}
{Overbeek}, D., {McNaught}, R.~H., {Whitelock}, P., {Cragg}, T., \& {Verdenet},
  M. 1987, \iaucirc, 4395

\bibitem[{{Page} {et~al.}(2010){Page}, {Osborne}, {Evans}, {Wynn}, {Beardmore},
  {Starling}, {Bode}, {Ibarra}, {Kuulkers}, {Ness}, \&
  {Schwarz}}]{2010MNRAS.401..121P}
{Page}, K.~L., {Osborne}, J.~P., {Evans}, P.~A., {et~al.} 2010, \mnras, 401,
  121

\bibitem[{{Page} {et~al.}(2015){Page}, {Osborne}, {Kuin}, {Henze}, {Walter},
  {Beardmore}, {Bode}, {Darnley}, {Delgado}, {Drake}, {Hernanz}, {Mukai},
  {Nelson}, {Ness}, {Schwarz}, {Shore}, {Starrfield}, \&
  {Woodward}}]{2015MNRAS.454.3108P}
{Page}, K.~L., {Osborne}, J.~P., {Kuin}, N.~P.~M., {et~al.} 2015, \mnras, 454,
  3108

\bibitem[{{Pagnotta} \& {Schaefer}(2014)}]{2014ApJ...788..164P}
{Pagnotta}, A., \& {Schaefer}, B.~E. 2014, \apj, 788, 164

\bibitem[{{Pagnotta} {et~al.}(2015){Pagnotta}, {Schaefer}, {Clem}, {Landolt},
  {Handler}, {Page}, {Osborne}, {Schlegel}, {Hoffman}, {Kiyota}, \&
  {Maehara}}]{2015ApJ...811...32P}
{Pagnotta}, A., {Schaefer}, B.~E., {Clem}, J.~L., {et~al.} 2015, \apj, 811, 32

\bibitem[{{Piascik} {et~al.}(2014){Piascik}, {Steele}, {Bates}, {Mottram},
  {Smith}, {Barnsley}, \& {Bolton}}]{2014SPIE.9147E..8HP}
{Piascik}, A.~S., {Steele}, I.~A., {Bates}, S.~D., {et~al.} 2014, in Society of
  Photo-Optical Instrumentation Engineers (SPIE) Conference Series, Vol. 9147,
  Society of Photo-Optical Instrumentation Engineers (SPIE) Conference Series,
  8

\bibitem[{{Pickering} \& {Fleming}(1896)}]{1896ApJ.....4..369P}
{Pickering}, E.~C., \& {Fleming}, W.~P. 1896, \apj, 4, doi:10.1086/140291

\bibitem[{{Pietsch}(2010)}]{2010AN....331..187P}
{Pietsch}, W. 2010, Astronomische Nachrichten, 331, 187

\bibitem[{{Pietsch} {et~al.}(2005){Pietsch}, {Fliri}, {Freyberg}, {Greiner},
  {Haberl}, {Riffeser}, \& {Sala}}]{2005A&A...442..879P}
{Pietsch}, W., {Fliri}, J., {Freyberg}, M.~J., {et~al.} 2005, \aap, 442, 879

\bibitem[{{Pietsch} {et~al.}(2011){Pietsch}, {Henze}, {Haberl}, {Hernanz},
  {Sala}, {Hartmann}, \& {Della Valle}}]{2011A&A...531A..22P}
{Pietsch}, W., {Henze}, M., {Haberl}, F., {et~al.} 2011, \aap, 531, A22

\bibitem[{{Pietsch} {et~al.}(2007){Pietsch}, {Haberl}, {Sala}, {Stiele},
  {Hornoch}, {Riffeser}, {Fliri}, {Bender}, {B{\"u}hler}, {Burwitz}, {Greiner},
  \& {Seitz}}]{2007A&A...465..375P}
{Pietsch}, W., {Haberl}, F., {Sala}, G., {et~al.} 2007, \aap, 465, 375

\bibitem[{{Poole} {et~al.}(2008){Poole}, {Breeveld}, {Page}, {Landsman},
  {Holland}, {Roming}, {Kuin}, {Brown}, {Gronwall}, {Hunsberger}, {Koch},
  {Mason}, {Schady}, {vanden Berk}, {Blustin}, {Boyd}, {Broos}, {Carter},
  {Chester}, {Cucchiara}, {Hancock}, {Huckle}, {Immler}, {Ivanushkina},
  {Kennedy}, {Marshall}, {Morgan}, {Pandey}, {de Pasquale}, {Smith}, \&
  {Still}}]{2008MNRAS.383..627P}
{Poole}, T.~S., {Breeveld}, A.~A., {Page}, M.~J., {et~al.} 2008, \mnras, 383,
  627

\bibitem[{{Pottasch}(1967)}]{1967BAN....19..227P}
{Pottasch}, S.~R. 1967, \bain, 19, 227

\bibitem[{{Pravec} {et~al.}(1994){Pravec}, {Hudec}, {Sold{\'a}n}, {Sommer}, \&
  {Schenkl}}]{1994ExA.....5..375P}
{Pravec}, P., {Hudec}, R., {Sold{\'a}n}, J., {Sommer}, M., \& {Schenkl}, K.~H.
  1994, Experimental Astronomy, 5, 375

\bibitem[{{Prialnik} \& {Kovetz}(1995)}]{1995ApJ...445..789P}
{Prialnik}, D., \& {Kovetz}, A. 1995, \apj, 445, 789

\bibitem[{{Prialnik} {et~al.}(1978){Prialnik}, {Shara}, \&
  {Shaviv}}]{1978A&A....62..339P}
{Prialnik}, D., {Shara}, M.~M., \& {Shaviv}, G. 1978, \aap, 62, 339

\bibitem[{{R Development Core Team}(2011)}]{R_manual}
{R Development Core Team}. 2011, R: A Language and Environment for Statistical
  Computing, R Foundation for Statistical Computing, Vienna, Austria, {ISBN}
  3-900051-07-0

\bibitem[{{Raman}(1928)}]{raman}
{Raman}, C.~V. 1928, Indian J. Phys., 2, 387

\bibitem[{{Rauch} {et~al.}(2010){Rauch}, {Orio}, {Gonzales-Riestra}, {Nelson},
  {Still}, {Werner}, \& {Wilms}}]{2010ApJ...717..363R}
{Rauch}, T., {Orio}, M., {Gonzales-Riestra}, R., {et~al.} 2010, \apj, 717, 363

\bibitem[{{Ribeiro} {et~al.}(2013){Ribeiro}, {Bode}, {Darnley}, {Barnsley},
  {Munari}, \& {Harman}}]{2013MNRAS.433.1991R}
{Ribeiro}, V.~A.~R.~M., {Bode}, M.~F., {Darnley}, M.~J., {et~al.} 2013, \mnras,
  433, 1991

\bibitem[{{Ribeiro} {et~al.}(2011){Ribeiro}, {Darnley}, {Bode}, {Munari},
  {Harman}, {Steele}, \& {Meaburn}}]{2011MNRAS.412.1701R}
{Ribeiro}, V.~A.~R.~M., {Darnley}, M.~J., {Bode}, M.~F., {et~al.} 2011, \mnras,
  412, 1701

\bibitem[{{Ribeiro} {et~al.}(2009){Ribeiro}, {Bode}, {Darnley}, {Harman},
  {Newsam}, {O'Brien}, {Bohigas}, {Echevarr{\'{\i}}a}, {Bond}, {Chavushyan},
  {Costero}, {Coziol}, {Evans}, {Eyres}, {Le{\'o}n-Tavares}, {Richer},
  {Tovmassian}, {Starrfield}, \& {Zharikov}}]{2009ApJ...703.1955R}
{Ribeiro}, V.~A.~R.~M., {Bode}, M.~F., {Darnley}, M.~J., {et~al.} 2009, \apj,
  703, 1955

\bibitem[{{Ritchey}(1917)}]{1917PASP...29..210R}
{Ritchey}, G.~W. 1917, \pasp, 29, 210

\bibitem[{{Roming} {et~al.}(2005){Roming}, {Kennedy}, {Mason}, {Nousek}, {Ahr},
  {Bingham}, {Broos}, {Carter}, {Hancock}, {Huckle}, {Hunsberger}, {Kawakami},
  {Killough}, {Koch}, {McLelland}, {Smith}, {Smith}, {Soto}, {Boyd},
  {Breeveld}, {Holland}, {Ivanushkina}, {Pryzby}, {Still}, \&
  {Stock}}]{2005SSRv..120...95R}
{Roming}, P.~W.~A., {Kennedy}, T.~E., {Mason}, K.~O., {et~al.} 2005, \ssr, 120,
  95

\bibitem[{{Rosino} \& {Iijima}(1987)}]{1987rorn.conf...27R}
{Rosino}, L., \& {Iijima}, T. 1987, in RS Ophiuchi (1985) and the Recurrent
  Nova Phenomenon, ed. M.~F. {Bode} (Utrecht: VNU Science Press), 27

\bibitem[{{Rupen} {et~al.}(2008){Rupen}, {Mioduszewski}, \&
  {Sokoloski}}]{2008ApJ...688..559R}
{Rupen}, M.~P., {Mioduszewski}, A.~J., \& {Sokoloski}, J.~L. 2008, \apj, 688,
  559

\bibitem[{{Sanford}(1949)}]{1949ApJ...109...81S}
{Sanford}, R.~F. 1949, \apj, 109, 81

\bibitem[{{Scargle}(1982)}]{1982ApJ...263..835S}
{Scargle}, J.~D. 1982, \apj, 263, 835

\bibitem[{{Schaefer}(2005)}]{2005ApJ...621L..53S}
{Schaefer}, B.~E. 2005, \apjl, 621, L53

\bibitem[{{Schaefer}(2010)}]{2010ApJS..187..275S}
---. 2010, \apjs, 187, 275

\bibitem[{{Schaefer} {et~al.}(2010){Schaefer}, {Pagnotta}, {Xiao}, {Darnley},
  {Bode}, {Harris}, {Dvorak}, {Menke}, {Linnolt}, {Templeton}, {Henden},
  {Pojma{\'n}ski}, {Pilecki}, {Szczygie{\l}}, \&
  {Watanabe}}]{2010AJ....140..925S}
{Schaefer}, B.~E., {Pagnotta}, A., {Xiao}, L., {et~al.} 2010, \aj, 140, 925

\bibitem[{{Schaefer} {et~al.}(2011){Schaefer}, {Pagnotta}, {LaCluyze},
  {Reichart}, {Ivarsen}, {Haislip}, {Nysewander}, {Moore}, {Oksanen},
  {Worters}, {Sefako}, {Mentz}, {Dvorak}, {Gomez}, {Harris}, {Henden}, {Guan
  Tan}, {Templeton}, {Allen}, {Monard}, {Rea}, {Roberts}, {Stein}, {Maehara},
  {Richards}, {Stockdale}, {Krajci}, {Sjoberg}, {McCormick}, {Revnivtsev},
  {Molkov}, {Suleimanov}, {Darnley}, {Bode}, {Handler}, {Lepine}, \&
  {Shara}}]{2011ApJ...742..113S}
{Schaefer}, B.~E., {Pagnotta}, A., {LaCluyze}, A.~P., {et~al.} 2011, \apj, 742,
  113

\bibitem[{{Schatzman}(1949)}]{1949AnAp...12..281S}
{Schatzman}, E. 1949, Annales d'Astrophysique, 12, 281

\bibitem[{{Schatzman}(1951)}]{1951AnAp...14..294S}
---. 1951, Annales d'Astrophysique, 14, 294

\bibitem[{{Schmid}(1989)}]{1989A&A...211L..31S}
{Schmid}, H.~M. 1989, \aap, 211, L31

\bibitem[{{Schwarz} {et~al.}(2011){Schwarz}, {Ness}, {Osborne}, {Page},
  {Evans}, {Beardmore}, {Walter}, {Helton}, {Woodward}, {Bode}, {Starrfield},
  \& {Drake}}]{2011ApJS..197...31S}
{Schwarz}, G.~J., {Ness}, J.-U., {Osborne}, J.~P., {et~al.} 2011, \apjs, 197,
  31

\bibitem[{{Shafter}(2015)}]{2015ATel.7968....1S}
{Shafter}, A.~W. 2015, The Astronomer's Telegram, 7968, 1

\bibitem[{{Shafter}(2016)}]{2016arXiv160602358S}
---. 2016, ArXiv e-prints, arXiv:1606.02358

\bibitem[{{Shafter} {et~al.}(2011{\natexlab{a}}){Shafter}, {Bode}, {Darnley},
  {Misselt}, {Rubin}, \& {Hornoch}}]{2011ApJ...727...50S}
{Shafter}, A.~W., {Bode}, M.~F., {Darnley}, M.~J., {et~al.} 2011{\natexlab{a}},
  \apj, 727, 50

\bibitem[{{Shafter} {et~al.}(2012){Shafter}, {Hornoch}, {Ciardullo}, {Darnley},
  \& {Bode}}]{2012ATel.4503....1S}
{Shafter}, A.~W., {Hornoch}, K., {Ciardullo}, J.~V.~R., {Darnley}, M.~J., \&
  {Bode}, M.~F. 2012, The Astronomer's Telegram, 4503, 1

\bibitem[{{Shafter} {et~al.}(2015{\natexlab{a}}){Shafter}, {Horst}, \&
  {Darnley}}]{2015ATel.7967....1S}
{Shafter}, A.~W., {Horst}, J.~C., \& {Darnley}, M.~J. 2015{\natexlab{a}}, The
  Astronomer's Telegram, 7967, 1

\bibitem[{{Shafter} \& {Irby}(2001)}]{2001ApJ...563..749S}
{Shafter}, A.~W., \& {Irby}, B.~K. 2001, \apj, 563, 749

\bibitem[{{Shafter} {et~al.}(2009){Shafter}, {Rau}, {Quimby}, {Kasliwal},
  {Bode}, {Darnley}, \& {Misselt}}]{2009ApJ...690.1148S}
{Shafter}, A.~W., {Rau}, A., {Quimby}, R.~M., {et~al.} 2009, \apj, 690, 1148

\bibitem[{{Shafter} {et~al.}(2011{\natexlab{b}}){Shafter}, {Darnley},
  {Hornoch}, {Filippenko}, {Bode}, {Ciardullo}, {Misselt}, {Hounsell},
  {Chornock}, \& {Matheson}}]{2011ApJ...734...12S}
{Shafter}, A.~W., {Darnley}, M.~J., {Hornoch}, K., {et~al.} 2011{\natexlab{b}},
  \apj, 734, 12

\bibitem[{{Shafter} {et~al.}(2015{\natexlab{b}}){Shafter}, {Henze}, {Rector},
  {Schweizer}, {Hornoch}, {Orio}, {Pietsch}, {Darnley}, {Williams}, {Bode}, \&
  {Bryan}}]{2015ApJS..216...34S}
{Shafter}, A.~W., {Henze}, M., {Rector}, T.~A., {et~al.} 2015{\natexlab{b}},
  \apjs, 216, 34

\bibitem[{{Shara} {et~al.}(2016){Shara}, {Doyle}, {Lauer}, {Zurek}, {Neill},
  {Madrid}, {Welch}, \& {Baltz}}]{2016arXiv160200758S}
{Shara}, M.~M., {Doyle}, T., {Lauer}, T.~R., {et~al.} 2016, ArXiv e-prints,
  arXiv:1602.00758

\bibitem[{{Shore} {et~al.}(2014){Shore}, {De Gennaro Aquino}, {Scaringi}, \&
  {van Winckel}}]{2014A&A...570L...4S}
{Shore}, S.~N., {De Gennaro Aquino}, I., {Scaringi}, S., \& {van Winckel}, H.
  2014, \aap, 570, L4

\bibitem[{{Shore} {et~al.}(1996){Shore}, {Kenyon}, {Starrfield}, \&
  {Sonneborn}}]{1996ApJ...456..717S}
{Shore}, S.~N., {Kenyon}, S.~J., {Starrfield}, S., \& {Sonneborn}, G. 1996,
  \apj, 456, 717

\bibitem[{{Shore} {et~al.}(1991){Shore}, {Sonneborn}, {Starrfield}, {Hamuy},
  {Williams}, {Cassatella}, \& {Drechsel}}]{1991ApJ...370..193S}
{Shore}, S.~N., {Sonneborn}, G., {Starrfield}, S.~G., {et~al.} 1991, \apj, 370,
  193

\bibitem[{{Skopal} {et~al.}(2008){Skopal}, {Pribulla}, {Buil}, {Vittone}, \&
  {Errico}}]{2008ASPC..401..227S}
{Skopal}, A., {Pribulla}, T., {Buil}, C., {Vittone}, A., \& {Errico}, L. 2008,
  in Astronomical Society of the Pacific Conference Series, Vol. 401, RS
  Ophiuchi (2006) and the Recurrent Nova Phenomenon, ed. A.~{Evans}, M.~F.
  {Bode}, T.~J. {O'Brien}, \& M.~J. {Darnley}, 227

\bibitem[{{Skrutskie} {et~al.}(2006){Skrutskie}, {Cutri}, {Stiening},
  {Weinberg}, {Schneider}, {Carpenter}, {Beichman}, {Capps}, {Chester},
  {Elias}, {Huchra}, {Liebert}, {Lonsdale}, {Monet}, {Price}, {Seitzer},
  {Jarrett}, {Kirkpatrick}, {Gizis}, {Howard}, {Evans}, {Fowler}, {Fullmer},
  {Hurt}, {Light}, {Kopan}, {Marsh}, {McCallon}, {Tam}, {Van Dyk}, \&
  {Wheelock}}]{2006AJ....131.1163S}
{Skrutskie}, M.~F., {Cutri}, R.~M., {Stiening}, R., {et~al.} 2006, \aj, 131,
  1163

\bibitem[{{Sokoloski} {et~al.}(2008){Sokoloski}, {Rupen}, \&
  {Mioduszewski}}]{2008ApJ...685L.137S}
{Sokoloski}, J.~L., {Rupen}, M.~P., \& {Mioduszewski}, A.~J. 2008, \apjl, 685,
  L137

\bibitem[{{Soraisam} {et~al.}(2016){Soraisam}, {Gilfanov}, {Wolf}, \&
  {Bildsten}}]{2016MNRAS.455..668S}
{Soraisam}, M.~D., {Gilfanov}, M., {Wolf}, W.~M., \& {Bildsten}, L. 2016,
  \mnras, 455, 668

\bibitem[{{Sostero} \& {Guido}(2006{\natexlab{a}})}]{2006IAUC.8673....3S}
{Sostero}, G., \& {Guido}, E. 2006{\natexlab{a}}, \iaucirc, 8673

\bibitem[{{Sostero} \& {Guido}(2006{\natexlab{b}})}]{2006IAUC.8681....4S}
---. 2006{\natexlab{b}}, \iaucirc, 8681

\bibitem[{{Sostero} {et~al.}(2006){Sostero}, {Guido}, \&
  {West}}]{2006CBET..502....2S}
{Sostero}, G., {Guido}, E., \& {West}, J.~D. 2006, Central Bureau Electronic
  Telegrams, 502

\bibitem[{{Stark} {et~al.}(1992){Stark}, {Gammie}, {Wilson}, {Bally}, {Linke},
  {Heiles}, \& {Hurwitz}}]{1992ApJS...79...77S}
{Stark}, A.~A., {Gammie}, C.~F., {Wilson}, R.~W., {et~al.} 1992, \apjs, 79, 77

\bibitem[{{Starrfield} {et~al.}(2008){Starrfield}, {Iliadis}, \& {Hix}}]{Sta08}
{Starrfield}, S., {Iliadis}, C., \& {Hix}, W.~R. 2008, in {Classical Novae, 2nd
  Edition}, ed. M.~F. {Bode} \& A.~{Evans} (Cambridge: Cambridge University
  Press), 77--101

\bibitem[{{Starrfield} {et~al.}(2016){Starrfield}, {Iliadis}, \&
  {Hix}}]{2016PASP..128e1001S}
{Starrfield}, S., {Iliadis}, C., \& {Hix}, W.~R. 2016, \pasp, 128, 051001

\bibitem[{{Starrfield} {et~al.}(2012){Starrfield}, {Iliadis}, {Timmes}, {Hix},
  {Arnett}, {Meakin}, \& {Sparks}}]{2012BASI...40..419S}
{Starrfield}, S., {Iliadis}, C., {Timmes}, F.~X., {et~al.} 2012, Bulletin of
  the Astronomical Society of India, 40, 419

\bibitem[{{Starrfield} {et~al.}(1985){Starrfield}, {Sparks}, \&
  {Truran}}]{1985ApJ...291..136S}
{Starrfield}, S., {Sparks}, W.~M., \& {Truran}, J.~W. 1985, \apj, 291, 136

\bibitem[{{Starrfield} {et~al.}(1972){Starrfield}, {Truran}, {Sparks}, \&
  {Kutter}}]{1972ApJ...176..169S}
{Starrfield}, S., {Truran}, J.~W., {Sparks}, W.~M., \& {Kutter}, G.~S. 1972,
  \apj, 176, 169

\bibitem[{{Steele} {et~al.}(2004){Steele}, {Smith}, {Rees}, {Baker}, {Bates},
  {Bode}, {Bowman}, {Carter}, {Etherton}, {Ford}, {Fraser}, {Gomboc}, {Lett},
  {Mansfield}, {Marchant}, {Medrano-Cerda}, {Mottram}, {Raback}, {Scott},
  {Tomlinson}, \& {Zamanov}}]{2004SPIE.5489..679S}
{Steele}, I.~A., {Smith}, R.~J., {Rees}, P.~C., {et~al.} 2004, in Society of
  Photo-Optical Instrumentation Engineers (SPIE) Conference Series, Vol. 5489,
  Ground-based Telescopes, ed. J.~M. {Oschmann}, Jr., 679--692

\bibitem[{{Stone}(1977)}]{1977ApJ...218..767S}
{Stone}, R.~P.~S. 1977, \apj, 218, 767

\bibitem[{{Strope} {et~al.}(2010){Strope}, {Schaefer}, \&
  {Henden}}]{2010AJ....140...34S}
{Strope}, R.~J., {Schaefer}, B.~E., \& {Henden}, A.~A. 2010, \aj, 140, 34

\bibitem[{{Tang} {et~al.}(2013){Tang}, {Cao}, \&
  {Kasliwal}}]{2013ATel.5607....1T}
{Tang}, S., {Cao}, Y., \& {Kasliwal}, M.~M. 2013, The Astronomer's Telegram,
  5607, 1

\bibitem[{{Tang} {et~al.}(2014){Tang}, {Bildsten}, {Wolf}, {Li}, {Kong}, {Cao},
  {Cenko}, {De Cia}, {Kasliwal}, {Kulkarni}, {Laher}, {Masci}, {Nugent},
  {Perley}, {Prince}, \& {Surace}}]{2014ApJ...786...61T}
{Tang}, S., {Bildsten}, L., {Wolf}, W.~M., {et~al.} 2014, \apj, 786, 61

\bibitem[{{Tarasova}(2009)}]{2009ARep...53..203T}
{Tarasova}, T.~N. 2009, Astronomy Reports, 53, 203

\bibitem[{{Tatischeff} \& {Hernanz}(2007)}]{2007ApJ...663L.101T}
{Tatischeff}, V., \& {Hernanz}, M. 2007, \apjl, 663, L101

\bibitem[{{Taylor} {et~al.}(1989){Taylor}, {Davis}, {Porcas}, \&
  {Bode}}]{1989MNRAS.237...81T}
{Taylor}, A.~R., {Davis}, R.~J., {Porcas}, R.~W., \& {Bode}, M.~F. 1989,
  \mnras, 237, 81

\bibitem[{{Tody}(1993)}]{1993ASPC...52..173T}
{Tody}, D. 1993, in Astronomical Society of the Pacific Conference Series,
  Vol.~52, Astronomical Data Analysis Software and Systems II, ed. R.~J.
  {Hanisch}, R.~J.~V. {Brissenden}, \& J.~{Barnes}, 173

\bibitem[{{van Rossum}(2012)}]{2012ApJ...756...43V}
{van Rossum}, D.~R. 2012, \apj, 756, 43

\bibitem[{{van Rossum} \& {Ness}(2010)}]{2010AN....331..175V}
{van Rossum}, D.~R., \& {Ness}, J. 2010, Astronomische Nachrichten, 331, 175

\bibitem[{{Wallerstein} \& {Garnavich}(1986)}]{1986PASP...98..875W}
{Wallerstein}, G., \& {Garnavich}, P.~M. 1986, \pasp, 98, 875

\bibitem[{{Walter} {et~al.}(2012){Walter}, {Battisti}, {Towers}, {Bond}, \&
  {Stringfellow}}]{2012PASP..124.1057W}
{Walter}, F.~M., {Battisti}, A., {Towers}, S.~E., {Bond}, H.~E., \&
  {Stringfellow}, G.~S. 2012, \pasp, 124, 1057

\bibitem[{{Wang} \& {Han}(2012)}]{2012NewAR..56..122W}
{Wang}, B., \& {Han}, Z. 2012, \nar, 56, 122

\bibitem[{{Watanabe} {et~al.}(2012){Watanabe}, {Takahashi}, {Sato}, {Watanabe},
  {Fukuhara}, {Hamamoto}, \& {Ozaki}}]{2012SPIE.8446E..2OW}
{Watanabe}, M., {Takahashi}, Y., {Sato}, M., {et~al.} 2012, in \procspie, Vol.
  8446, Ground-based and Airborne Instrumentation for Astronomy IV, 84462O

\bibitem[{{Whelan} \& {Iben}(1973)}]{1973ApJ...186.1007W}
{Whelan}, J., \& {Iben}, Jr., I. 1973, \apj, 186, 1007

\bibitem[{{White} {et~al.}(1995){White}, {Giommi}, {Heise}, {Angelini}, \&
  {Fantasia}}]{1995ApJ...445L.125W}
{White}, N.~E., {Giommi}, P., {Heise}, J., {Angelini}, L., \& {Fantasia}, S.
  1995, \apjl, 445, L125

\bibitem[{{Williams} {et~al.}(2004){Williams}, {Garcia}, {Kong}, {Primini},
  {King}, {Di Stefano}, \& {Murray}}]{2004ApJ...609..735W}
{Williams}, B.~F., {Garcia}, M.~R., {Kong}, A.~K.~H., {et~al.} 2004, \apj, 609,
  735

\bibitem[{{Williams}(2012)}]{2012AJ....144...98W}
{Williams}, R. 2012, \aj, 144, 98

\bibitem[{{Williams}(1992)}]{1992AJ....104..725W}
{Williams}, R.~E. 1992, \aj, 104, 725

\bibitem[{{Williams} {et~al.}(1994){Williams}, {Phillips}, \&
  {Hamuy}}]{1994ApJS...90..297W}
{Williams}, R.~E., {Phillips}, M.~M., \& {Hamuy}, M. 1994, \apjs, 90, 297

\bibitem[{{Williams} {et~al.}(2014){Williams}, {Darnley}, {Bode}, {Keen}, \&
  {Shafter}}]{2014ApJS..213...10W}
{Williams}, S.~C., {Darnley}, M.~J., {Bode}, M.~F., {Keen}, A., \& {Shafter},
  A.~W. 2014, \apjs, 213, 10

\bibitem[{{Williams} {et~al.}(2013){Williams}, {Darnley}, {Bode}, \&
  {Shafter}}]{2013ATel.5611....1W}
{Williams}, S.~C., {Darnley}, M.~J., {Bode}, M.~F., \& {Shafter}, A.~W. 2013,
  The Astronomer's Telegram, 5611

\bibitem[{{Williams} {et~al.}(2016{\natexlab{a}}){Williams}, {Darnley}, {Bode},
  \& {Shafter}}]{2016ApJ...817..143W}
---. 2016{\natexlab{a}}, \apj, 817, 143

\bibitem[{{Williams} {et~al.}(2016{\natexlab{b}}){Williams}, {Darnley},
  {Henze}, {Hornoch}, \& {Shafter}}]{2016WillLN}
{Williams}, S.~C., {Darnley}, M.~J., {Henze}, M., {Hornoch}, K., \& {Shafter},
  A.~W. 2016{\natexlab{b}}, in preparation, for submission to MNRAS

\bibitem[{{Williams} {et~al.}(2015){Williams}, {Shafter}, {Hornoch}, {Henze},
  \& {Darnley}}]{2015ATel.8234....1W}
{Williams}, S.~C., {Shafter}, A.~W., {Hornoch}, K., {Henze}, M., \& {Darnley},
  M.~J. 2015, The Astronomer's Telegram, 8234

\bibitem[{{Wilms} {et~al.}(2000){Wilms}, {Allen}, \&
  {McCray}}]{2000ApJ...542..914W}
{Wilms}, J., {Allen}, A., \& {McCray}, R. 2000, \apj, 542, 914

\bibitem[{{Wolf} {et~al.}(2013){Wolf}, {Bildsten}, {Brooks}, \&
  {Paxton}}]{2013ApJ...777..136W}
{Wolf}, W.~M., {Bildsten}, L., {Brooks}, J., \& {Paxton}, B. 2013, \apj, 777,
  136

\bibitem[{{Worters} {et~al.}(2007){Worters}, {Eyres}, {Bromage}, \&
  {Osborne}}]{2007MNRAS.379.1557W}
{Worters}, H.~L., {Eyres}, S.~P.~S., {Bromage}, G.~E., \& {Osborne}, J.~P.
  2007, \mnras, 379, 1557

\bibitem[{{Woudt} \& {Ribeiro}(2014)}]{2014ASPC..490.....W}
{Woudt}, P.~A., \& {Ribeiro}, V.~A.~R.~M., eds. 2014, Astronomical Society of
  the Pacific Conference Series, Vol. 490, {Stella Novae: Past and Future
  Decades} (San Francisco: Astronomical Society of the Pacific)

\bibitem[{{Wright} \& {Barlow}(1975)}]{1975MNRAS.170...41W}
{Wright}, A.~E., \& {Barlow}, M.~J. 1975, \mnras, 170, 41

\bibitem[{{Yaron} {et~al.}(2005){Yaron}, {Prialnik}, {Shara}, \&
  {Kovetz}}]{2005ApJ...623..398Y}
{Yaron}, O., {Prialnik}, D., {Shara}, M.~M., \& {Kovetz}, A. 2005, \apj, 623,
  398

\bibitem[{{Zacharias} {et~al.}(2013){Zacharias}, {Finch}, {Girard}, {Henden},
  {Bartlett}, {Monet}, \& {Zacharias}}]{2013AJ....145...44Z}
{Zacharias}, N., {Finch}, C.~T., {Girard}, T.~M., {et~al.} 2013, \aj, 145, 44

\end{thebibliography}

\appendix

\section{Visible and Near Infrared Photometry}\label{app:optical_photometry}

\subsection{Liverpool Telescope photometry}\label{lt_photometry}

Significant maintenance work was carried out on the LT between the 2014 and 2015 eruptions, including re-aluminization of both the primary and secondary mirrors.  This work, coupled with additional improvements achieved approximately a doubling of the throughput of the telescope (averaged across all wavelengths).  The improvements were significantly greater in the blue, with a $\sim225\%$ increase in $u'$-band throughput.  These improvements, along with the realization of the extreme blue nature of the eruptions of \novak\ (\otwok), motivated us to amend our LT strategy from 2014 and include the $u'$ filter for monitoring the 2015 eruption.  To achieve a more complete coverage of the spectral energy distribution (SED) of the eruption, we also included the $z'$ and $H$-band filters in our follow-up program.  Unlike the 2014 eruption, H$\alpha$ observations were not employed this year as, unusually for novae, \novak\ faded beyond detectability in H$\alpha$ before it did in the broad-band $R/r'$ filters (see \otwok).

A pre-planned broadband $u'$, $B$, $V$, $r'$, $i'$,and $z'$-band photometry program, employing the IO:O detector, was initiated on the LT immediately following the LCOGT detection of the 2015 eruption of \novak.  

The LT observing strategy again involved taking $3\times120$\,s exposures through each of the six filters for every epoch. The LT robotic scheduler was initially requested to repeat these observations with a minimum interval (between repeat observations) of 1\,h. This minimum interval was increased to 1 day from the night beginning 2015 Sep.\,3 UT. To counter the signal-to-noise losses as the nova faded, the exposure time was increased to $3\times300$\,s in the $u'$-band and $3\times180$\,s in all other filters from Aug.\,30.5 UT onward.  The exposure times were subsequently increased to $3\times300$\,s in all filters from Sep.\,2.5 UT onward, and then to $3\times450$\,s in $u'$, $B$, $V$, and $r'$ from Sep.\,3.5 onward -- as the nova faded.  From Sep.\,11.5 UT onward the $r'$, $i'$, and $z'$-band eruption monitoring ceased due to consecutive non-detections on previous nights.  There were no $V$-band observations after Sep.\,15 UT, and no $B$ observations after Sep.\,17.  LT observations following the 2015 eruption formally ended on Sep.\,22 UT, the $r'$-band monitoring campaign for the next eruption (following the strategy described in Section~\ref{sec:monitor_and_detect}) had begun on 2015 Sep.\,11.

The LT data were pre-processed at the telescope and then further processed using standard routines within Starlink \citep{1982QJRAS..23..485D} and IRAF \citep{1993ASPC...52..173T}.  PSF fitting was performed using the Starlink {\tt photom} (v1.12-2) package.  Photometric calibration was achieved using 17 stars from \citet{2006AJ....131.2478M} within the IO:O field (see Table~\ref{tab_calib}; expanded from the original version in \otwok). Transformations from \citet{2005AJ....130..873J} were used to convert these calibration stars from $UBVRI$ to $u'g'r'i'z'$.  In all cases, uncertainties from the photometric calibration are not the dominant source of error.  

\begin{sidewaystable}
\caption{Photometry calibration stars in the field of \novak\ employed with the LT, LCOGT, and MLO observations.\label{tab_calib}}
\begin{center}
\begin{tabular}{lll|lllllll|lllll}
\tableline\tableline
\# & R.A.\ (J2000)$^\dag$ & Dec.\ (J2000)$^\dag$ & $U$ & $B$ &  $V$ &  $R$ &  $I$ &  $J$ &  $H$ &  $u'$ &  $g'$ &  $r'$ &  $i'$ &  $z'$\\
\tableline
1  & $0^{\mathrm{h}}45^{\mathrm{m}}11\fs73$ & $+41\degr53\arcmin52\farcs2$ & 19.098& 18.635 & 17.759 & 17.270 & 16.782 & 16.001 & 15.789 & 19.887 &18.165 & 17.501 & 17.257 & 17.072 \\
2  & $0^{\mathrm{h}}45^{\mathrm{m}}12\fs71$ & $+41\degr54\arcmin48\farcs5$ & 18.166& 17.711 & 16.873 & 16.423 & 16.010 & 15.495 & 14.979 & 18.968 & 17.256 & 16.631 & 16.455 & 16.331 \\
3  & $0^{\mathrm{h}}45^{\mathrm{m}}14\fs35$ & $+41\degr55\arcmin5\farcs4$ & 20.765& 19.600 & 18.239 & 17.412 & 16.635 & 15.708 & 15.185 & 21.557 & 18.936 & 17.777 & 17.270 & 16.851 \\
4  & $0^{\mathrm{h}}45^{\mathrm{m}}15\fs43$ & $+41\degr54\arcmin6\farcs9$ & 18.953 & 18.963 & 18.319 & 17.953 & 17.566 & \nodata & \nodata &19.703 & 18.585 & 18.159 & 18.006 & 17.903 \\
5  & $0^{\mathrm{h}}45^{\mathrm{m}}18\fs25$ & $+41\degr54\arcmin38\farcs3$ & 19.047 & 18.933 & 18.200 & 17.778 & 17.353 & \nodata & \nodata &19.796 & 18.520 & 18.002 & 17.815 & 17.681\\
6  & $0^{\mathrm{h}}45^{\mathrm{m}}19\fs69$ & $+41\degr56\arcmin5\farcs9$ & 17.808 & 17.740 & 17.068 & 16.680 & 16.290 & 15.783 & 15.625 &18.568 & 17.351 & 16.896 & 16.741 & 16.635 \\
7  & $0^{\mathrm{h}}45^{\mathrm{m}}22\fs59$ & $+41\degr53\arcmin37\farcs5$ & 17.097 & 16.352 & 15.607 & 15.197 & \nodata & 14.351 & 14.018 &18.018 & 15.934 & 15.404 & \nodata & \nodata \\
8  & $0^{\mathrm{h}}45^{\mathrm{m}}22\fs75$ & $+41\degr55\arcmin6\farcs6$ & 21.744 & 20.532 & 19.087 & 18.183 & 17.233 & 16.582 & 15.783 &22.515 & 19.834 & 18.590 & 17.926 & 17.366 \\
9  & $0^{\mathrm{h}}45^{\mathrm{m}}25\fs24$ & $+41\degr55\arcmin32\farcs6$ & 19.634 & 19.121 & 18.233 & 17.742 & 17.278 & 16.721 & 16.116 &20.432 & 18.646 & 17.970 & 17.748 & 17.582 \\
10 & $0^{\mathrm{h}}45^{\mathrm{m}}27\fs48$ & $+41\degr55\arcmin30\farcs4$ & 17.998 & 17.606 & 16.785 & 16.331 & 15.911 & 15.307 & 14.849 &18.789 & 17.158 & 16.550 & 16.368 & 16.238 \\
11 & $0^{\mathrm{h}}45^{\mathrm{m}}28\fs55$ & $+41\degr54\arcmin51\farcs7$ & 19.527 & 19.162 & 18.349 & 17.876 & 17.358 & \nodata & \nodata & 20.314 & 18.717 & 18.118 & 17.846 & 17.637\\
12 & $0^{\mathrm{h}}45^{\mathrm{m}}30\fs01$ & $+41\degr53\arcmin20\farcs9$ & 18.772 & 17.991 & 16.945 & 16.318 & \nodata & 14.938 & 14.445 &19.582 & 17.453 & 16.616 & \nodata & \nodata \\
13 & $0^{\mathrm{h}}45^{\mathrm{m}}30\fs20$ & $+41\degr56\arcmin4\farcs8$ & 18.535 & 18.362 & 17.640 & 17.230 & 16.833 & 16.299 & 15.148 &19.305 & 17.953 & 17.447 & 17.285 & 17.174 \\
14 & $0^{\mathrm{h}}45^{\mathrm{m}}30\fs50$ & $+41\degr55\arcmin11\farcs9$ & 15.588 & 15.410 & 14.738 & 14.367 & \nodata & 13.574 & 13.270 &16.379 & 15.021 & 14.566 & \nodata & \nodata \\
15 & $0^{\mathrm{h}}45^{\mathrm{m}}34\fs14$ & $+41\degr55\arcmin4\farcs1$ & 18.490 & 18.002 & 17.030 & 16.496 & 15.964 & 15.288 & 14.742 &19.248 & 17.493 & 16.732 & 16.448 & 16.227 \\
16 & $0^{\mathrm{h}}45^{\mathrm{m}}39\fs98$ & $+41\degr55\arcmin32\farcs0$ & 18.341 & 17.452 & 16.416 & 15.810 & 15.300 &14.549 & 14.068 &19.186 & 16.918 & 16.091 & 15.827& 15.624 \\
17 & $0^{\mathrm{h}}45^{\mathrm{m}}46\fs80$ & $+41\degr54\arcmin0\farcs0$ & 18.139 & 18.074 & 17.375 & 16.949 & 16.546 & 15.881 & 15.580 &18.888 & 17.674 & 17.191 & 17.025 & 16.908 \\
\tableline
 & $0^{\mathrm{h}}45^{\mathrm{m}}32\fs50$ & $+41\degr54\arcmin43\farcs3$ & \multicolumn{12}{c}{PSF Star}\\
 \tableline
 \end{tabular}
 \end{center}
 \catcode`\&=12
 \tablenotetext{\dag}{Astrometry based on that published by \citet{2006AJ....131.2478M}, and does not take into account any corrections as reported in \citet{2016arXiv160400112M}.}
 \tablecomments{Updated calibration photometry from that presented in \citep{2015A&A...580A..45D}.  Astrometry and $UBVRI$ photometry from \citet{2006AJ....131.2478M,2011AJ....141...28M}, $JH$ photometry from 2MASS \citep{2006AJ....131.1163S}, and Sloan $u'g'r'i'z'$ photometry computed via the transformations in \citet[see their Table~1]{2005AJ....130..873J}.  A finding chat showing the position of \novak\ and the position of these 17 calibration stars is shown in Figure~\ref{finder}.}
\end{sidewaystable}

For the 2015 eruption we also employed the newly commissioned IO:I near-infrared imager \citep{2016JATIS...2a5002B}, a Teledyne $2,048\times2,048$ Hawaii-2RG HgCdTe Array, providing a $6^{\prime}\!.27\times6^{\prime}\!.27$ field of view.  The IO:I instrument provides a fixed $H$-band filter, and each observation  comprised of $9\times60$\,s exposures using a 9 pointing ($3\times3$) dither pattern with $14^{\prime\prime}$ spacings between each pointing.  The IO:I data were reduced by a pipeline running at the telescope, this included bias subtraction, correlated double sampling, non-linearity correction, flat fielding, sky subtraction, registration, and alignment \citep[see][for details]{2016JATIS...2a5002B}.  Photometry was performed on the reduced data as described above for IO:O.  Photometric calibration was carried out using sources in the Two Micron All Sky Survey \citep[2MASS;][see Table~\ref{tab_calib}]{2006AJ....131.1163S}.

\subsection{LCOGT 2-meter photometry}\label{ftn_photometry}

The LCOGT 2\,m observing strategy was identical to that of the LT.  Here we observed through $u'$, $B$, $V$, $r'$, and $i'$-band filters using the Spectral CCD camera.  Due to weather and scheduling constraints, LCOGT observations of the 2015 eruption were only obtained on the night of 2015 Aug.\,28 UT.  The LCOGT data were pre-processed at the telescope and then reduced in an identical fashion to the LT IO:O data.  

\subsection{Ond\v{r}ejov Observatory 0.65\,m photometry}

Photometric observations at Ond\v{r}ejov started shortly after maximum brightness of the 2015 eruption of the nova on 2015 August 29.814 UT. We used the 0.65\,m telescope at the Ond\v{r}ejov Observatory (operated partly by Charles University, Prague) equipped with a Moravian Instruments G2-3200 CCD camera (using a Kodak KAF-3200ME sensor and standard $BVRI$ photometric filters) mounted at the prime focus. For each epoch, a series of numerous 90\,s exposures were taken (see Table~\ref{optical_photometry} for total exposure times for each epoch). Standard reduction procedures for raw CCD images were applied (bias and dark-frame subtraction and flat field correction) using APHOT\footnote{A synthetic aperture photometry and astrometry software package developed by M.~Velen and P.~Pravec at the Ond\v{r}ejov Observatory} \citep{1994ExA.....5..375P}. Reduced images within the same series were co-added to improve the signal-to-noise ratio and the gradient of the galaxy background was flattened using a spatial median filter via the SIPS\footnote{\url{http://ccd.mii.cz}} program. Photometric measurements of the nova were then performed using aperture photometry in APHOT. Five nearby secondary standard stars (including \#9 and \#11 listed in Table~\ref{tab_calib}) from \citet{2006AJ....131.2478M} were used to photometrically calibrate the magnitudes.  The photometry was reported in \citet{2015ATel.7969....1H} and \citet{2015ATel.7976....1H}, and is presented in Table~\ref{optical_photometry}.

\subsection{Mount Laguna Observatory 1.0\,m photometry}\label{mlo_photometry}

Photometric observations of \novak\ were carried out on 2015 Aug.\ 29 UT over a 7-hour period between Aug.\ 29.238 to Aug.\ 29.511 (within a day of the discovery of the 2015 eruption), and on Aug.\ 30.292 and 30.312, using the MLO 1\,m reflector. Exposures were taken through each of the Johnson-Cousins $B$, $V$, $R$, and $I$ filters \citep[see][$I$-band only on Aug.\ 30]{1990PASP..102.1181B}, and imaged on a Loral $2,048\times2,048$ pixel CCD camera.  The data were initially processed (bias subtracted and flat-fielded) using standard routines in the IRAF software package. The individual images for a given filter were subsequently aligned to a common coordinate system and averaged forming master $B$, $V$, $R$, and $I$-band images. Calibrated $B$, $V$, $R$, and $I$ magnitudes for \novak\ were then determined by comparing the instrumental magnitudes for the nova with those of several nearby secondary standard stars ($\#9-12$ and \#14; see Table~\ref{tab_calib}) using the IRAF {\tt apphot} package. The resulting magnitudes were reported in \citet{2015ATel.7967....1S}, \citet{2015ATel.7968....1S}, and \citet{2015ATel.7976....1H}, they are presented in Table~\ref{optical_photometry}.

\subsection{Bolshoi Teleskop Alt-azimutalnyi 6.0\,m photometry}

Additional photometry of \novak\ was collected by the Russian BTA 6\,m telescope at the Special Astrophysical Observatory in the Caucasus Mountains in the south of the Russian Federation.  The observations were conducted using the SCORPIO instrument \citep{2005AstL...31..194A} on 2015 Sep.\ 4.85 UT and Sep.\ 6.03.  Photometric calibration was conducted using stars from the \citet{2006AJ....131.2478M} catalog.  The photometry was reported in \citet{2015ATel.8033....1F} and \citet{2015ATel.8038....1H} and these data are included in Table~\ref{optical_photometry}.

\subsection{Corona Borealis Observatory 0.3\,m Telescope photometry}

Observations of \novak\ were conducted on 2015 Aug.\ 28 and 30 UT using the 0.3\,m telescope at the CBO in Kunsha Town, Ngari, Tibet, China.  On each night, a series of three $4\times600$\,s observations were taken through a $V$-band filter.  The subsequent photometry was calibrated using reference stars from the UCAC-4 catalogue \citep{2013AJ....145...44Z}.  The CBO photometry was reported in \citet{2015ATel.7974....1C} and is contained within Table~\ref{optical_photometry}.

\subsection{Nayoro Observatory of Hokkaido University 1.6\,m photometry}

\novak\ was observed by the 1.6\,m Pirka telescope at the Nayoro Observatory, Faculty of Science, Hokkaido University, Japan on the night of 2015 Aug.\ 28.  A pair of $V$-band exposures were obtained using the MSI multispectral imager \citep{2012SPIE.8446E..2OW}.  These observations were reported by \citet{2015ATel.7979....1A} and are recorded in Table~\ref{optical_photometry}.

\subsection{Okayama Astrophysical Observatory 0.5\,m photometry}

A pre-eruption upper limit for \novak\ was obtained by the OAO 0.5\,m MITSuME Telescope \citep{2005NCimC..28..755K}, equipped with an Apogee Alta U6 camera, on 2015 Aug.\ 27.677 UT.  The MITSuME observation was published in \citet{2015ATel.7979....1A} and is included in Table~\ref{optical_photometry}.

\subsection{iTelescope.net T24 photometry}

Photometric observations of \novak\ were carried out remotely with iTelescope.net utilizing the T24 telescope, a Planewave 24\,inch CDK Telescope f/6.5 and a FLI PL-9000 CCD camera, at the hosting site in Sierra Remote Observatory (SRO), Auberry, CA USA.  $V$-band observations were taken at 2015 Aug.\ 30.2041 UT, they were reported by \citet{2015ATel.7979....1A} and are presented in Table~\ref{optical_photometry}.

\section{Visible Spectroscopy}\label{app:optical_spectroscopy}

\subsection{Liverpool Telescope SPRAT spectroscopy}\label{LT_spectra}

Spectroscopy of the 2015 eruption of \novak\ was obtained on five successive nights from 2015 Aug.\ 28--Sep.\ 02 using the SPRAT spectrograph \citep{2014SPIE.9147E..8HP} in the blue optimized mode on the LT.  A slit width of $1\farcs8$ was used, yielding a spectral resolution of $\sim20$\,\AA, and a velocity resolution of $\sim1000$\,km\,s$^{-1}$ at the central wavelength of 5850\,\AA.

On the night of 2015 Aug.\ 28, following the detection of the 2015 eruption, the LT made four separate spectroscopic visits attempting $3\times900$\,s exposures each time.  The first visit occurred at Aug.\ 28.95 UT just half a day after the discovery of the eruption.  The second and third epochs at Aug.\ 29.06 UT and Aug.\ 29.13 both suffered significantly from variable, but thick, cloud and the effects of a bright, full, and nearby moon, and the data were subsequently discarded due to low signal-to-noise; the fourth visit took place at Aug.\ 29.24.  Significant spectral evolution, see Section~\ref{sec:vis_spec}, was seen between the first and fourth visits, so these spectra were not combined.  

On each of the nights of 2015 Aug.\ 29 and 30, the LT made two separate spectroscopic observations with $3\times900$\,s exposures each.  As the nova had faded substantially, and no significant evolution was seen between the spectra, the six exposures from each of these two nights were combined into a pair of spectra.  On each of the nights of 2015 Sep.\ 01 and 02, the LT made a single spectroscopic observation with $3\times1200$\,s exposures each.  The exposure time was increased from previous nights to counter the decreasing luminosity of the nova.  Spectroscopic observations on subsequent nights were not attempted as \novak\ had faded below the useful brightness of the instrument.  The nights of 2015 Aug.\ 29 and 31, Sep.\ 01 and 02 were photometric, Aug.\ 28 suffered from thick cloud, and Aug.\ 30 light cloud.  A log of the spectroscopic observations subsequently used for analysis is provided in Table~\ref{spec_log}.

Following bias subtraction, flat fielding, and cosmic ray removal, data reduction was carried out using the Starlink {\tt figaro} \citep[v5.6-6;][]{1988igbo.conf..448C} package.  Sky subtraction was accomplished in the 2D images via a linear fit of the variation of the sky emission in the spatial direction (parallel to the slit).  Following this, a simple extraction of the spectra was carried out.  No trace of residual sky emission could be detected in the extracted spectra. The extracted spectra were then wavelength calibrated using observations of a Xe arc lamp obtained directly after each exposure (rms residual $\sim1$\,\AA).  Following wavelength calibration the spectra were re-binned to a uniform wavelength scale of 6.46\,\AA\ per pixel between 4200 and 7500\,\AA.  The spectra were then co-added as described in the previous paragraphs.

The co-added spectra were flux calibrated using observations of the spectrophotometric standard \object{BD+33 2642} \citep{1977ApJ...218..767S} obtained at 2015 Aug.\ 29.90 UT (with the same spectrograph configuration and slit width), and are therefore presented in units of $F_\nu$ (mJy).  Comparison of imaging observations between the calibration night and the LT spectra show zero-point differences of $<0.1$ magnitude (i.e.\ $<10\%$). The greatest uncertainties in the flux calibration will therefore be due to slit losses caused by seeing variations and misalignment of the object with the slit.  We measure this from our repeated observations of the source on the same night to be $\sim15\%$.  Hence, we estimate a total flux uncertainty of $\sim 20\%$.

\subsection{LCOGT 2-meter spectroscopy}

We obtained a pair of spectra of the 2015 eruption of \novak\ using the Floyds spectrograph mounted on the LCOGT 2\,m, Hawaii. Floyds uses a low dispersion grating (235 lines per mm) and a cross-dispersed prism in concert to work in first and second order simultaneously, allowing for 3200--10000\,\AA\ wavelength coverage in a single exposure.  Wavelength calibration is accomplished with a HgAr lamp, and flat fielding with a combination of a Tungsten-Halogen lamp.

These spectra were reduced using the PyRAF-based {\tt floydsspec} pipeline, which rectifies, trims and extracts the spectra, performs cosmic ray removal, fringe correction, wavelength calibration, flux calibration (using a library sensitivity function and observations of a standard star observed the second night) and telluric correction.

\subsection{Kitt Peak National Observatory 4m spectroscopy}

A spectrum of \novak\ was obtained on 2015 Aug.\ 29.38 UT with the KOSMOS (Kitt Peak Ohio State Multi Object Spectrograph)on the KPNO 4m telescope. We used the blue VPH grism, a $0^{\prime\prime}\!\!.9$ slit,and imaged the spectrum onto a E2V CCD detector.  An FeAr hollow cathode lamp was used for the wavelength calibration.  The resulting spectrum, which has an integration time of 1200\,s, a wavelength range of 3806--6628\,\AA, and a dispersion of 0.689\,\AA\,pixel$^{-1}$, was processed and extracted with standard IRAF software.

\newpage

\section{\novak\ finder chart}

\begin{figure}[ht]
\begin{center}\includegraphics[width=0.88\textwidth]{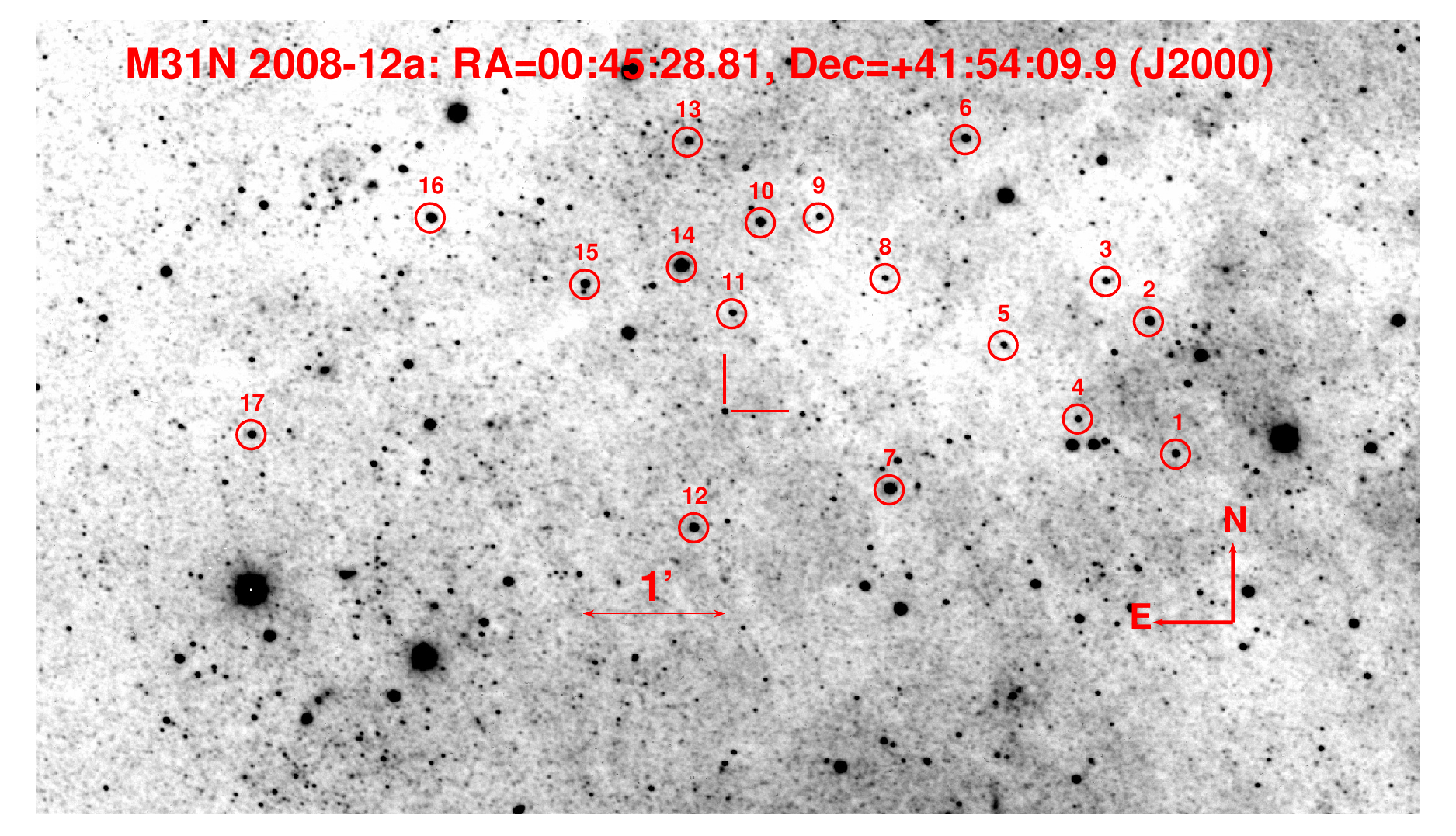}\end{center}
\caption{Eruption finding chart for the \novak\ also indicating the 17 photometry calibration stars used throughout and summarised in Table~\ref{tab_calib}.\label{finder}}
\end{figure}

In Figure~\ref{finder} we provide a finder chart indicating the position of \novak\ and the 17 photometric calibration stars (see Table~\ref{tab_calib}).  The chart is approximately 10$^\prime$ wide and 5$^\prime$ height, with North at the top, and East to the left.

\section{Observations of the 2015 eruption of \novak}

The following Tables~\ref{optical_photometry}--\ref{tab:xspec} provide full details of the observations and X-ray spectral modelling of the 2015 eruption of \novak.

\begin{table}[h]
\caption{Visible and near infrared photometric observations of the 2015 eruption of \novak.\label{optical_photometry}}
\begin{center}
\begin{tabular}{lllllllll}
\hline\hline
Date & $\Delta t$ & \multicolumn{2}{c}{MJD 57\,000+} & Telescope \& & Exposure & Filter & SNR & Photometry \\
(UT) & (days) & Start & End & Instrument & (secs) &  &  & \\
\hline
2015-08-28.971 & 0.691 & 262.969 & 262.973 & LT IO:O & $3\times120$ & $B$ & 114.3 & $18.726\pm0.011$\tablenotemark{a}\\
2015-08-29.192 & 0.912 & 263.190 & 263.195 & LT IO:O & $3\times120$ & $B$ & 148.1 & $18.654\pm0.009$\\
2015-08-29.301 & 1.921 & \multicolumn{2}{c}{263.301} & MLO 1.0m & 1200 & $B$ & \ldots & $18.71\pm0.06$\tablenotemark{b}\\
2015-08-29.405 & 1.125 & \multicolumn{2}{c}{263.405} & MLO 1.0m & 1200 & $B$ & \ldots & $18.71\pm0.06$\tablenotemark{b}\\
2015-08-29.462 & 1.182 & 263.460 & 263.465 & LCOGT Spectral & $3\times120$ & $B$ & 25.6 & $18.966\pm0.044$\\
\hline
\end{tabular}
\end{center}
\tablerefs{(a)~\citet{2015ATel.7965....1D},
(b)~\citet{2015ATel.7967....1S},
(c)~\citet{2015ATel.7969....1H},
(d)~\citet{2015ATel.7980....1D},
(e)~\citet{2015ATel.7984....1H},
(f)~\citet{2015ATel.8029....1D},
(g)~\citet{2015ATel.7979....1A},
(h)~\citet{2015ATel.7974....1C},
(i)~\citet{2015ATel.7976....1H},
(j)~\citet{2015ATel.7968....1S},
(k)~\citet{2015ATel.7964....1D},
(l)~\citet{2015ATel.8033....1F},
(m)~\citet{2015ATel.8038....1H}.}
\tablecomments{Table~\ref{optical_photometry} is published in its entirety in the machine-readable format.  A portion is shown here for guidance regarding its form and content.}
\end{table}

\begin{table}
\caption{Color evolution of the 2015 eruption of \novak.\label{colour_table}}
\begin{center}
\begin{tabular}{lllllll}
\hline\hline
Date & $t-t_{\mathrm{max}}$ & \multicolumn{2}{c}{JD 2\,456\,000.5+} & Telescope \& & Filters & Color \\
(UT) & (days) & Start & End & Instrument &  &  \\
\hline
2015-08-28.906  &       0.626   &       262.802 &       263.009 &       \swift UVOT / LT IO:I   &       $(\mathrm{uvw1}-u')$    &       $-1.066 \pm     0.190$  \\
2015-08-29.505  &       1.225   &       263.398 &       263.612 &       \swift UVOT / LT IO:I   &       $(\mathrm{uvw1}-u')$    &       $-1.100 \pm     0.176$  \\
2015-08-29.921  &       1.641   &       263.796 &       264.046 &       \swift UVOT / LT IO:I   &       $(\mathrm{uvw1}-u')$    &       $-0.848 \pm     0.151$  \\
2015-08-30.248  &       1.968   &       264.200 &       264.296 &       \swift UVOT / LT IO:I   &       $(\mathrm{uvw1}-u')$    &       $-0.542 \pm     0.102$  \\
2015-08-30.920  &       2.640   &       264.895 &       264.944 &       \swift UVOT / LT IO:I   &       $(\mathrm{uvw1}-u')$    &       $-0.566 \pm     0.301$  \\
\hline
\end{tabular}
\end{center}
\tablecomments{Table~\ref{colour_table} is published in its entirety in the machine-readable format.  A portion is shown here for guidance regarding its form and content.}
\end{table}

\begin{table}
\caption{\swift observations of nova \nova following the 2015 eruption.\label{tab:obs_swift}}
\begin{center}
\begin{tabular}{lrrrrrrrr}
\hline\hline
ObsID$^a$ & Exp$^b$ & Date$^c$ & MJD$^c$ & $\Delta t^d$ & \multicolumn{2}{c}{UV$^e$ (mag)} & Rate &  L$_{0.2-1.0}$ $^f$\\
 & (ks) & (UT) & (d) & (d) & uvm2 & uvw1 & (\power{-2} ct s$^{-1}$) & (\power{38} erg s$^{-1}$)\\
\hline
00032613104\_1 & 0.11 & 2015-08-28.01 & 57262.01 & -0.27 & $>18.9$ & \nodata & $<8.0$ & $<6.0$ \\
00032613096 & 0.7 & 2015-08-28.01 & 57262.01 & -0.27 & \nodata & $>20.2$ & $<2.0$ & $<1.5$ \\
00032613104\_2 & 0.23 & 2015-08-28.40 & 57262.40 & 0.12 & $17.3\pm0.2$ & \nodata & $<4.2$ & $<3.2$ \\
00032613097 & 0.8 & 2015-08-28.41 & 57262.41 & 0.13 & \nodata & $17.6\pm0.1$ & $<2.9$ & $<2.2$ \\
00032613104\_3 & 0.12 & 2015-08-28.60 & 57262.60 & 0.32 & $17.0\pm0.2$ & \nodata & $<7.0$ & $<5.2$ \\
\hline
\end{tabular}
\end{center}
\tablenotetext{a}{ObsIDs 104 and 105 consisted of four short exposures each immediately prior to ObsIDs 096--103.}
\tablenotetext{b}{Dead-time corrected XRT exposure time}
\tablenotetext{c}{Start date of the observation}
\tablenotetext{d}{Time in days after the eruption on 2015-08-28.28 UT (MJD 57262.28; see Section~\ref{sec:time})}
\tablenotetext{e}{\swift UVOT filter was \texttt{uvw1} (central wavelength 2600\,\AA) throughout except for one initial \texttt{uvm2} (2250\,\AA) observation consisting of four snapshots}
\tablenotetext{f}{X-ray luminosities (unabsorbed, black body fit, 0.2--10.0\,keV) and upper limits were estimated according to Section~\ref{sec:time}.}
\tablecomments{Table~\ref{tab:obs_swift} is published in its entirety in the machine-readable format.  A portion is shown here for guidance regarding its form and content.}
\end{table}

\begin{table}
\caption{X-ray model parameters for the 9 groups of spectra shown in Figure~\ref{fig:xray_group_spec} and the high/low state spectra shown in Figure~\ref{fig:xray_dip}c.}
\label{tab:xspec}
\begin{center}
\begin{tabular}{llllllllll}
\hline\hline
Group$^a$ & \multicolumn{3}{c}{Black body only} & \multicolumn{6}{c}{Black body plus emission lines}\\
\hline
ID: Days & \nh & kT & cstat & \nh & kT & \multicolumn{3}{c}{Line energies$^b$ (keV)} & cstat\\
\small{Color} & (\footnotesize{\ohcm{21}}) & (eV) & dof & (\footnotesize{\ohcm{21}}) & (eV) & \multicolumn{3}{c}{(Prominent nearby lines$^c$)} & dof\\
\hline
1: 6-7 & $0.6^{+0.7}_{-0.6}$ & $65^{+13}_{-9}$ & 66 & & & & &\\
black & & & 88 & & & & &\\[0.2cm]
2: 8 & $1.1^{+0.9}_{-1.0}$ & $82^{+18}_{-12}$ & 66 & & & &\\
orange & & & 101 & & & & &\\[0.2cm]
3: 9 & $4.0^{+1.5}_{-1.1}$ & $69^{+9}_{-9}$ & 148 & $0.6^{+0.2}_{-0.5}$ & $123^{+24}_{-13}$ & $0.52^{+0.03}_{-0.03}$ & $0.68^{+0.04}_{-0.04}$ & & 115\\
purple & & & 195 & & & \small{(N\,{\sc vii} $\alpha$)} & \small{(O\,{\sc viii} $\alpha$)} & & 173\\[0.2cm]
4: 10 & $5.5^{+2.0}_{-0.5}$ & $75^{+8}_{-12}$ & 185 & $0.3^{+0.4}_{-0.3}$ & $123^{+37}_{-35}$ & $0.55^{+0.05}_{-0.04}$ & $0.70^{+0.05}_{-0.06}$ & $0.86^{+0.04}_{-0.03}$ & 139\\
red & & & 222 & & & \small{(O\,{\sc vii} $\alpha$)} & \small{(O\,{\sc viii} $\alpha$)} & \small{(O\,{\sc viii}?)} & 183\\[0.2cm]
5: 11 & $0.9^{+0.6}_{-0.6}$ & $123^{+16}_{-12}$ & 175 & $0.7^{+0.3}_{-0.3}$ & $119^{+16}_{-31}$ & $0.41^{+0.04}_{-0.04}$ & $0.62^{+0.04}_{-0.03}$ & $0.85^{+0.03}_{-0.03}$ & 115\\
blue & & & 229 & & & \small{(N\,{\sc vi} $\alpha$)} & \small{(N\,{\sc vii} $\gamma$)} & \small{(O\,{\sc viii}?)} & 173\\[0.2cm]
6: 12 & $2.7^{+1.3}_{-0.7}$ & $96^{+9}_{-13}$ & 172 & $0.7^{+0.4}_{-0.3}$ & $110^{+6}_{-10}$ & $0.67^{+0.04}_{-0.04}$ & $0.76^{+0.05}_{-0.08}$ & $0.89^{+0.06}_{-0.05}$ & 115\\
red & & & 212 & & & \small{(O\,{\sc viii} $\alpha$)} & \small{(O\,{\sc viii} $\beta$)} & \small{(Ne\,{\sc ix} $\alpha$)} & 173\\[0.2cm]
7: 13-14 & $3.0^{+1.3}_{-1.1}$ & $87^{+15}_{-12}$ & 136 & $0.7^{+0.4}_{-0.3}$ & $103^{+14}_{-34}$ & $0.52^{+0.03}_{-0.06}$ & $0.73^{+0.03}_{-0.03}$ & $0.92^{+0.08}_{-0.06}$ & 107\\
purple & & & 166 & & &\small{(N\,{\sc vii} $\alpha$)} & \small{(O\,{\sc vii}?)} & \small{(Ne\,{\sc ix} $\alpha$)} & 133\\[0.2cm]
8: 15 & $1.5^{+1.0}_{-0.8}$ & $106^{+20}_{-15}$ & 108 & $0.8^{+0.9}_{-0.7}$ & $101^{+18}_{-17}$ & & $0.72^{+0.04}_{-0.05}$ & & 95\\
orange & & & 135 & & & & \small{(O\,{\sc vii}?)} & & 124\\[0.2cm]
9: 16-17 & $0.6^{+0.9}_{-0.6}$ & $88^{+17}_{-14}$ & 70 & & & & &\\
black & & & 79 & & & & &\\
\hline
high & $4.3^{+0.8}_{-1.1}$ & $82^{+10}_{-6}$ & 254 & $1.1^{+0.2}_{-1.1}$ & $118^{+5}_{-7}$ & $0.56^{+0.04}_{-0.03}$ & $0.72^{+0.03}_{-0.04}$ & $0.85^{+0.03}_{-0.03}$ & 193\\
red & & & 310 & & & \small{(O\,{\sc vii} $\alpha$)}& \small{(O\,{\sc vii}?)} & \small{(O\,{\sc viii}?)} & 265\\[0.2cm]
low & $0.0^{+0.4}_{-0.0}$ & $143^{+11}_{-15}$ & 98 & $1.5^{+3.4}_{-0.6}$ & $87^{+11}_{-69}$ & $0.39^{+0.07}_{-0.19}$ & $0.78^{+0.03}_{-0.05}$ & $0.91^{+0.06}_{-0.05}$ & 70\\
blue & & & 136 & & & \small{(N\,{\sc vi} $\alpha$)} & \small{(O\,{\sc viii} $\beta$)} & \small{(Ne\,{\sc ix} $\alpha$)} & 88\\
\hline
\end{tabular}
\end{center}
\tablenotetext{a}{Spectral groups are identified by their number and color in Figure \ref{fig:xray_group_spec} and the associated time-span in days post-eruption.}
\tablenotetext{b}{The quoted errors combine the statistical uncertainties, as estimated in \texttt{XSPEC}, and the calibration precision of the \swiftk/XRT energy scale \citep[$\sim0.02$\,keV, see][]{2016Min}.}
\tablenotetext{c}{Known H-like (C\,{\sc vi}, N\,{\sc vii}, O\,{\sc viii}, Ne\,{\sc x}) and He-like (N\,{\sc vi}, O\,{\sc vii}, Ne\,{\sc ix}) transitions close to the potential emission line energies. These are \textit{not} clear identifications but first tentative suggestions.}
\end{table}

\end{document}